# Recent advances in hydrogen production using sulfide-based photocatalysts

Suresh Chandra Baral[1], Dilip Sasmal[1], Mitali Hupele[1], Sradhanjali Lenka[1] and Somaditya Sen[1*]

[1]Department of Physics, Indian Institute of Technology Indore, Indore, 453552, India

*Corresponding author: sens@iiti.ac.in (SS)

## Abstract

Sulfide-based photocatalysts (PC) are promising materials for efficiently producing hydrogen ($H_2$). This chapter aims to provide a detailed survey of the recent advancements in sulfide-based photocatalysts and emphasize their enhanced performance and pathways to efficient $H_2$ production. A detailed summary has been given, including several metal sulfides, such as cadmium sulfide (CdS), zinc sulfide (ZnS), molybdenum disulfide ($MoS_2$), tungsten disulfide ($WS_2$), lead sulfide (PbS), nickel sulfides ($NiS/NiS_2$), iron disulfide ($FeS_2$), copper sulfides ($CuS/Cu_2S$), cobalt sulfides ($CoS/CoS_2$), tin disulfide ($SnS_2$), Indium sulfide ($In_2S_3$), bismuth sulfide ($Bi_2S_3$), zinc cadmium sulfide ($Zn_xCd_{1-x}S$), manganese cadmium sulfide ($Mn_xCd_{1-x}S$), zinc indium sulfide ($ZnIn_2S_4$), and Cadmium indium sulfide ($CdIn_2S_4$). This chapter will focus on the latest advancements in metal-sulfide-based materials for photocatalytic hydrogen evolution reactions (HER), taking its accelerated growth and excellent research into account. After briefly outlining the basic properties, the chapter will showcase the cutting-edge strategies and recent research progress, including the construction of heterojunctions, defect engineering, co-catalyst loading, elemental doping, and single-atom engineering, which improve the electronic structure and charge separation capabilities of metal sulfides for photocatalytic hydrogen production. A future perspective and outlook have been proposed, focusing on some key points and a standard protocol. With this knowledge, we hope sulfide-based photocatalysts can be modified and engineered to improve their efficiency and stability in future research.

1. **Introduction**

Fossil fuels have caused the global average surface temperature to reach 1.2 °C above pre-industrial levels. Limiting global warming to 1.5 °C requires cutting carbon dioxide ($CO_2$) emissions [1]. As renewable energy becomes a viable solution, the intermittent nature of solar and wind energy inconsistencies makes it challenging for existing electricity grids [2]. Hence, integrating large-scale energy storage is necessary to manage supply and power demand. Chemical energy storage is a good option due to its high energy density and capacity. However, there are challenges in large-scale energy storage, such as the need for efficient and cost-effective storage systems [3]. Hydrogen, a sustainable and clean energy fuel, is presently a topic of applied research. This fuel can be synthesized using various chemical processes, from coal or liquid fossil fuels as sources, as summarized in Figure 1 [4], [5]. However, hydrogen generated from water splitting using renewable energy sources can be a clean fuel used in fuel cells, internal combustion engines, and energy storage [6]. There are several ways to produce green hydrogen from water using renewable energy. Photovoltaic-assisted electrochemical, photoelectrochemical, and photocatalytic water-splitting systems can produce solar hydrogen from water [7]. However, photocatalytic water splitting is the most suited and simplified technique for large-scale $H_2$ production. The process includes the dispersion of semiconducting particles dispersed in solution with incident sunlight to produce hydrogen and oxygen. [8], [9].

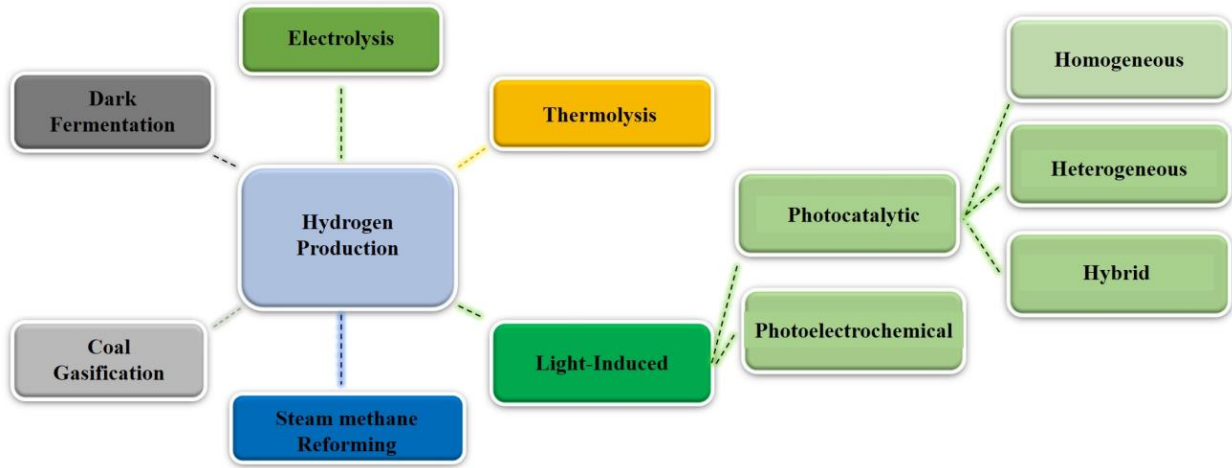

*Figure 1 shows the most common routes used to produce hydrogen. Since photocatalytic water splitting evolves, Hydrogen offers high efficiency, is environmentally benign, is cost-effective, and is environmentally benign, among other approaches.*

### 1.1. Design and Mechanisms

The photocatalytic water splitting using a semiconductor photocatalyst can be understood as follows [4]:

$$PC + h\nu\ (\geq E_g) \rightarrow PC^* + h^+ + e^- \quad (1)$$

Water Oxidation: $\quad H_2O + 2h^+ \rightarrow \tfrac{1}{2} O_2 + 2H^+ \quad\quad (+0.82\ V\ vs\ NHE,\ pH = 7) \quad (2)$

Water Reduction: $\quad 2H^+ + 2e^- \rightarrow H_2 \quad\quad\quad\quad\quad (-0.41\ V\ vs\ NHE,\ pH = 7) \quad (3)$

Overall: $\quad H_2O \rightarrow H_2 + \tfrac{1}{2} O_2 \quad\quad\quad\quad\quad\quad \Delta G = 238\ kJ/mol \quad (4)$

In general, as demonstrated in Equations (1) – (4), photocatalytic water splitting involves three crucial steps: First is the generation of excitonic pairs by photon absorption. Secondly, the electron-hole pairs separation and migration to the photocatalyst surface, and last, the splitting of water molecules via charge carriers on the surface of the photocatalyst. The essential condition to carry out the oxidation reaction is that the valence band (VB) of the photocatalyst must be more positive than the Normal hydrogen electrode (NHE) oxidation potential (+0.82 V vs. NHE) [Equation (3)]. In contrast, a more negative conduction band (CB) is necessary as compared to the hydrogen evolution potential (−0.41 V vs. NHE) to drive the reduction reactions [Equation (4)] [10]. As a result, this process produces clean and renewable hydrogen without emitting harmful gasses. In practice, a single photocatalyst must fulfill band structure and energy gap criteria to produce hydrogen and oxygen simultaneously from water splitting. Different catalysts, such as homogeneous (where the reactants and products are in the same phases, i.e., the catalyst is molecularly dispersed in the solution), heterogeneous (where the reactants and products are in different phases, i.e., a phase boundary exists between the catalyst and reactants), and hybrid [which can be categorized in different processes as (1) a catalyst activating another catalyst, (2) more than one catalyst simultaneously activating a substrate, (3) stepwise activation of a substrate by more than one catalyst, (4) tandem catalysis, (5) heterogeneous catalysis comprising more than one component, (6) combinations of two distinct modalities, such as electrochemical/transition metal catalysis and enzymatic/chemical catalysis], have been developed [11], [12]. Among all others, heterogeneous photocatalysts are known for excellent chemical stability in liquid media, high recycling or robustness, efficient isolation from the reaction media, non-hazardous nature, low material and operational cost, selective product formation, etc [13], [14], [15], [16], [17], [18], [19], [20], [21], [22], [23], [24].

Compared to metal oxides (MOs), MOs have extremely wide band gap energy because their valence band is composed entirely of deep 2p oxygen orbitals due to their confined nature, thus suffering from low mobility as hole carriers have a relatively heavy effective mass. On the other hand, Metal sulfides (MSs) usually have a visible absorption and higher charge carrier mobilities [25], [26]. A particularly notable fact is that MSs are different from MOs in chemical composition, where the main differences in chemical reactivity are attributed to atomic numbers and size differences between oxygen (O) and sulfur (S) atoms and corresponding negative divalent anions (the ionic radius of the $O^{2-}$ ranges between 1.35–1.42 Å, while the ionic radius of $S^{2-}$ is 1.84 Å) [27]. As a result, S is characterized by higher average polarizability ($α_S = 2.90 \times 10^{-24}$ cm$^3$) with respect to O ($α_O = 0.802 \times 10^{-24}$ cm$^3$) [28]. In this regard, the electronegativity of S (2.5) is significantly lower than O (3.5), which indicates that the M-S bond is more covalent than M-O [29]. Additionally, MSs possess suitable electronic band gaps, band positions, exposed active sites, high photosensitivity, large specific capacities, low redox potential, low melting point, nanocrystalline morphology, and a long lifetime compared to MOs [30]. MSs are compounds in which a sulfur anion is combined with a cationic metal or semimetal to form $M_xS_y$ with stoichiometric compositions like MS, $M_2S$, $M_3S_4$, and $MS_2$.

Recently, MSs have attracted significant attention due to their superior physical and chemical properties [31], [32], [33], [34]. MSs consist of S elements, and one or more metal elements are categorized as binary MSs (CdS, CuS, etc.), ternary MSs ($ZnIn_2S_4$, $CuInS_2$, etc.), and poly-nary MSs ($Cu_2ZnSnS_4$, etc.). In Figure 2, we have summarized the most utilized metal sulfide compounds for photocatalytic hydrogen production. Most of these compounds have been reported to have a good visible light response, sufficient active sites, and appropriate reduction potential. Moreover, emerging quantum size effects help further tunability towards fast charge transfer, enhanced excited state lifetime, etc. [35], [36]. Thus, in the last decade, tremendous work has been devoted to developing MS photocatalysts for improving solar hydrogen generation [37], [38], [39], [40], [41], [42]. MS, especially with reduced dimensions, for example, 2D nanosheets, flakes, and anisotropic structures, is the central theme of current investigations. Moreover, these low-dimensional MS structures are a growing research topic concerning increased rates of hydrogen production [43], and for example, using metals as dopants or other compounds (i.e., see Figure 2) as co-catalysts is a prominent way of enhancing photocatalytic activity.

*Figure 2: Chemical elements with sulfide compounds widely used in photocatalytic hydrogen generation (blue boxes). Elements employed as dopants (purple) and whose oxides or other compounds (red) were used to construct heterostructures with metal sulfides for hydrogen evolution are also marked. Please note that lanthanides and actinides are not shown.*

This chapter will highlight the recent advancements regarding nanostructured MS photocatalysts for hydrogen evolution as an alternative energy source. Following this general introduction, we will briefly discuss the fundamental properties of MS-based photocatalytic systems. The chapter also discusses some specific sulfide systems that have been the focus of current research on photocatalytic hydrogen production. Each subsection will present the possible and most promising routes to photocatalytic enhancement.

Specific parameters are essential to improve photocatalytic reaction yield in addition to general environmental and industrial requirements. Indeed, tremendous ongoing research focuses on developing excellent photocatalysts to meet all requirements, resulting in versatile photocatalysts and the corresponding reaction routes. All the catalytic processes involve fundamental steps like adsorption, interaction, desorption, etc., which determine the overall reaction mechanism, products, and yield. On the other hand, perfect tuning of these factors enables the design of an effective photocatalyst [Figure 3].

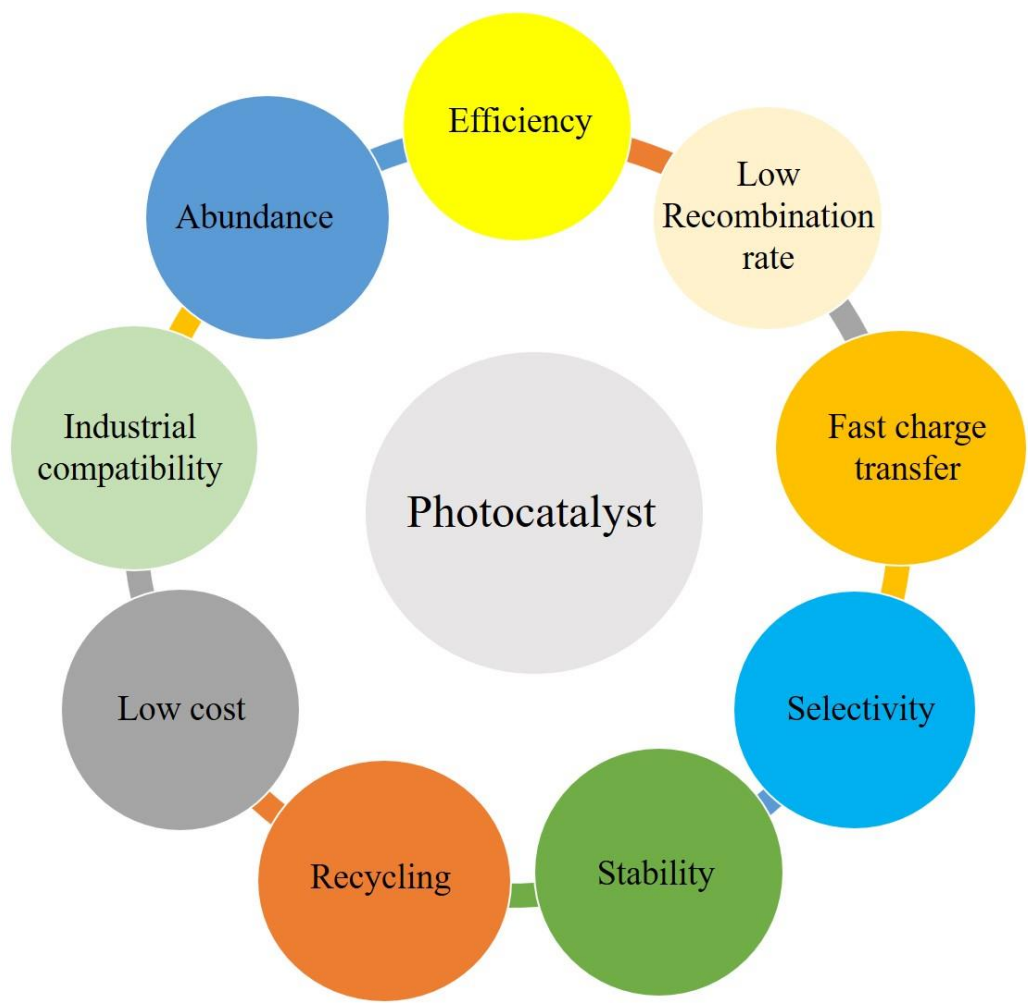

*Figure 3: General requirements for designing highly efficient photocatalysts.*

Although MS semiconductors as photocatalysts are considered economical, environmentally benign, renewable, and clean technology, their usefulness is limited due to low solar energy utilization and the rapid recombination rate of excitonic pairs after photoexcitation.

Several strategies have been developed to overcome these inherent limitations of MS photocatalysts, which we will discuss in the following parts. Among these, some general strategies and their most employed combination for metal sulfide-based photocatalysts are listed in Figure 4. Recently, semiconducting photocatalysts have been designed with new morphologies and decorations to enhance their light spectrum absorption. Research and development regarding photocatalytic hydrogen production focuses on hollow structures, porous morphologies, nanorods (NRs), 1D nanowires (NWs), 0D quantum dots (QDs), 2D sheets, and layered structures. In this regard, it has been demonstrated that the formation of mesoporous structure enhances the transport of reactants and reaction products [44]. Additionally, the porous medium enhances solar light absorption by increasing the light wave's reflection within the porous network. [45].

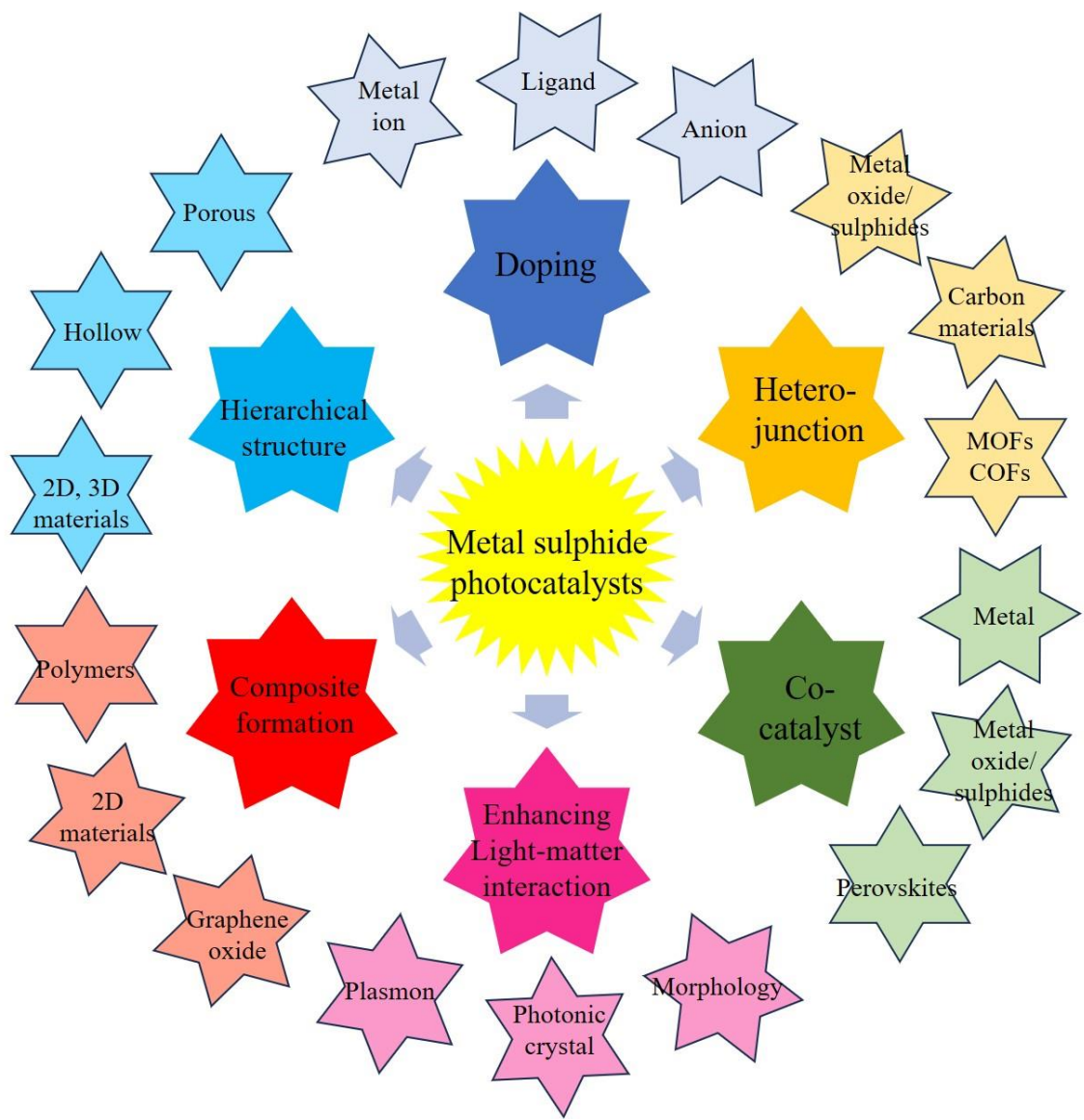

*Figure 4: Most employed strategies and representative components for enhanced hydrogen production over the MS photocatalysts. The figure covers all the techniques based on recent literature discussed in this review.*

Furthermore, porous structures have a high Brunauer-Emmett-Teller (BET) surface area and more active sites, facilitating the photocatalytic reaction. Photocatalytic enhancement requires improved solar energy utilization and efficiency to keep photogenerated charge carriers separated, which can be accomplished using low-dimensional nanostructures. As a result of size reduction, the surface areas of the particles that touch each other are increased, and interparticle interactions are strengthened, facilitating rapid charge transportation excited by the light photons. Moreover, strong electronic interaction due to emerging quantum size effects results in fast charge transfer. Also, the formed discrete and trap states at quantum particles further reduce the charge recombination rate, improving the electron-hole separation [46], [47], [48], [49]. The robustness and recycling iterations are improving with more rugged materials and protective surface coatings.

In addition, the MS photocatalysts coupled with plasmonic metallic structures have been intensively investigated over the last few years. Combining noble metal (e.g., Au, Ag, Pt) Nanoparticles (NPs) to the semiconductor nanostructures, photogenerated electrons from the CB of the semiconductor can be transferred to the adjacent metals due to the lower Fermi energy level of metals. The Schottky barrier formed at the metal/semiconductor interface further promotes electron/hole separation in photocatalysis, while the metal acts as an electron sink. The photoinduced electrons in noble metals and holes in the VB of semiconductors participate in subsequent redox reactions. The whole process increases charge carrier lifetime and improves photocatalytic efficiency. Furthermore, the unique localized surface plasmon resonance (LSPR) of noble metal NPs can boost photocatalytic activity by increasing the incoming photon absorption [50], [51]. The facile tunability of LSPR (i.e., by adjusting size and shape) allows for a broader spectrum of light absorption [52] , enhancing the photocatalytic hydrogen production rate. The noble metal, non-precious metals, metal oxides/sulfides, and metal phosphides have also been utilized [53], [54], [55], [56], [57], [58], [59], [60], [61], [62]. Again, depending on the types of co-catalyst and their size and loading methods, the performance can be different [63].

A highly active crystallographic orientation or facet is another valuable parameter for achieving a higher hydrogen evolution rate (HER). In this context, zinc-blende CdS nano-cubes with a well-controlled charge flow rate and selective transfer of photogenerated electrons and holes to the (111) and (100) facets, respectively, exhibited improved separation of charge carrier and enhancement in photocatalytic $H_2$ production [64]. A facet and morphology-dependent photocatalytic hydrogen evolution with CdS nanoflowers using a novel mixed solvothermal strategy was investigated more recently. The photocatalytic $H_2$ production increased with the increase of the exposed (0 0 2) facet, which suggested that the (0 0 2) facet of CdS played a critical role in improving the photocatalytic activity [65]. A superior photocatalytic activity was also observed by constructing hollow nano-octahedrons with exposed {101} active facets, facilitating efficient light harvesting and photoinduced charge separation of CdS photocatalysts elsewhere [66].

High-performance photocatalytic materials such as $TiO_2$ and ZnO, absorb light exclusively in the near-UV region ($\lambda = 400$ nm) due to their large band gaps. On the other hand, low-bandgap photocatalysts, such as CdS, retain photo activity for a short time due to photo corrosion as the sulfur ions are oxidized to elemental sulfur when attacked by photogenerated holes. A study on metal chalcogenide photocatalysts revealed the limiting role of hole transport in these structures. The isolation of photoinduced holes from the reaction site strategy was adopted, and positive results were obtained [67], [68], [69]. Heterojunction photocatalysts frequently have a modified band structure at the interface, better charge carrier separation, and limited charge carrier recombination compared to the photocatalysts with only one type of semiconductor. In a typical MS, Sulfide atoms near metal ions act as a catalytic site for the association and dissociation of H atoms. Therefore, increasing the number of Sulfur sites is another parameter for enhanced photocatalytic H₂ production. It is worth noting that amorphous metal sulfides often exhibit more unsaturated sulfur atoms as compared to crystalline metal sulfides, for excessive chemical reactivity, functioning as active sites for better H₂ production. Further, the band gap energies of several MS with respect to the water oxidation and reduction potential are summarized in Figure 5.

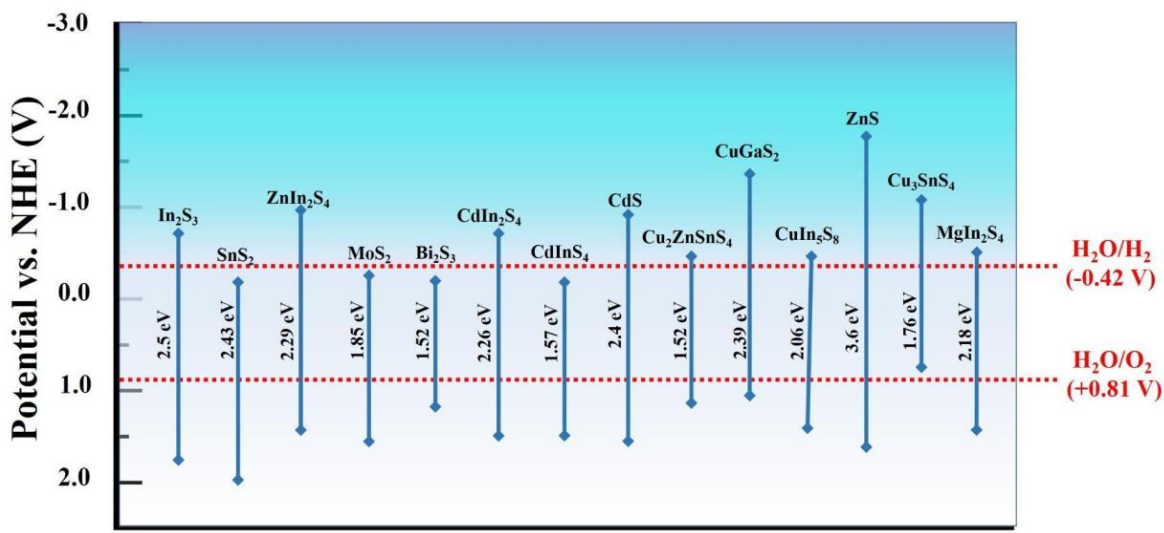

*Figure 4: The band gap positions of the selected MS semiconductors, and the required redox potentials for water splitting. Data obtained from, In$_2$S$_3$ [70], SnS$_2$ [71], ZnIn$_2$S$_4$ [72], MoS$_2$ [73], Bi$_2$S$_3$ [74], CdIn$_2$S$_4$ [75], CuInS$_2$ [76], CdS [77], Cu$_2$ZnSnS$_4$ [78], CuGaS$_2$ [79], CuIn$_5$S$_8$ [80], ZnS [81], Cu$_3$SnS$_4$ [82], MgIn$_2$S$_4$ [83].*

Morphological modifications and heterojunctions construction are the two most employed approaches to significantly increase the photocatalytic activity. The proximity between the two nanosheet subunits induces a strong interaction, facilitating the separation and transfer of charge carriers. In this regard, a reverse cation exchange-mediated and a $Cu_{2-x}S$ nano-cube template-assisted growth strategy for fabricating hollow multi-nary metal sulfide including binary compounds (CdS, ZnS, $Ag_2S$, PbS, SnS), ternary compounds ($CuInS_2$, $Zn_xCd_{1-x}S$), and quaternary compounds (single-atom platinum-attached $Zn_xCd_{1-x}S$; $Zn_xCd_{1-x}S$-$Pt_1$) was reported [84].

Again, the charge carrier recombination rate on the semiconducting photocatalytic surface is ~$2 \times 10^7$ fs, which is three times faster than the chemical oxidation/reduction reactions ($10^{10}$–$10^{11}$ fs) [85]. As a result, single-component photocatalysts are often ineffective for hydrogen generation reactions. To overcome this shortcoming, metal sulfides are combined with metal oxide catalysts. With this strategy, charge carrier recombination can be reduced in MSs, and the absorbing capacity of wide-bandgap metal oxides can be further enhanced in composite photocatalysts.

The reaction pathway follows different routes depending on the selected co-catalyst or heterojunction, mainly dictated by the chemical potential of the components [86], [87], [88]. Literature on metal sulfide photocatalysts provides multiple routes for achieving this. Figure 6 illustrates the most pronounced types of photocatalytic reactions. In type-I heterojunction, a photogenerated charge carrier establishes a built-in electric field inside the large band gap semiconductor before transmitting into the narrower band gap semiconductors; hence, electron-hole pairs cannot be spatially separated. On the contrary, rapidly generating a built-in electric field between two semiconductors in the type II heterojunctions induced a prolonged lifetime of photogenerated charge carriers.

Consequently, it is essential to facilitate type-II interfacial contact of the photocatalytic system with effective migration pathways for enhanced charge carrier separation in heterojunction (i.e., see Figure 6). Recently, a type-II hexagonal CdS/MoS$_2$ heterostructure was synthesized and tested for enhanced photocatalytic hydrogen production [89]. A NiSe$_2$/Cd$_{0.5}$Zn$_{0.5}$S type-II heterojunction photocatalyst was also synthesized, and enhanced photocatalytic hydrogen production was observed [90]. On the other hand, the Van der Waals broken band gap heterojunctions of Type-III heterostructures are less studied in the application of photocatalytic hydrogen production in two-dimensional (2D) materials. Since the photogenerated charge carriers are not synergistically coupled, electron-hole pairs decay fast. In the S-scheme heterojunction, photogenerated holes and electrons present in the valence and conduction bands, respectively participate in various redox reactions.

Meanwhile, the internal electric field and Coulomb gravity recombine the weak photogenerated carriers. Ultimately, more efficient electrons can be used for photocatalytic hydrogen reduction [91], [92], [93]. Because of their strong reduction ability, NPs of noble and transition metals (Ag, Au, Rh, Ru, Cu, Pt, etc.) serve as electron conductors/mediators between two SCs in all-solid-state Z-scheme photocatalysts. After the pioneering work of the anisotropic CdS-Au-TiO$_2$ nano junction SCs, this type of heterojunction has been extensively studied in the past decade [94]. Two photoexcitation steps are required in a direct Z-scheme photocatalytic process miming natural photosynthesis. VB of SC II generates a hole-rich region by pooling photogenerated holes, and CB of SC I forms an electron-rich region by pooling photogenerated electrons. This reduces bulk electron-hole recombination. Meanwhile, the photogenerated electrons in PC II's CB and holes in PC I's VB with less redox power recombine.

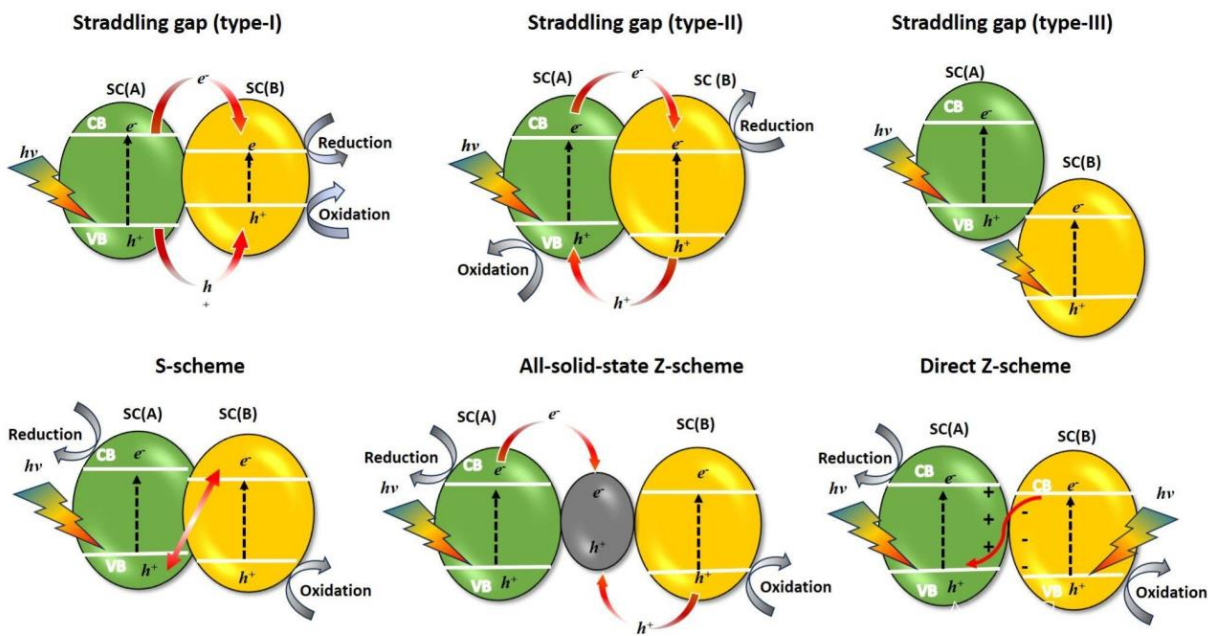

*Figure 6 illustrates the most frequently reported charge transfer mechanisms for photocatalytic systems. SC (A) and SC (B) are the two different semiconductor components of the corresponding photocatalyst.*

In recent years, sulfur vacancy engineering has been one of the effective strategies for improving photocatalytic performance by avoiding the rapid recombination of photogenerated charge carriers and severe photo corrosion [95]. The improvement has been associated with their optical and electrical properties. In a recent review, sulfur vacancy engineering in several common metal sulfide photocatalysts (e.g., binary sulfides, ternary sulfides, and other sulfides) and their reaction mechanisms in promoting charge separation and inhibiting photo corrosion are summarized [96]. Additionally, it has been reported that sulfur vacancy can modify the optical harvesting range, dynamics of charge carriers, and photoinduced surface chemical reactions, which are deeply ascribed to overall water-splitting applications [97]. Nevertheless, as we go along, we will discuss various sulfides, their heterojunctions, composites, and their derivatives separately for photocatalytic hydrogen production.

## 2. Hydrogen Generation as an Energy Material

### 2.1. Binary Metal Sulfide Compounds

MSs contain one metal element, and sulfur is classified as a binary MS. The chemical formula can be generalized as $M_xS_y$ with stoichiometric compositions like MS, $M_2S$, $M_3S_4$, and $MS_2$. Depending upon the type of metal ions and their numbers and the corresponding sulfur ions, several binary metal sulfide (BMS) compounds have been reported. A recent review summarized the details of several BMSs with different crystalline structures and the types of conductivity (N-type and P-type) [98]. Up to now, several BMS compounds, such as PbS [99], [100], CdS [101], ZnS [102], $MoS_2$ [103], $SnS_2$ [71], [104], [105], [106], [107], $Bi_2S_3$ [108], [109], $In_2S_3$ [110], $Cu_2S$ [111], [112], $NIS/NiS_2$ [113], [114], [115], and $CoS_2$ [116] as well as their derivatives and heterostructures have been widely used for photocatalytic hydrogen production. Below, we will discuss each BMS separately to make the chapter more convenient for the reader.

#### 2.1.1. Cadmium sulfide (CdS)

Cadmium sulfide (CdS) has attracted considerable research attention due to its negative conduction band position in comparison to hydrogen evolution potential, narrow band gap for visible-light response, and strong driving force for photocatalytic hydrogen production [117], [118]. Again, photo corrosion and low activity for HER reactions due to electron-hole pairs recombination are issues related to CdS that limit its practical application [119]. Previously, comprehensive materials and technology progress has been documented focusing on the CdS-based photocatalysts for applications in photocatalytic H2 generation. [120]. Several strategies are well summarized, including noble metal-CdS heterojunction, transition metal-based cocatalyst loading on CdS, CdS-based solid solutions, and CdS quantum dot photocatalysts. We have also added a detailed summary of CdS-based photocatalysts for $H_2$ production in Table 1 for the readers. Several strategies have been developed to overcome the limitation of CdS-based photocatalysts for practical applications, while loading both an $H_2$-generation cocatalyst and an $O_2$-generation cocatalyst on the surface of CdS would be an effective strategy to accelerate the photogenerated charge carrier transfer and migration. Additionally, highly efficient CdS-based Z-scheme photocatalytic water-splitting systems can inhibit backward reactions, which is a demanding opportunity for future researches [120]. More recently, an S-Scheme heterojunction was prepared using a hierarchical photocatalyst composed of a few-layer violet phosphorene (VP) and CdS NPs. In addition, a Pd single atom (SAs) co-catalyst was added, which helped enhance light-harvesting capacity and optimization of proton adsorption thermodynamics for photocatalytic hydrogen production. Ultimately, a record-breaking visible-light hydrogen production rate of 82.5 mmol $h^{-1}g^{-1}$ was attained by an optimal one wt% Pd and 5 wt% VP on CdS photocatalyst, which manifests a 54-fold increase with respect to that of CdS NPs, in addition to an apparent quantum efficiency (AQE) of 25.7% at 420 nm [121]. Several other S-scheme heterojunctions with CdS-based photocatalysts for hydrogen production can be found elsewhere [122], [123], [124]. Sun *et al.* synthesized a hollow porous CdS photocatalyst by using microporous zeolites as a host and a hard template. Additionally, ultrasmall Pd and PdS nanoparticles have been anchored separately onto the inner and outer surfaces of a hollow CdS structure. The metallic Pd pulls the photoexcited electrons away from CdS, while PdS pushes the holes for more thorough oxidation of the sacrificial agent while effectively restraining electron-hole recombination. The final Pd@CdS/PdS product exhibits a superior visible-light-driven photocatalytic $H_2$ evolution rate of up to 144.8 mmol $h^{-1}g^{-1}$. This was among the highest values of all the reported CdS-based catalysts [125]. Cheng *et al.* prepared a FeP/CdS composite photocatalyst for photo-generating $H_2$ in aqueous lactic acid solution under visible light irradiation and also got a record-breaking 202 mmol $h^{-1}g^{-1}$ for the first 5 h with visible light irradiation (light resource: $\lambda > 420$ nm, LED: $30 \times 3$ W, 28 mW $cm^{-2}$) and a 106 mmol $h^{-1}g^{-1}$ under natural solar irradiation [126]. More recently, a NiS-ReS$_2$ dual co-catalyst on CdS was tested, creating dual Z-scheme paths to improve photocatalytic hydrogen production. NiS-ReS$_2$/CdS heterostructures provided an excellent hydrogen production rate of up to 139 mmol $g^{-1}h^{-1}$ [113]. Also, double Z-scheme heterojunction ternary heterostructures of ZnS/CdS/Zn-Cd-MOF were studied elsewhere. The heterojunctions claimed to have a directional carrier migration channel that promoted electron transfer and effectively separated photogenerated carriers, which significantly improved the photocatalytic activity with a rate of hydrogen production of 27.65 mmol $g^{-1}h^{-1}$ with good stability [127]. WS$_2$ has been an excellent co-catalyst and has been reported to have several advantages when combined with CdS photocatalysts. In this regard, Wang *et al.* reported a cocatalyst 1T-WS$_2$-assisted Prussian blue derivatives Ni-CdS to enhance photocatalytic hydrogen production driven by visible light. The Ni-CdS/1T-WS$_2$ composite materials exhibited a hydrogen production rate of 62.65 mmol $g^{-1}h^{-1}$ with optimal conditions [128]. Verma *et al.* have prepared phosphorus-doped phase-modulated WS$_2$ nanosheets on CdS nanorods and achieved a high rate of hydrogen evolution of 262.12 mmol $g^{-1}h^{-1}$ [129]. Additionally, several other pieces of literature can be found in Table 1 related to CdS/WS$_2$ photocatalysts [130], [131]. Recently, Cobalt, Nickel, and Iron-based sulfides, selenides, and phosphides have emerged as excellent co-catalysts for CdS-based material for overall good photocatalytic hydrogen production [126], [132].

*Table 1: Summary of CdS-based photocatalytic for $H_2$ production*

| Photocatalysts | Co-Catalyst/ | Light source | Sacrificial reagent | $H_2$ evolution Activity ($\mu mol\ h^{-1}\ g^{-1}$) | Quantum yield (%) | Ref. |
|---|---|---|---|---|---|---|
| NiS-ReS$_2$/CdS | | 300 W Xenon lamp ($\lambda > 420$ nm) | Lactic acid | 138900 | 1.34 (420 nm) | [113] |
| VP/CdS | 1 wt% Pd | 300 W Xe lamp ($\lambda > 420$ nm) | 20 vol % lactic acid | 82500 | 25.7 (420 nm) | [121] |
| CdS | Pd and 1% PdS | 300 W Xe lamp ($\lambda > 400$ nm) | Na$_2$S/Na$_2$SO$_3$ | 144800 | 11.65 (420 nm) | [125] |
| CdS | 5 wt% FeP | Natural Solar light | 10 vol% Lactic acid | 202000 | 35 (520 nm) | [126] |
| ZnS/CdS/Zn-Cd-MOF | | 300 W Xe lamp ($\lambda > 420$ nm) | Na$_2$S/Na$_2$SO$_3$ | 27650 | 20.37 (420 nm) | [127] |
| Ni-CdS | 1T-WS$_2$ | 5 W LED | Lactic acid | 62650 | | [128] |
| CdS/WS$_2$-P | | Simulated solar irradiation (AM 1.5G) | Lactic acid | 262120 | 98.4 (425 nm) | [129] |
| CdS | MoS$_2$ | 300 W Xe lamp ($\lambda > 420$ nm) | Na$_2$S/Na$_2$SO$_3$ | 70050 | 2.104 (450 nm) | [130] |
| CdS NRs | MoS$_2$/exfoliated RGO | 150 W Xe lamp ($\lambda > 425$ nm) | lactic acid | 234000 | 47.4 | [131] |
| CdS | MoS$_2$-CoSe$_2$ | Solar light | lactic acid | 191500 | | [132] |
| CdS NPs | 2% Cu and 1% Ni  3% Cu (SPR) | Sunlight ($\geq 420$) | 5% Ethanol | 14160  12500 | 72  63.86 | [133] |
| CdS NRs | NiCoP and NiCoPi | 300 W Xe lamp (>420 nm) | 10 vol % lactic acid | 80800 | | [59] |
| CdS NPs | 5% Bi (SPR) | Visible light | | 1501.52 | | [134] |
| CdS NPs | Co(OH)$_2$ and 10% graphene | 300 W Xe lamp (≥ 420) | TEOA | 1173.53 | | [135] |
| CdS NSPs | NiMo | visible light (λ ≥ 400 nm) | 10 vol% TEOA | 2523 | | [136] |
| CdS HCs | Ni-Mo-S | 300 W Xe lamp | 10 vol% TEOA | 838.17 | | [137] |
| CdS | Hollow Au Nanosphere (SPR) Rh/Rh$_2$S$_3$ | visible light illumination (λ = 400-700 nm) | Na$_2$S/Na$_2$SO$_3$ | 2610 | 8.2% (320 nm) | [138] |
| CdS-H170 | | 300 W Xe lamp (≥ 420) | Na$_2$S/Na$_2$SO$_3$ | 26500 | 35.33% | [139] |
| CdS | 2% Cu & 1% Ag 3% Cu (SPR) 3% Ag | sunlight ≥ 420 nm | 5% Ethanol | 18930 12630 8630 | 45.04 32.51 20.71 | [140] |
| CdS | NiS and DUT-67 | 300 W Xe lamp (≥ 420) | Na$_2$S/Na$_2$SO$_3$ | 9618 | | [141] |
| CdS | 3 wt% Ni$_3$FeN | 300 W Xe lamp (≥ 420) | Na$_2$S/Na$_2$SO$_3$ | 4600 | 13.2% | [142] |
| CdS | Ni | 3 W blue-LED lamps (λ = 450 nm) | 0.1 M methanol | 10083 | | [143] |
| CdS | Ni(OH)$_2$ and Ni | 300 W Xe lamp (λ > 420 nm) | 20 vol % TEOA | 30510 | | [144] |
| CdS | Ni(OH)$_2$ | 300 W Xe lamp (λ > 420 nm) | Na$_2$S and Na$_2$SO$_3$ | 20140 | | [145] |
| Ni-doped CdS NRs | Ni | 300 W Xe lamp (λ > 420 nm) | 18 vol % lactic acid | 20600 | | [146] |
| CdS NRs | Ni-Co–S | 300 W Xe lamp (λ > 420 nm) | 10 vol % TEOA | 6560 | | [147] |
| CdS NSs | Ni@graphene core-shell | 300 W Xe lamp (λ > 420 nm) | 15 vol % TEOA | 5900 | | [148] |
| CdS | g-C$_3$N$_4$ and Ni | 300 W Xe lamp (λ > 420 nm) | 10 vol % TEOA | 4130 | | [149] |
| CdS | NiCS$_3$ | 300 W Xe lamp (λ > 420 nm) | 10 vol% Lactic acid | 61900 | | [150] |
| CdS | 10 wt% CoP | 300 W Xe lamp (λ > 420 nm) | 30 vol% Lactic acid | 23590 | | [151] |
| CdS NWs | NiSe$_2$ | Visible-light illumination (λ≥400 nm) | lactic acid | 7610.1 | 41.68 | [152] |
| CdS NRs | MoS$_2$-SnS$_2$ | Solar light | Lactic acid | 185360 | | [153] |
| CdS NRs | MoS$_2$ | | | 123000 | | |
| CdS NRs | MoS$_2$-SeS$_2$ | Solar light | Lactic acid | 168930 | | [154] |
| CdS | MoS$_2$ | 300 W Xe lamp (λ > 420 nm) | Na$_2$S/Na$_2$SO$_3$ | 19340 | | [155] |
| CdS NRs | MoS$_2$ | Natural Solar light | 20 vol% Lactic acid | 174000 | 38.7 (425 nm) | [156] |
| CdS | Ag-MoS$_2$ | 300 W Xe lamp (λ > 420 nm) | Lactic acid | 107000 | | [157] |
| CdS | MoS$_2$/rGO/CoP | Simulated Sunlight | Lactic acid | 83907 | 22.5 | [158] |
| 1D CdS/2D-NiFe LDH | | 150 W Xe lamp (λ ≥ 425 nm) | 20 vol % lactic acid | 72000 | 18.2 (425 nm) | [159] |
| CdS NRs | Pt | | | 34000 | | |
| CdSe/CdS | | 300 W Xenon lamp (λ > 420 nm) | 20 vol % TEOA | 16030 | 12.64 (420 nm) | [160] |
| ZnO/ZnS/CdS | 0.3 wt% Pt | 300 W Xenon lamp (λ > 420 nm) | Na$_2$S/Na$_2$SO$_3$ | 26400 | | [161] |
| CdS/ZIF-8 | Ni$_2$P | 1 kW Xe lamp | Na$_2$S/Na$_2$SO$_3$ | 21050 | 23.75 | [162] |

| CdS@CoSe$_2$/ WS$_2$ CdS/CoSe$_2$ CdS/WS$_2$ |  | AM 1.5G | 20 vol % Lactic acid | 246980 82500 190250 | 76.7 (425 nm) | [163] |
| Co$_2$P/CdS |  | Solar light | Lactic acid | 262160 | 2.26 (700 nm) | [164] |
| CdS | CoSe$_2$ | Simulated solar radiation | Lactic acid | 82500 | 29.2 (425 nm) | [165] |
| CdS | 2% CoP | 300 W Xenon lamp | 10% Ethanol | 56300 |  | [166] |
| Ni$_2$P/CdS |  | 300 W Xenon lamp | 10 vol % Lactic acid | 199700 | 5.3 (450 nm) | [167] |
| CdS | NiCS$_3$ | 300 W Xenon lamp (λ > 420 nm) | 10 vol % Lactic acid | 61900 |  | [150] |

*NPs = Nanoparticles, NRs = Nanorods, NWs = Nanowires, NSs = Nanosheets, NSPs = Nanospheres, HCs = Hollow cubes, and HNRs = hetero-nanorods*

### 2.1.2. Zinc sulfide (ZnS)

ZnS is one of the most widely investigated photocatalysts owing to its electron-hole pair generation capability under photoexcitation and exhibits a relatively high H$_2$ production activity under UV radiations due to the highly negative reduction potentials of excited electrons [168], [169]. Several attempts have been made to extend the optical absorption of ZnS into the visible region by doping it with transition-metal ions (Au, Ni, Cu) [170], [171]. Recently, a NiMoP$_2$ co-catalyst-modified Cu-doped ZnS was studied for enhanced photocatalytic hydrogen evolution. Interestingly, it exhibited a superior visible-light-driven photocatalytic H$_2$ evolution rate of up to 5.13 mmol h$^{-1}$g$^{-1}$. It is worth mentioning that NiMoP$_2$ acted as a non-precious metal-based co-catalyst providing active site for the reduction reaction, hence accelerating the transfer efficiency of photogenerated electrons [55]. Additionally, S-vacancies can help extend the optical absorption of ZnS into the visible region, and photocatalytic H$_2$ production under visible light has been explored earlier [172]. Nevertheless, the most reported ZnS photocatalysts suffer from relatively low charge separation efficiency because of their wide band gap (about 3.7 eV), which results in the incompetent use of solar energy. Also, ZnS suffers exceedingly from photo-corrosive decomposition [173]. To solve this inherent problem of the ZnS component, CdS is considered a promising co-constituent for photocatalytic $H_2$ production because of its low band gap (2.4 eV) and high visible light activity. Several reports can be found on ZnS/CdS-based photocatalysts for hydrogen production [161], [174], [175], [176], [177]. However, CdS still poses serious problems in photo corrosion and charge recombination under visible light. To avoid such, a $MoS_2$ cocatalyst modified CdS/ZnS core-shell nanorod heterostructures was tested for hydrogen evolution under visible-light irradiation and got a good enhancement in $H_2$ evolution rate of up to 24.1 mmol $h^{-1}g^{-1}$ [178]. Despite CdS, the heterojunction of ZnS with several other sulfides has been reported to have enhanced performance towards photocatalytic applications [173], [179], [180], [181], [182]. Recently, a ZnS/ZnSe photocatalyst has been synthesized and showed a remarkable $H_2$ generation of around 84.8 mmol $h^{-1}g^{-1}$ [183]. In addition to several binary composites, ternary composites, including ZnS, have emerged as a good strategy for overall enhancement in performance [184]. More recently, broad light absorption and multichannel charge transfer mediated by the topological surface state in the $CdS/ZnS/Bi_2Se_3$ ternary composite was tested under simulated solar light irradiation, which displayed a robust photocatalytic hydrogen production rate of 7.13 mmol $h^{-1}$ $g^{-1}$ [182]. Further, a detailed summary of ZnS-based photocatalysts for $H_2$ production can be found in Table 2.

*Table 2: Summary of ZnS-based photocatalysts for $H_2$ production*

| Photocatalysts | Co-Catalyst/ | Light source | Sacrificial reagent | $H_2$ evolution Activity (μmol $h^{-1}$ $g^{-1}$) | Quantum yield (%) | Ref. |
|---|---|---|---|---|---|---|
| ZnO/ZnS/CdS | Pt | 300 W Xenon lamp (λ > 420 nm) | Na$_2$S/Na$_2$SO$_3$ | 2640 | | [161] |
| CdS/ZnS | | 300 W Xenon lamp (λ > 420 nm) | Na$_2$S/Na$_2$SO$_3$ | 22050 | 4.68 (420 nm) | [174] |
| CdS/ZnS | Pt | 300 W Xenon lamp | Na$_2$S/Na$_2$SO$_3$ | 24100 | 9.3 (420 nm) | [177] |
| CdS/ZnS | MoS$_2$ | 350 W Xenon lamp (λ > 420 nm) | Na$_2$S/Na$_2$SO$_3$ | 13589 | 6.3 (420 nm) | [178] |
| ZnS/ZnSe | | Simulated solar radiation | Na$_2$S/Na$_2$SO$_3$ | 84800 | | [183] |
| Cu$_2$O@CdS/ZnS | | 350 W Xenon lamp (λ > 420 nm) | Na$_2$S/Na$_2$SO$_3$ | 2770 | | [185] |
| ZnS | 1 wt% Pt and 2 wt% GO | 350 W Xenon lamp (λ > 420 nm) | 20 vol % Lactic acid | 1082 | | [186] |
| CdS/ZnS | | 350 W Xenon lamp (λ > 420 nm) | Na$_2$S/Na$_2$SO$_3$ | 46630 | 9.5 (400 nm) | [187] |
| ZnO@ZnS | | 350 W Xenon lamp (λ > 420 nm) | Na$_2$S/Na$_2$SO$_3$ | 2400 | 2.58 | [188] |
| ZnS-Graphene | | 300 W mercury lamp (UV light) | Na$_2$S/Na$_2$SO$_3$ | 11600 | | [189] |

| ZnS (5%) / CoMoO$_4$ | | 5 W LED lamp (λ > 420 nm) | 10 vol % TEOA | 1233 | 1.1 (420 nm) | [190] |
| ZnS/polyaniline | | UV light | Na$_2$S/Na$_2$SO$_3$ | 6570 | | [191] |
| dark ZnS | | 300 W Xe-lamp (solar simulator) | Na$_2$S/Na$_2$SO$_3$ | 5202.4 | | [192] |
| ZnIn$_2$S$_4$@ZnS | | 350 W Xenon lamp (λ > 400 nm) | Na$_2$S/Na$_2$SO$_3$ | 2873 | | [193] |

### 2.1.3. Molybdenum disulfide (MoS$_2$)

Among different BMS, molybdenum disulfide (MoS$_2$) is extensively studied due to its unique electrical and optical properties [194], [195]. MoS$_2$ has excellent stability against photo corrosion in solution due to its antibonding state formed from an interaction between molybdenum, $d_z^2$, and sulfur $p_z$ orbitals at the top of the valence band [196]. However, the conduction band position of bulk MoS$_2$ with an indirect bandgap of 1.3 eV is slightly more positive than that for the HER, indicating that bulk MoS$_2$ cannot evolve hydrogen under light illumination without a negative bias or quantum confinement. However, as a layered structured material, the band gap of MoS$_2$ varies with the number of its layers [197]. As the thickness of MoS$_2$ decreased to the size of a monolayer, a transformation from the indirect band gap (1.3 eV) to the direct band gap (1.8-1.9 eV) was observed [198], [199], [200], [201], [202], [203]. So, due to its metallic type behavior, MoS$_2$ has been extensively studied as a noble metal-free co-catalyst for enhanced hydrogen evolution reactions [53], [54], [203]. Several other reports exist already describing the unique catalytic properties of MoS$_2$ [204], [205], [206], [207], [208], [209], [210]. A recent review summarizes the various synthesis procedures and heterojunction formations for hydrogen production, referring to several existing pieces of literature already [211]. We have also added a summary of MoS$_2$-based material in Table 3 with recent updates. To learn more details about

MoS$_2$, we suggest that the reader refer to the above paper. More recently, Li *et al.* have synthesized a MOCS cocatalyst consisting of numerous highly-dispersed MoO$_2$ NPs embedded in carbon NRs matrix (MoO$_2$-C) as cores and few-layer MoS$_2$ nanosheets (NSs) as wings grown on MoO$_2$ surface. The metallic MoO$_2$ NPs could induce electron accumulation to generate the localized surface plasmon resonance effects. As a result, the 20 wt% MOCS/CdS system showed an excellent quasi-full-spectrum driven photocatalytic H$_2$ evolution activity of 48.41 mmol h$^{-1}$g$^{-1}$ [212]. Zeng *et al.* developed an effective strategy for regulating the basal plane properties of MoS$_2$ via oxygen doping. They used oxygen-doped MoS$_2$ as a cocatalyst for OMoS/CdS heterojunction photocatalyst, showing a stronger built-in electric field and improved photocatalytic H$_2$ production activity rate of 58.47 mmol h$^{-1}$g$^{-1}$ [213]. Also, a fast carrier separation induced by the metal-like O-doped MoS$_2$/CoS cocatalyst for achieving high photocatalytic activity has been reported elsewhere [214]. The 2D/2D stacked metallic O doped MoS$_2$/CoS nanosheets exhibited a significantly improved H$_2$ production activity of 95.5 mmol g$^{-1}$h$^{-1}$ under 420 nm mono wavelength irradiation. In Table 3, we have summarized MoS$_2$-based photocatalysts for H$_2$ production.

*Table 3: Summary of MoS$_2$-based photocatalysts for H$_2$ production*

| Photocatalysts | Co-Catalyst/ | Light source | Sacrificial reagent | H$_2$ evolution Activity (μmol h$^{-1}$ g$^{-1}$) | Quantum yield (%) | Ref. |
|---|---|---|---|---|---|---|
| Zn$_{0.1}$Cd$_{0.9}$S | MoS$_2$/CoS | Xe lamp AM 1.5 G | 10 vol % Lactic acid | 95500 | 39.5 (420 nm) | [214] |
| MoS$_2$ | | 300 W Xenon lamp (λ > 420 nm) | 15 vol % TEOA | 35000 | | [215] |
| MoS$_2$/Au | | 300 W Xenon lamp (Simulated solar radiation) | Na$_2$S | 285000 | | [216] |
| MoS$_2$/CdS | | 300 W Xenon lamp (λ > 420 nm) | Lactic acid | 29000 | 77.2 (420 nm) | [217] |
| Mo$_x$S@TiO$_2$@Ti$_3$C | | 300 W Xenon lamp | TEOA | 10505.8 | | [218] |
| MoS$_2$/N–ZnO | | Sunlight | Na$_2$S/Na$_2$SO$_3$ | 17300 | | [219] |
| Zn$_{0.5}$Cd$_{0.5}$S/MoS$_2$ | | 10 W LED lamp | 10 vol % Lactic acid | 19400 | | [220] |
| MoS$_2$/Ti$_3$C$_2$/CdS | | 300 W Xenon lamp (Simulated solar radiation) | 10 vol % Lactic acid | 14100 | | [221] |
| Zn$_x$Cd$_{1-x}$S/MoS$_2$/NiS | | Sunlight | Na$_2$S/Na$_2$SO$_3$ | 41290 | 19 (435 nm) | [222] |
| MoS$_2$/Zn$_x$Cd$_{1-x}$S@Graphene | | 300 W Xenon lamp (λ > 420 nm) | Na$_2$S/Na$_2$SO$_3$ | 39500 | | [223] |
| MoS$_2$/CNTs/CdS | | 300 W Xenon lamp (λ > 420 nm) | Lactic acid | 101180 | | [224] |
| MoS$_{2+x}$ | | 300 W Xenon lamp (λ > 420 nm) | 10 vol% TEOA | 1923.5 | | [225] |
| MoS$_2$/CdS | | 300 W Xenon lamp ($\lambda$ > 420 nm) | Lactic acid | 11750 | | [226] |
| ZnO/ZnS/MoS$_2$ | | 100 W Xenon lamp | Na$_2$S/Na$_2$SO$_3$ | 10420 | | [227] |
| MoS$_2$/Cd$_{0.6}$Zn$_{0.4}$S | | 300 W Xenon lamp ($\lambda$ > 420 nm) | Na$_2$S/Na$_2$SO$_3$ | 13466.5 | 21.81 (365 nm) | [228] |
| MoS$_2$/Ni@NiO/g-C$_3$N$_4$ | | 300 W Xenon lamp ($\lambda$ > 420 nm) | 20 vol % Lactic acid | 7980 | 67.25 | [229] |
| ZnCdS/MoS$_2$ | | 10 W LED lamp | 10 vol % Lactic acid | 19400 | | [220] |

### 2.1.4. Tungsten disulfide (WS$_2$)

Similar to MoS$_2$, bulk WS$_2$ is not practical for photocatalytic hydrogen production due to its very small indirect bandgap (1-1.3 eV) and its band position compared to water redox potentials [230], [231]. Again, WS$_2$ in the monolayer form usually exhibits a bandgap around ~2 eV and is favorable for photocatalytic hydrogen evolution but demonstrates slightly less efficient electron transfer compared to MoS$_2$ [231], [232], [233], [234], [235], [236]. While the intrinsic activity at edge sites of WS$_2$ is generally lower than that of MoS$_2$, certain modifications or doping strategies can improve its performance. In this regard, Wu *et al.* have synthesized WS$_2$ nanosheets through a simple and highly reproducible mechanical activation strategy by using WO$_3$ and S as the starting materials and showed it could outperform the commonly used MoS$_2$ (JDC) catalyst in terms of activity [237]. In recent years, WS$_2$ has been utilized as a catalyst and co-catalyst with heterojunctions to several other sulfides and organic materials for enhanced photocatalytic hydrogen evolution [238], [239], [240], [241], [242], [243], [244], [245], [246], [247], [248], [249]. Recently, Li *et al.* studied interface-engineering-derived $WS_2$/S-g-$C_3N_4$ nanosheet Z-scheme heterostructures towards enhanced photocatalytic activity of 1.38 mmol $h^{-1}g^{-1}$ which was 3.7 times that of S-g-$C_3N_4$ nanosheets [250]. Seo *et al.* fabricated edge-rich 3D structuring photocatalysts involving vertically grown TMDs ($MoS_2$ and $WS_2$) nanosheets on a 3D porous graphene framework. The size-tailored heterostructure showed a superior hydrogen generation rate of 6.51 mmol $h^{-1}g^{-1}$ for $MoS_2$/3D graphene and 7.26 mmol $h^{-1}g^{-1}$ for $WS_2$/3D graphene, respectively, which were 3.59 and 3.76 times greater than that of $MoS_2$ and $WS_2$ samples [251]. Cao *et al.* prepared a few-layered 1T-$WS_2$/g-$C_3N_4$ heterostructures with a one-step calcination route and found an $H_2$ evolution rate of 10.08 mmol $h^{-1}g^{-1}$, which was 180 and 2.44 times larger than that of pure, pristine g-$C_3N_4$ and g-$C_3N_4$ (Pt), respectively [252]. Also, a First-principles study of electronic, optical adsorption, and photocatalytic water-splitting properties of a strain-tuned SiC/$WS_2$ heterojunction has been carried out, and the 4% strain on the material can show good performance in photocatalytic water splitting [253]. Nevertheless, $MoS_2$ is typically preferred for photocatalytic hydrogen production due to its superior catalytic activity and better compatibility with common photocatalysts. However, if long-term stability in harsh or oxidative conditions is a significant concern, $WS_2$ can be a viable alternative with potentially lower hydrogen production rates. Even $WS_2$ can outperform $MoS_2$ in some cases. For an overview of $WS_2$-based photocatalysts, we have added summarized details in Table 4.

*Table 4: Summary of $WS_2$-based photocatalysts for $H_2$ production*

| Photocatalysts | Co-Catalyst/ | Light source | Sacrificial reagent | $H_2$ evolution Activity ($\mu mol\ h^{-1}\ g^{-1}$) | Quantum yield (%) | Ref. |
|---|---|---|---|---|---|---|
| $CdS/WS_2/g\text{-}C_3N_4$ | | 300 W Xenon lamp ($\lambda > 420$ nm) | 10 vol% TEOA | 1174.5 | 5.4 (400 nm) | [243] |
| $CdS/WS_2/g\text{-}C_3N_4$ | | 500 W Xenon lamp ($\lambda > 420$ nm) | $Na_2S/Na_2SO_3$ | 1842 | 21.2 (420 nm) | [246] |
| $TiO_2/1T\text{-}WS_2$ | | 150 W Xe arc lamp (Simulated solar radiation) | Methanol | 2570 | | [247] |
| $WS_2$/3D graphene | | 150 W Xe arc lamp (Simulated solar radiation) | $Na_2S/Na_2SO_3$ | 3760 | | [251] |
| $1T\text{-}WS_2/g\text{-}C_3N_4$ | | 300 W Xenon lamp ($\lambda > 420$ nm) | 10 vol% TEOA | 10080 | 28 (420 nm) | [252] |
| $WS_2/g\text{-}C_3N_4$ | | 30 W LED lamp | 10 vol% TEOA | 2428.7 | | [254] |
| $WS_2/WSe_2$ | | 300 W Xenon lamp ($\lambda > 420$ nm) | $Na_2S/Na_2SO_3$ | 3856.7 | | [255] |
| $WO_3\text{-}WS_2$ | | 30 W LED lamp ($\lambda = 390\text{–}840$ nm) | 10 vol% TEOA | 5998.6 | | [256] |
| $CdS/WS_2\text{-}MoS_2$ $CdS/MoS_2$ $CdS/WS_2$ $CdS/Pt$ | | Simulated solar radiation | Lactic acid | 209790 123310 169820 34980 | 51.4 (420 nm) | [257] |

| MoSe$_2$-WS$_2$ | | 100 W LED lamp | Na$_2$S/Na$_2$SO$_3$ | 1600.2 | | [258] |
|---|---|---|---|---|---|---|
| WS$_2$/CdS | | 300 W Xenon lamp (Simulated solar radiation) | 20 vol% Lactic acid | 17730 | 67 (420 nm) | [259] |

### 2.1.5. Lead sulfides (PbS)

Lead sulfide (PbS) photocatalysts are emerging as promising materials for hydrogen production due to their unique optical and electronic properties, particularly in the visible to near-infrared (NIR) range. The bulk PbS exhibits a narrow band gap (0.41 eV), and the Bohr radius is 18 nm. The preparation methods can tune the band gap and morphology of the PbS due to the quantum confinement effect [260]. Surface modifications, doping, and forming PbS nanostructures or heterostructures with other oxides/sulfides are standard methods to boost photocatalytic activity and stability, enabling enhanced performance for hydrogen evolution reactions [261], [262], [263], [264]. In this regard, Ganapathy *et al.* prepared a heterojunction using amorphous SrTiO$_3$ and crystalline PbS for efficient photocatalytic hydrogen production. Interestingly, the amorphous-crystalline heterojunction inhibits the self-trapped excitons recombination by suppressing the electron transfer resistance, ultimately showing an efficient photocatalytic hydrogen production of 5.9 mmol h$^{-1}$g$^{-1}$ under ultraviolet irradiation [265]. More recently, experimental and DFT insights have been explored on hydrothermally synthesized PbS-doped bismuth titanate perovskites. However, it was noticed that 0.6% PbS-doped bismuth titanate (BT) revealed the highest hydrogen production of 202.43 μmol/g in the presence of EDTA [266]. Chang *et al.* prepared PbS@Cu$_2$S nanocomposites, transforming lead-acid battery waste to an efficient H$_2$ production photocatalyst with excellent H$_2$ production activity of 3.824 mmol h$^{-1}$g$^{-1}$ [267]. Nevertheless, PbS can undergo oxidation or sulfur loss, particularly in non-inert environments.

Again, its toxicity, due to lead content, poses environmental and safety concerns. Strategies to mitigate these include encapsulation techniques or replacing PbS with less toxic alternatives in hybrid systems. Although not much literature has been reported related to PbS-based photocatalysts for $H_2$ production, we have summarized a few for the reader's convenience in Table 5.

*Table 5: Summary of PbS-based photocatalysts for $H_2$ production*

| Photocatalysts | Co-Catalyst/ | Light source | Sacrificial reagent | $H_2$ evolution Activity ($\mu mol\ h^{-1}\ g^{-1}$) | Quantum yield (%) | Ref. |
|---|---|---|---|---|---|---|
| $WO_{3-x}$/PbS/Au | | 300 W Xenon lamp ($\lambda > 420$ nm) | $Na_2S/Na_2SO_3$ | 1530 | 1.5 (800 nm) | [263] |
| PbS@CuS | | 300 W Xenon lamp | $Na_2S/Na_2SO_3$ | 1736 | | [264] |
| $SrTiO_3$–PbS | | UV light | $Na_2S/Na_2SO_3$ | 5900 | | [265] |
| PbS–$K_2La_2Ti_3O_{10}$ | | 300 W mercury lamp | $Na_2S/Na_2SO_3$ | 42396 | | [268] |
| CuS/PbS/ZnO | | 300 W lamp (Simulated solar radiation) | $Na_2S/Na_2SO_3$ | 6654 | | [269] |

### 2.1.6. Nickel sulfides (**NiS/NiS$_2$**)

Among the family of metal sulfides, nickel sulfide is an appealing class that demonstrates good catalytic activity in hydrogen evolution reactions [270], [271], [272]. Nickel sulfides have various phases, such as NiS, $NiS_2$, $Ni_3S_2$, $Ni_7S_6$, and $Ni_9S_8$ [273]. However, NiS and $NiS_2$ have been studied more than others in catalytic activity, while $NiS_2$ can efficiently work as a HER catalyst and NiS as an OER catalyst [274]. Nevertheless, it is a conductive or semi-metallic material with a narrow bandgap, often used as a co-catalyst in photocatalytic systems. At the same time, $NiS_2$ is a more stable semiconductor with a moderate bandgap and pyrite structure, suitable for a broader range of electronic and photocatalytic applications. As far as photocatalytic hydrogen production is concerned, we will describe both with their heterojunctions. As mentioned before, NiS with a small band gap of 0.5 eV has been reported as a good cocatalyst for hydrogen evolution in water splitting [275], [276], [277], [278]. NiS often performs best when combined with other semiconductors, such as g-$C_3N_4$ and $TiO_2$ [279], [280], [281], [282], [283], [284]. NiS can act as a co-catalyst, improving charge separation and transferring electrons more efficiently to facilitate the hydrogen evolution reaction (HER). Several other heterojunctions of NiS with rGO [285], CdS [115], [286], [287], [288], and ZnO/ZnS [289] have been reported earlier. More recently, an excellent review has detailed the research progress of NiS Cocatalysts in Photocatalytic Applications [290]. As reported before, NiS-$ReS_2$/CdS photocatalysts have been reported to have $H_2$ evolution at a rate of 139 mmol $h^{-1}g^{-1}$ [113]. Recently, Khan *et al.* synthesized a NiS/CdS-P in which the surface phosphorus (P) atoms are incorporated in CdS NRs to create an imbalanced charge distribution and a localized built-in electric field in the crystal structure of CdS. As a result, the built-in electric field not only separates the photogenerated excitonic pairs but also accelerates the photogenerated electrons from the CdS-P conduction band to the NiS surface for rapid hydrogen reduction rather than hydrogen adsorption to avoid unfavorable H desorption. Consequently, the prepared photocatalytic displayed a significant $H_2$-generation rate of 44.39 mmol $h^{-1}g^{-1}$ and an apparent quantum efficiency of 41% at 420 nm [291]. Zhang *et al.* prepared S-vacancy-rich $Cd_{0.6}Zn_{0.4}S$ NRs, modified them with Ni-Co PBA-derived NiSx NPs as cocatalysts, and tested them for photocatalytic hydrogen production. The optimized CZS/$NiS_x$-N-1.5 hybrids (CZS/$NiS_x$-N, where N represents nickel nitrate, which was used as a nickel source) exhibited outstanding stability and activity; the average rate of $H_2$ was up to 192.82 mmol $h^{-1}g^{-1}$ in deionized water medium, which was 201.4% higher than that of CZS. Besides, $NiS_x$-A and $NiS_x$-C cocatalysts with different particle sizes were obtained using nickel acetate (A) and nickel chloride (C) as nickel sources, respectively. The CZS/$NiS_x$-C-1.5 and CZS/$NiS_x$-A-1.5 exhibited the $H_2$ evolution rate of 180.70 and 201.93 mmol $h^{-1}g^{-1}$ in a deionized water medium, respectively [292]. Another report on the adsorption of $Ni^{2+}$ on the surface of CdS by photo-deposition to obtain NiS/CdS composites has also been reported. It improved visible light response while providing abundant active sites for photocatalytic reactions. As a result, NiS/CdS exhibited a superior photocatalytic $H_2$ evolution rate of up to 7.557 mmol $h^{-1}g^{-1}$, which is approximately 4.9 times higher than that of CdS with AQE, which is 19.6% at 420 nm [293]. NiS was also deposited onto a covalent triazine framework composite, CTF-$ES_{200,}$ under xenon lamp irradiation to form heterostructures and got a hydrogen production rate as high as 22.98 mmol $h^{-1}g^{-1}$ [294]. More recently, $MS_2$/$TiO_2$(M=Ni/Co/Sn) nanocomposites have been synthesized and tested under simulated solar irradiation for photocatalytic hydrogen production. Interestingly, hydrogen productions of 48.11, 24.94, and 15.04 mmol/g were obtained with $NiS_2$/$TiO_2$, $CoS_2$/$TiO_2$, and $SnS_2$/$TiO_2$ nanomaterials in seawater [295].

### 2.1.7. Iron disulfide ($FeS_2$)

$FeS_2$ has drawn attention as a photocatalyst for hydrogen production due to its abundance, non-toxicity, and good light absorption properties [296], [297], [298], [299], [300], [301]. $FeS_2$ has a narrow bandgap of around 0.95 eV, which allows it to absorb a significant portion of visible light [302], [303]. This makes it a potentially efficient photocatalyst under sunlight. Several attempts have also been made to engineer the bandgap of $FeS_2$ [302], [304]. Other attempts have also been reported to enhance the hydrogen evolution activity of the given material [305]. $FeS_2$ suffers from rapid recombination of photogenerated electrons and holes, limiting its photocatalytic efficiency. In this regard, doping $FeS_2$ with other elements (e.g., nickel or cobalt) can promote electron transfer to improve charge separation [306], [307], [308]. To enhance its photocatalytic activity further, $FeS_2$ is often combined with other materials like $BaZrO_3$ [309], $Bi_2S_3$ [310], $TiO_2$ [296], [311], CdS [312] to improve charge separation, reduce recombination of electron-hole pairs, and stabilize the surface to improve hydrogen production efficiency. Recently, He *et al.* prepared a hollow core-shell $FeS_2/CuCo_2O_4$ Z-scheme heterostructure and tested it for photocatalytic hydrogen production. The $FeS_2$ could generate heat under near-infrared light to provide the elevated temperature required for the pyroelectric effect of $CuCo_2O_4$ (CCO) under the influence of $\Delta T$. With this thermoelectric and pyroelectric effect, CCO could release surface charges, generating spontaneous polarization, changing the electron ($e_g$) filling number on Co surface sites, reversing the spin electron state, thus controlling the carriers' migration direction, enhancing their separation and migration rate, and finally prolonging their lifetime. As a result, the as-designed FS/CCO-15 composites yielded up to 19.5 mmol $h^{-1}g^{-1}$ extraordinary photocatalytic performance, i.e., 10.2 and 6.45-fold of the pure $FeS_2$ (1.92 mmol $h^{-1}g^{-1}$) and CCO (3.02 mmol $h^{-1}g^{-1}$) under simulated sunlight, respectively [313]. Another S-scheme $FeS_2/Cd_{0.3}Mn_{0.7}S$ heterostructure has been prepared, and via solid interaction at the interface, a boosted photocatalytic hydrogen evolution rate has been reported. The heterojunction achieved a photocatalytic hydrogen evolution activity of 11.88 mmol $h^{-1}g^{-1}$, which was 66.7 and 1.5 times higher than that of $FeS_2$ and $Cd_{0.3}Mn_{0.7}S$, respectively, under the full spectrum of a xenon lamp. The enhancement in photocatalytic performance was attributed to forming the S-scheme heterojunction, which boosts the separation efficiency of charge carriers through interface engineering, thereby improving light energy utilization. Additionally, FeS$_2$ served as an efficient catalyst, providing active sites for photocatalytic hydrogen production [314]. Similar photothermal and pyroelectric effects in hollow FeS$_2$/CdS nanocomposites have been explored, but enhancement in photocatalytic hydrogen evolution rate was 154.0 μmol h$^{-1}$g$^{-1}$, which was low [315]. On the other hand, another study reported that a noble-metal-free FeS$_2$/Mn$_{0.5}$Cd$_{0.5}$S heterojunction could generate H$_2$ at a rate of 6.1 mmol h$^{-1}$g$^{-1}$. The enhancement in photocatalytic production H$_2$ activity was attributed to good carrier separation and light absorption of the optimal composite [316]. Zhang *et al.* fabricated FeS$_2$/Mn$_{0.3}$Cd$_{0.7}$S S-scheme heterojunction for enhanced photothermal-assisted photocatalytic H$_2$ evolution under full-spectrum light. Ultimately, the photocatalytic hydrogen production rate of the optimal composite was 52.017 mmol h$^{-1}$g$^{-1}$ [317]. We have summarized several reports related to FeS$_2$-based photocatalysts for H$_2$ production in Table 6.

*Table 6: Summary of FeS$_2$-based photocatalysts for H$_2$ production*

| Photocatalysts | Co-Catalyst/ | Light source | Sacrificial reagent | H$_2$ evolution Activity (μmol h$^{-1}$ g$^{-1}$) | Quantum yield (%) | Ref. |
|---|---|---|---|---|---|---|
| BaZrO$_3$–FeS$_2$ | | 450 W lamp (Simulated solar radiation) | | 4490 | | [309] |
| FeS$_2$/Bi$_2$S$_3$ | | 300 W Xenon lamp (Simulated solar radiation) | Na$_2$S/Na$_2$SO$_3$ | 16800 | 14.7 (375 nm) | [310] |
| FeS$_2$/TiO$_2$ | | UV-lamps (λ = 365 nm) | aq. Methanol | 11200 | | [296], [311] |
| FeS$_2$–TiO$_2$ | | 300 W Xenon lamp or 400 mercury arc lamps | 50 % v/v Methanol | 331 | | [296], [311] |
| CdS/FeS$_2$ | | 300 W Xenon lamp | Na$_2$S/Na$_2$SO$_3$ | 10756 | 40 | [312] |
| FeS$_2$/CuCo$_2$O$_4$ | | Simulated sunlight | | 19500 | 19.8 | [313] |
| FeS$_2$/Cd$_{0.3}$Mn$_{0.7}$S | | 300 W Xenon lamp | Na$_2$S/Na$_2$SO$_3$ | 11880 | 1.05 (365 nm) | [314] |
| FeS–FeSe$_2$ | | 300 W Xenon lamp | Na$_2$S/Na$_2$SO$_3$ | 2071.1 | | [318] |

### 2.1.8. Copper sulfides (CuS/Cu$_2$S)

Cu$_2$S (copper(I) sulfide) and CuS (copper (II) sulfide) are both essential copper sulfides with distinct properties and potential applications, including in photocatalysis and materials science, for there is a vast literature, including both materials. For the reader's convenience, we will discuss both materials separately.

### 2.1.8.1. CuS

Copper sulfide (CuS) is an intriguing material for photocatalytic activity due to its narrow bandgap, good electrical conductivity, and tunable optoelectronic properties [319], [320], [321], [322], [323], [324], [325], [326]. These characteristics make CuS particularly suitable for capturing visible light, which is a significant part of solar energy [327], [328], [329], [330]. Most CuS has been used as a co-catalyst for enhanced photocatalytic hydrogen evolution [279]. However, fast recombining electron-hole pairs in CuS limits its overall photocatalytic efficiency. Additionally,

CuS undergoes photo corrosion, which degrades under continuous illumination, reducing its long-term stability and efficiency [331]. Combining CuS with materials like TiO$_2$ [331], [332], [333], [334], [335], [336], ZnO [269], [337], [338], [339], NiO [340] and graphitic carbon nitride (g-C$_3$N$_4$) [341], [342], [343], [344] could help to improve stability and charge separation, mitigating recombination losses. Several other heterojunctions have also been reported with ZnS [179], [345], [346], [347], [348], [349], CdS [350], [351], [352], [353], MoS$_2$ [354], MXene [355], and In$_2$S$_3$ [356]. Recently, a CuS/TiO$_2$ P-N heterojunction bifunctional photocatalyst for converting benzylamine to hydrogen got an H$_2$ production rate of 881.3 μmol h$^{-1}$ g$^{-1}$ [357]. Du *et al.* prepared a CuS/ZnCdS Z-scheme heterojunction using a rapid microwave for efficient photocatalytic hydrogen evolution and got an H$_2$ production rate of 7.58 mmol h$^{-1}$ g$^{-1}$ [358]. Another is 2D CuS/CdS porous hierarchical heterostructures synthesized using a facile in-situ cation-exchange method. The unique properties of intimate interfaces and widened-and-shorted charge transfer paths for CuS/CdS 2D heterostructures synergistically improve charge transfer and separation. Besides, the enhanced light absorption caused by localized surface plasmon resonance of CuS provides more electrons to attend the reaction of H$^+$ reduction into H$_2$. Owing to those merits endowed by CuS/CdS heterostructures, the photocatalytic H$_2$ evolution reaction (HER) activity was significantly boosted to 21.5 mmol h$^{-1}$ g$^{-1}$, which was 13 times CdS [359]. Shen *et al.* prepared Cu ions doped in Mn$_{0.5}$Cd$_{0.5}$S (MCS) NRs, resulting in S-rich vacancies. Consequently, the H$_2$ production rate of Cu: MCS@CuS NRs in seawater under visible light reached 40.6 mmol h$^{-1}$g$^{-1}$ [360]. Recently, a novel CuS/TiO$_2$ catalyst has been synthesized for efficient ethanol dehydrogenation to co-produce hydrogen and acetaldehyde under mild conditions. Interestingly, the CuS/TiO$_2$ catalyst showed a H$_2$ evolution rate of 51.61 mmol h$^{-1}$ g$^{-1}$ [361]. *Table 7 Summarizes CuS-based photocatalysts for H$_2$ production.*

*Table 7: Summary of CuS-based photocatalysts for H$_2$ production*

| Photocatalysts | Co-Catalyst/ | Light source | Sacrificial reagent | $H_2$ evolution Activity ($\mu mol\ h^{-1}\ g^{-1}$) | Quantum yield (%) | Ref. |
|---|---|---|---|---|---|---|
| CuS-TiO$_2$ | | UV-LED lamp | 20 vol% v/v methanol | 2950 | 3.2 (365 nm) | [331] |
| TiO$_2$@CuS | | Full-spectrum radiation | Na$_2$S/Na$_2$SO$_3$ | 2467 | 13.4 (420 nm) | [332] |
| CuS/TiO$_2$ | | 300 W xenon arc lamp | Na$_2$S/Na$_2$SO$_3$ | 1262 | | [335] |
| CuS–ZnO/rGO/CdS | | 300 W Xenon lamp ($\lambda > 420$ nm) | Na$_2$S/Na$_2$SO$_3$ | 1073 | | [337] |
| ZnO@CuS | | 450 W arc lamp ($\lambda > 400$ nm) | Na$_2$S/Na$_2$SO$_3$ | 10113.59 | 22.3 | [339] |
| CuS-NiO | | Simulated solar (UV-visible) light irradiation | lactic acid | 52300 | 70.1 | [340] |
| CuS/ZnS/g-C$_3$N$_4$ | | 250 W halogen lamp | Na$_2$S/Na$_2$SO$_3$ | 9868 | | [344] |
| Cu$_2$O/CuS/ZnS | | Blue LED light irradiation | Na$_2$S | 1109 | | [346] |
| CeO$_2$/ZnS-CuS | | 300 W Xenon lamp | aq. Methanol | 13470 | | [347] |
| CuS–ZnS/CNTF | | 300W high-pressure mercury lamp | Na$_2$S/Na$_2$SO$_3$/NaCl | 1213.5 | | [362] |
| CuS/ZnS | | 300 W Xenon lamp ($\lambda > 420$ nm) | Na$_2$S/Na$_2$SO$_3$ | 4147 | 20 (420 nm) | [179] |
| CuS/ZnCdS | | 1000W high-pressure mercury lamp | Na$_2$S/Na$_2$SO$_3$ | 3520 | | [350] |
| CdS@CuS | | 300 W Xenon lamp (λ > 420 nm) | lactic acid | 1654.53 | 6.51 (450 nm) | [351] |
| CuS/CdS(H)/CdS(C) | | 300 W Xenon lamp (λ > 420 nm) | Na$_2$S/Na$_2$SO$_3$ | 2042 | | [353] |
| CuS–MoS$_2$ | | 300 W Xe arc lamp | Na$_2$S/Na$_2$SO$_3$ | 9648.7 | | [354] |
| CuS/In$_2$S$_3$ | | 300 W Xenon lamp (λ > 420 nm) | Na$_2$S/Na$_2$SO$_3$ | 14950 | 9.3 (420 nm) | [356] |
| CuS-TiO$_2$ | | 300 W Xe lamp (λ = 320–780 nm) | DMF/DI/benzyl amine | 881.3 | | [357] |
| CuS/ZnCdS | | 300 W Xenon lamp (λ > 420 nm) | Na$_2$S/Na$_2$SO$_3$ | 7580 | 10.07 (420 nm) | [358] |
| CuS/CdS | | 300 W Xenon lamp (λ > 420 nm) | Na$_2$S/Na$_2$SO$_3$ | 21500 | 16.76 (420 nm) | [359] |
| Cu:Mn$_{0.5}$Cd$_{0.5}$S@CuS | | 300 W Xenon lamp (λ > 420 nm) | Na$_2$S/Na$_2$SO$_3$ | 40600 | 29.5 (400 nm) | [360] |
| CuS/ TiO$_2$ | | 300 W Xenon lamp (λ > 420 nm) | Ethanol | 51610 | | [361] |

### 2.1.8.2.  Cu$_2$S

Cu$_2$S is often favored for hydrogen production due to its lower bandgap and higher efficiency in visible light absorption [111], [363]. Cu$_2$S has been utilized as a heterojunction catalyst with other photocatalysts like TiO$_2$ [112], [364], [365], [366], Cu$_2$O [367], CdS [368], [369], [370], [371], [372], [373], [374], [375], [376], K$_4$Nb$_6$O$_{17}$ [377], g-C$_3$N$_4$ [378], PbS [267], ZnS [379], [380], ZnO-ZnS [381], MoS$_2$ [382], [383], MOF [384], NiAl-LDH [385]. More recently,

Ag$_2$S/Cu$_2$S co-catalysts deposited on CdZnS and observed the 1% Ag$_2$S/Cu$_2$S on CdZnS to produce hydrogen with a 9.76 mmol g$^{-1}$h$^{-1}$ rate. Which was 3 times higher than bare CdZnS (2.91 mmol g$^{-1}$h$^{-1}$) [386]. Another, Cu$_2$S@CdS P-N junction heterostructures and Cu$_x$Cd$_{1-x}$S nanorods have been synthesized to boost solar-driven hydrogen production rates. Interestingly, the optimized Cu$_2$S@CdS composites showed an H$_2$ evolution rate of 10 mmol g$^{-1}$h$^{-1}$, while the optimized Cu$_x$Cd$_{1-x}$S composites showed a hydrogen rate of 12.1 mmol g$^{-1}$h$^{-1}$ [387]. Again, we have summarized Cu$_2$S-based photocatalysts for H$_2$ production in Table 8 for the readers.

Cu$_2$S has higher efficiency in terms of hydrogen production than CuS. CuS tends to be more stable under various conditions than Cu$_2$S, which may be prone to oxidation and other degradation processes. Nevertheless, both materials have been widely utilized in the domain of photocatalytic hydrogen production over the years. Table 8 Summarizes Cu$_2$S-based photocatalysts for H$_2$ production.

*Table 8: Summary of Cu$_2$S-based photocatalysts for H$_2$ production*

| Photocatalysts | Co-Catalyst/ | Light source | Sacrificial reagent | H$_2$ evolution Activity (μmol h$^{-1}$ g$^{-1}$) | Quantum yield (%) | Ref. |
|---|---|---|---|---|---|---|
| TiO$_2$/Cu$_2$S | | 300 W Xenon lamp (λ > 400 nm) | Na$_2$S/Na$_2$SO$_3$ | 1280 | | [111] |
| Cu$_2$S NS/TiO$_2$ | | 300 W Xenon lamp | methanol | 1430.4 | | [112] |
| CdS/Cu$_2$S | | 300 W Xenon lamp | methanol | 640.95 | 1.4 (456 nm) | [363] |
| Cu$_2$S/TiO$_2$ | | 300 W Xenon lamp (λ > 420 nm) | Na$_2$S/Na$_2$SO$_3$ | 45600 | | [364] |
| Cu$_2$S@TiO$_2$ | | 260 W Xenon lamp | H$_2$S containing wastewater + Na$_2$S/Na$_2$SO$_3$ | 41600 | 48.6 (768.6 nm) | [365] |
| b-TiO$_2$/MoS$_2$/Cu$_2$S | | 300 W Xenon lamp ($\lambda > 400$ nm) | 20 vol% v/v methanol | 3376.7 | | [366] |
| Cu$_2$S/CdS-DETA | | 300 W Xenon lamp ($\lambda > 420$ nm) | Na$_2$S/Na$_2$SO$_3$ | 9000 | 26.87 (420 nm) | [368] |
| Cu$_2$S/CdS | | 300 W Xenon lamp ($\lambda > 400$ nm) | lactic acid | 4760 | 23.6 (420 nm) | [369] |
| 2% Cu$_2$S/CZS | | 5 W LED lamp | Na$_2$S/Na$_2$SO$_3$ | 5904 | 2.13 (400 nm) | [370] |
| Cu$_2$S/Zn$_{0.5}$Cd$_{0.5}$S | | 300 W Xenon lamp ($\lambda > 400$ nm) | Na$_2$S/Na$_2$SO$_3$ | 4923.5 | 30.2 (420 nm) | [372] |
| Cu$_2$S/Ta$_2$O$_5$/CdS | | 250W QTH lamp | 10 vol% v/v lactic acid | 131000 | | [373] |
| CdS/Cu$_2$S/SiO$_2$ | | 300 W Xenon lamp ($\lambda > 420$ nm) | Na$_2$S/Na$_2$SO$_3$ | 1196.98 | | [375] |
| CdS/Cu$_2$S | | 300 W Xenon lamp ($\lambda > 420$ nm) | lactic acid | 14400 | 19.5 (420 nm) | [376] |
| Cu$_2$S/K$_4$Nb$_6$O$_{17}$ | | 300 W Xenon lamp ($\lambda > 400$ nm) | Na$_2$S/Na$_2$SO$_3$ | 2450 | | [377] |
| Cu$_2$S/ZnS | | 300 W Xenon lamp ($\lambda > 420$ nm) | Na$_2$S/Na$_2$SO$_3$ | 1000 | 17.6 (410 nm) | [379] |
| Cu$_2$S/ZnS | | Visible light | Na$_2$S/Na$_2$SO$_3$ | 2232 | | [380] |
| MoS$_2$/Cu$_2$S-TiO$_2$ | | 300 W Xenon lamp | 20 vol% v/v methanol | 1567.1 | | [382] |

| Ni-MOFs-P/Cu$_2$S | 5 W LED sunlight simulation | TEOA | 3122.76 | | [384] |
| 1 % Ag$_2$S/Cu$_2$S-CdZnS | Sunlight | Na$_2$S/Na$_2$SO$_3$ | 9760 | 32.02 (420 nm) | [386] |
| Cu$_2$S@CdS | 300 W Xenon lamp (λ > 420 nm) | 20 vol% v/v lactic acid | 10000 | 0.65 | [387] |

### 2.1.9. Cobalt sulfides (CoS, CoS$_2$)

Cobalt sulfides, both the monosulfide (CoS) and disulfide (CoS$_2$), have shown promise as a cocatalyst in photocatalytic hydrogen production, particularly due to their ability to enhance charge separation and catalytic activity when paired with primary photocatalysts [388], [389]. On the other hand, CoS$_2$ can act as an electron sink when combined with other photocatalysts, accepting electrons from the primary photocatalyst and preventing the recombination of electron-hole pairs. This extends the lifetime of charge carriers, which is crucial for increasing hydrogen evolution [390], [391], [392], [393]. In recent years, CoS$_2$ has been combined with several other photocatalysts like CdS [394], [395], [396], [397], [398], [399], [400], [401], ZnIn$_2$S$_4$ [402], [403], [404], [405], g-C$_3$N$_4$ [406], [407], [408], [409], [410], [411], [412], TiO$_2$ [413], CeO$_2$ [414], ZnS [116], Mn$_3$O$_4$ [415], and CoP [416] have been studied for photocatalytic hydrogen production. More recently, Liu *et al.* constructed a Z-scheme heterojunction and defect-engineered ZnS/CoS$_x$ nanospheres for excellent photocatalytic H$_2$ evolution performance. The enhancement of S-vacancy concentration could greatly strengthen the photocatalytic hydrogen evolution property, which can originate from the boosted direct Z-scheme charge-transfer process in ZnS/CoS$_x$ heterostructures. As a result, the ZnS/CoS$_x$ nanospheres with CoS$_x$ content of 5 wt% exhibited the optimal hydrogen production rate of 2546.6 μmol h$^{-1}$g$^{-1}$, which is 2.6 times higher than that of pure ZnS and superior to the most reported ZnS-based composites [417]. Another, ZIF-67-derived hollow CoS and $Mn_{0.2}Cd_{0.8}S$ has been synthesized to form a type-II heterojunction for boosting photocatalytic hydrogen evolution. the $CoS/Mn_{0.2}Cd_{0.8}S$ heterostructure showed a strong light-trapping ability, which enhanced charge separation and transfer due to its abundant reactive sites and heterostructure. Ultimately, the photocatalytic hydrogen production rate under visible light conditions was 43.7 mmol $h^{-1}g^{-1}$, which was 43.7 and 14.1 times higher than that of pure CoS and pure $Mn_{0.2}Cd_{0.8}S$, respectively [418]. Wang *et al.* prepared a CoNi-layered double hydroxide derived $CoS_2/NiS_2$ dodecahedron decorated with $ReS_2$ Z-scheme heterojunction for efficient hydrogen evolution. The construction of the Z-scheme heterojunction was found to facilitate the transfer and separation of photogenerated carriers while extending the lifetime of photogenerated electrons, thereby contributing to the enhancement of photocatalytic activity to a rate of 13.37 mmol $h^{-1}g^{-1}$ [419]. Another Z-scheme CoS@NC/CdS heterojunction with N-doped carbon (NC) as an electron mediator for enhanced photocatalytic $H_2$ evolution has also been reported. Interestingly, the composite exhibited the highest $H_2$ generation activity of 20.56 mmol $h^{-1}g^{-1}$ with an apparent quantum efficiency (AQE) of 23.98 %, which was 70 and 10-fold as high as that of CdS and CdS/CoS. Such improvement was attributed to the four synergisms of rapidly separating the electrons and holes by the NC electron mediator bridge in the Z-scheme system, retaining electrons at the higher position conduction band (CB) of the Z-scheme heterojunction more than that of the type-II heterojunction, increasing the specific surface area, and enhancing light absorption range and intensity [420]. As mentioned before, a fast carrier separation-induced metal-like O-doped $MoS_2/CoS$ cocatalyst for achieving high photocatalytic activity has also been reported, and the 2D/2D stacked composite exhibited a significantly improved $H_2$ production activity of 95.5 mmol $h^{-1}g^{-1}$ under 420 nm mono wavelength irradiation [214]. Another $MS_2/TiO$ nanocomposites has also been reported for photocatalytic hydrogen production. Interestingly, hydrogen productions of 48.11, 24.94, and 15.04 mmol/g were obtained with $NiS_2/TiO_2$, $CoS_2/TiO_2$, and $SnS_2/TiO_2$ nanomaterials in seawater [295]. In recent days, heterojunctions with ternary metal sulfides like $MgIn_2S_4$ [421] and $CaIn_2S_4$ [422] have also been reported. CuNi-LDH sheets and CoS nanoflakes decorated on graphitic carbon nitride heterostructure catalysts for efficient photocatalytic $H_2$ production have also been reported [423]. In-situ CoS grown on porous $g-C_3N_4$ for fluent charge transfer in photocatalytic hydrogen production. Consequently, the apparent quantum efficiency (AQE) of hydrogen evolution on 2% $CoS/g-C_3N_4$ at $\lambda = 420$ nm reaches 35.6%, and the maximum $H_2$-evolving rate reaches as high as 3.27 mmol $h^{-1}g^{-1}$ under the irradiation of natural light [424]. Another, CoS nanoparticles anchored on $g-C_3N_4$ via a facile and green strategy for boosting visible-light-driven photocatalytic hydrogen production has also been reported, but the $H_2$-evolving rate was only 532 µmol $h^{-1}g^{-1}$ under visible light [425]. In recent days, metal organic framework (MOF)-derived CoS for enhanced photocatalytic $H_2$ evolution has been reported as a new strategy. In this regard, a MOF-derived La/Gd-doped CoS has been reported that the incorporation of La or Gd in the bulk of CoS can create defect energy levels, reduce the band gap, and increase the number of active sites. Additionally, the small-sized ZIF-67 precursor can promise a smaller size of La-CoS and Gd-CoS nanoparticles with a larger specific surface area, which favors the migration of bulk phase charges. The $H_2$ evolution rates of 1.5 wt% La-CoS and 1.0 wt% Gd-CoS reach up to 3242.6 and 2916.7 µmol $h^{-1}g^{-1}$ with desired stability under artificial light source illumination; they are 3.63 times and 3.27 times higher than that of bulk CoS (893.1 µmol $h^{-1}g^{-1}$), respectively [426]. Table 8 Summarizes $CoS/CoS_2$-based photocatalysts for $H_2$ production.

*Table 8: Summary of $CoS/CoS_2$-based photocatalysts for $H_2$ production*

| Photocatalysts | Co-Catalyst/ | Light source | Sacrificial reagent | $H_2$ evolution Activity ($\mu$mol h$^{-1}$ g$^{-1}$) | Quantum yield (%) | Ref. |
|---|---|---|---|---|---|---|
| $CoS_2$/CdS | | 300 W Xenon lamp ($\lambda > 400$ nm) | $Na_2S/Na_2SO_3$ | 58000 | 39.6 (400 nm) | [394] |
| $CoS_2$/CdS | | 300 W Xenon lamp ($\lambda > 420$ nm) | $Na_2S/Na_2SO_3$ | 1688 | | [395] |
| $CoS_2$/CdS | | 300 W Xenon lamp ($\lambda > 400$ nm) | lactic acid | 3193.7 | 2.41 (420 nm) | [396] |
| $CoS_2$/CdS | | 300 W Xenon lamp | 10 % v/v lactic acid | 22150 | 14.69 (420 nm) | [397] |
| $CoS_2$/CdS | | 300 W Xe lamp ($\lambda = 420–780$ nm) | $Na_2S/Na_2SO_3$ | 2289.1 | 5.79 (380 nm) | [398] |
| $CoS_2$/CdS | | 5 W LED lamp, $\lambda \geq 420$ nm | 10 % v/v lactic acid | 25150 | 6.74 (420 nm) | [399] |
| $CoS_2$/CdS | | 300 W Xenon lamp ($\lambda > 420$ nm) | lactic acid | 65700 | 32.7 (420 nm) | [400] |
| CdS/$CoS_2$ | | Full solar spectrum irradiation | Ascorbic acid | 5540 | 10.2 (420 nm) | [401] |
| $CoS_x$–$ZnIn_2S_4$ | | 300 W Xenon lamp ($\lambda > 420$ nm) | 10 vol% TEOA | 1538.4 | 8.06 (405 nm) | [402] |
| $CoS_{1.097}$–$ZnIn_2S_4$ | | 300 W Xe lamp ($\lambda = 420–780$ nm) | 10 vol% TEOA | 2632.33 | | [403] |
| $CoS_2$–$ZnIn_2S_4$ | | 300 W Xenon lamp ($\lambda > 420$ nm) | TEOA | 22240 | 14.9 (420 nm) | [404] |
| $CoS_2$–$ZnIn_2S_4$ | | 300 W Xenon lamp ($\lambda > 350$ nm) | 20 vol% TEOA | 2768 | | [405] |
| $CoS_x$/g-$C_3N_4$ | | 300 W Xenon lamp | 10 vol% TEOA | 1596.01 | | [406] |
| CoS-CdS/g-$C_3N_4$ | | 300 W Xenon lamp ($\lambda > 420$ nm) | $Na_2S/Na_2SO_3$ | 2866 | | [407] |
| CoS/g-$C_3N_4$ | | 300 W Xenon lamp | 15 vol% TEOA | 9545 | | [408] |
| $CoS_2$/g-$C_3N_4$ | | 300 W Xenon lamp ($\lambda > 420$ nm) | 10 vol% TEOA | 1232 | | [409] |
| CoS/g-$C_3N_4$ | | 300 W Xenon lamp ($\lambda > 420$ nm) | 15 vol% TEOA | 1930 | 16.4 (420 nm) | [410] |
| $CoS_x$/g-$C_3N_4$ | | 350 W Xenon lamp ($\lambda > 400$ nm) | 20 vol% TEOA | 629 | | [411] |
| $CoS_2$-$MoS_2$/g-$C_3N_4$ | | 300 W Xenon lamp ($\lambda > 400$ nm) | 10 vol% TEOA | 315 | | [412] |
| $CoS_2$/$TiO_2$ | | 300 W Xenon lamp | Aq. Methanol | 2550 | | [413] |
| $CeO_2$/$CoS_2$ | | 5 W LED lamp | 10 vol% TEOA | 5172.20 | 3.61 (475 nm) | [414] |
| ZnS/$CoS_2$ | | Xe arc lamp (Simulated solar radiation) | $Na_2S/Na_2SO_3$ | 8001 | | [116] |
| $Mn_3O_4$/$CoS_2$ | | 5 W LED lamp ($\lambda > 420$ nm) | TEOA | 6360 | | [415] |
| $CoS_2$-CoP/g-$C_3N_4$ | | 80 W LED lamp | TEOA | 3780 | 13.1 | [416] |
| ZnS/$CoS_x$ | | 300 W Xenon lamp | $Na_2S/Na_2SO_3$ | 2546.6 | | [417] |
| CoS/$Mn_{0.2}Cd_{0.8}S$ | | 300 W Xenon lamp ($\lambda > 420$ nm) | $Na_2S/Na_2SO_3$ | 43700 | 21.55 (420 nm) | [418] |
| ReS$_2$/CoS$_2$/NiS$_2$ | | 300 W Xenon lamp | 10 vol% TEOA | 13370.6 | | [419] |
| CoS@NC/CdS | | 300 W Xenon lamp ($\lambda > 420$ nm) | 20 % v/v lactic acid | 20560 | 23.98 (420 nm) | [420] |
| Zn$_{0.1}$Cd$_{0.9}$S | MoS$_2$/CoS | Xe lamp AM 1.5 G | 10 vol % Lactic acid | 95500 | 39.5 (420 nm) | [214] |
| CoS$_2$/TiO$_2$ | | 300 W Xenon lamp | Simulated Seawater + Methanol | 9870 | | [295] |
| Cu$_1$Ni$_3$-LDH/CoS/g-C$_3$N$_4$ | | Simulated solar radiation | 10 vol% TEOA | 3981 | 20.1 | [423] |
| CoS/g-C$_3$N$_4$ | | 300 W Xenon lamp ($\lambda > 420$ nm) | TEOA | 3270 | 35.6 (420 nm) | [424] |
| 1.5 wt% La-CoS | | 300 W Xenon lamp ($\lambda > 420$ nm) | Na$_2$S/Na$_2$SO$_3$ | 3242.6 | | [426] |
| 1.0 wt% Gd-CoS | | 300 W Xenon lamp ($\lambda > 420$ nm) | Na$_2$S/Na$_2$SO$_3$ | 2916.7 | | [426] |
| CoS$_2$ | | 3 W LED lamp | TEOA | 1196 | | [427] |

### 2.1.10. Tin disulfide (SnS$_2$)

Tin disulfide (SnS$_2$) is gaining attention as a photocatalyst for hydrogen production due to its suitable band gap and high stability under light irradiation [428]. It has a band gap of around 2.2 eV, making it suitable for absorbing visible light, and a layered structure, which can enhance charge carrier mobility and separation [429]. Recent advancements in SnS$_2$ photocatalysis for hydrogen production have focused on improving its efficiency, stability, and overall performance. Strategy like doping metals like Cu [430], Cu or Ag [431], Ti [432] and composite materials synthesis like with SnS [39], $TiO_2$ [433], [434], g-$C_3N_4$ [435], [436], [437], COF [438], $SnO_2$ [439], $TiO_2$ [440], $ZrO_2$ [441], $NiWO_4$ [442], $Ag/AgVO_3$ [443], $Ag_2Mo_2O_7/CoMoO_4$ [444], CdS [105], [445], [446], [447], [448], ZnS [449], $ZnIn_2S_4$, [450], [451], $ZnIn_2S_4$-rGO [452], morphological engineering [106], [107], [453], [454], and surface modification like cocatalyst addition, Ni [455], core cell structure with $CdIn_2S_4$ [456] has been employed already. Additionally, using plasmonic nanoparticles to enhance light absorption and scattering in conjunction with $SnS_2$ has led to improved photocatalytic performance by utilizing a broader spectrum of sunlight [457]. Recently, $Zn_3In_2S_6@SnS_2$ heterostructure has been fabricated for efficient visible-light photocatalytic hydrogen evolution. Due to its unique composition advantages, the three-dimensional heterostructure $Zn_3In_2S_6@SnS_2$ exhibited a remarkable hydrogen production rate of 15443 μmol $h^{-1}g^{-1}$ without any co-catalyst, indicating high photocatalytic stability for water decomposition [458]. Cheng *et al.* constructed a layered $SnS_2$ and g-$C_3N_4$ nano-architectonics towards pollution degradation and $H_2$ generation. The optimized photocatalyst showed a hydrogen production rate of 1818.75 μmol $h^{-1}g^{-1}$ [459]. A $CuS_{1-x}/SnS_2$ Ohmic junction for enhanced photocatalytic $H_2$ evolution has also been reported. The 9% $CuS_{1-x}/SnS_2$ composite obtains the optimal $H_2$-production rate of 742.3 μmol $h^{-1}g^{-1}$ under visible light irradiation, which is 4.5 times higher than pure $SnS_2$ [460]. Xing *et al.* prepared a sulfur-vacancy-rich $SnS_2$ for g-$C_3N_4$/Vs-$SnS_2$/CdS heterojunctions and tested it for efficient photocatalytic hydrogen evolution. As a result, this catalyst showed an excellent photocatalytic hydrogen evolution rate of 2.3 mmol $h^{-1}g^{-1}$ [461]. A theoretical calculation has also suggested that a direct Z-scheme GaTe/$SnS_2$ van der Waals heterojunction with tunable electronic properties can be promising for highly efficient photocatalytic applications [462]. A cobalt-doped $SnS_2$ has been investigated as a catalyst for hydrogen production from sodium borohydride methanolysis, and a sample containing 10% Co exhibited the highest generation rate at 59382 mL min$^{-1}$g$^{-1}$ [463]. Additionally, Huang *et al.* have constructed a hollow cubic SnS$_2$/CdS nano-heterojunction for enhanced photocatalytic hydrogen evolution and an enhanced photocatalytic HER of 1424.56 μmol h$^{-1}$g$^{-1}$ reported [464]. Another green synthesis of 3D core-shell SnS$_2$/SnS-Cd$_{0.5}$Zn$_{0.5}$S multi-heterojunction for efficient photocatalytic H$_2$ evolution has also been reported. Interestingly, the optimal composite exhibited a remarkable hydrogen evolution rate of 168.85 mmol h$^{-1}$g$^{-1}$, which was 5.4 times higher than that of pristine twin CZS [465]. Lei *et al.* have reported an efficient photocatalytic H$_2$ evolution over SnS$_2$/twinned Mn$_{0.5}$Cd$_{0.5}$S hetero-homojunction with double S-scheme charge transfer routes. The synergistic impact of twinned homojunction and S-type heterojunction in 10 wt.% SnS$_2$/T-MCS composite contributed to a remarkable H$_2$ production rate of 182.82 mmol h$^{-1}$g$^{-1}$, which was 761.8 times higher than that achieved with SnS$_2$ alone (0.24 mmol mmol h$^{-1}$g$^{-1}$), as well as 5.8 times higher than that achieved with T-MCS alone (31.54 mmol mmol h$^{-1}$g$^{-1}$) [466]. A recent review can be further helpful for SnS$_2$ based photocatalysts for different photocatalytic applications including hydrogen production under visible light irradiation [467]. We have summarized the SnS$_2$-based photocatalysts for H$_2$ production in Table 9.

*Table 9: Summary of SnS$_2$-based photocatalysts for H$_2$ production*

| Photocatalysts | Co-Catalyst/ | Light source | Sacrificial reagent | H$_2$ evolution Activity (μmol h$^{-1}$ g$^{-1}$) | Quantum yield (%) | Ref. |
|---|---|---|---|---|---|---|
| 5% Cu/SnS$_{2-x}$ | | 300 W Xenon lamp (λ > 400 nm) | Na$_2$S/Na$_2$SO$_3$ | 1370 | 25.7 (425 nm) | [430] |
| Sn$_{0.7}$Ti$_{0.3}$S$_2$ | | UV-lamps (λ = 365 nm) | MeOH/H$_2$O | 49000 | | [432] |
| SnS$_2$/g-C$_3$N$_4$ | pt | 300 W Xenon lamp (λ > 420 nm) | 10 vol % TEOA | 972.6 | | [435] |
| SnS$_2$/Ag$_2$Mo$_2$O$_7$/CoMoO$_4$ | | 5 W LED light | TEOA | 1599 | | [444] |
| CdS/SnS$_2$ | | 150 W Xe lamp AM 1.5 G | 20 vol % Lactic acid | 20200 | | [445] |
| CdS/SnS$_2$ | | 300 W Xenon lamp (λ > 420 nm) | 10 vol % Lactic acid | 5180 | 59.3 (420 nm) | [446] |
| SnS$_2$/Zn$_{0.2}$Cd$_{0.8}$S | 1 wt% pt | 300 W Xenon lamp (λ > 420 nm) | TEOA | 12170 | 15.5 (420 nm) | [447] |
| CdS/MoS$_2$-SnS$_2$ | | 150 W Xe lamp AM 1.5 G | Lactic acid | 185360 | | [153] |
| ZnIn$_2$S$_4$/rGO/SnS$_2$ | | Visible light | Na$_2$S/Na$_2$SO$_3$ | 7847 | | [452] |
| SnS$_2$ | 1 wt% pt | 400 W mercury vapor lamp | Na$_2$S/Na$_2$SO$_3$ | 1306 | | [453] |
| SnS$_2$/g-C$_3$N$_4$ | | 300 W Xenon lamp (λ > 400 nm) | Na$_2$S/Na$_2$SO$_3$ | 6305.18 | | [107] |
| Ni-SnS$_2$ | 1 wt% pt | 400 W mercury vapour lamp | Na$_2$S/Na$_2$SO$_3$ | 1429.2 | 2.32 | [455] |
| Zn$_3$In$_2$S$_6$@SnS$_2$ | | 300 W Xenon lamp (λ > 420 nm) | TEOA | 15443 | 10.47 (420 nm) | [458] |
| SnS$_2$/g-C$_3$N$_4$ | | 300 W Xenon lamp | TEOA | 1818.75 | | [459] |

| g-C$_3$N$_4$/Vs-SnS$_2$/CdS | 3 wt% Pt | 300 W Xenon lamp | 20 vol % TEOA | 2300 | | [461] |
| --- | --- | --- | --- | --- | --- | --- |
| SnS$_2$/CdS | | 300 W Xenon lamp | Na$_2$S/Na$_2$SO$_3$ | 1424.56 | | [464] |
| SnS$_2$/SnS-Cd$_{0.5}$Zn$_{0.5}$S | | 300 W Xenon lamp ($\lambda$ > 420 nm) | Na$_2$S/Na$_2$SO$_3$ | 168850 | 24.78 (420 nm) | [465] |

### 2.1.11. Indium sulfide (In$_2$S$_3$)

Indium sulfide (In$_2$S$_3$) is a promising photocatalyst for hydrogen production due to its suitable band gap, good light absorption properties, and favorable electron transport characteristics [468]. In this regard, a recent review article summarized criteria for water splitting, the synthesis and morphological manipulations of In$_2$S$_3$, the synthesis of heterojunctions by coupling semiconductors to increase performance, and doping In$_2$S$_3$. In$_2$S$_3$-based heterojunctions, i.e., traditional type II, all-solid-state, and direct Z-scheme photocatalytic systems, show benefits such as larger charge separation, broad solar spectrum absorption, and amended conduction band and valence band edge potentials for maximum pollutant removal and H$_2$ production. The effect of dopant incorporation on electronic modulations of In$_2$S$_3$ has also been explained by the density functional theory [469]. However, we will try to report recent advancements related to In$_2$S$_3$-based photocatalysts in recent years. Recent advancements in In$_2$S$_3$ photocatalysts for hydrogen production have focused on enhancing their efficiency, stability, and overall performance. Strategies like doping In$_2$S$_3$ with various elements like transition metals like Ca [470], Cu [471], Pd [472], Bi, and Mo [473] or non-metals to improve its electronic properties and light absorption. Creating composites with other materials, such as carbon-based materials or other semiconductors WC [474], TiO$_2$ [475], [476], ZnO [477], In$_2$O$_3$ [110], [478], La$_2$Ti$_2$O$_7$ [479], Mo$_2$C [480], In(OH)$_3$–Cu$_2$O [481], Nb$_2$O$_5$/Nb$_2$C [482], CuS [356], CdS [471], MnS [483], MnS/PdS [484] has also been shown to enhance charge separation and reduce recombination. The synthesis of In$_2$S$_3$ in various nanostructured forms (like nanoparticles, nanorods, or nanosheets) has improved the surface area and photocatalytic activity [485]. These structures can facilitate better light absorption and increased active sites for hydrogen evolution and protective coatings for enhancing the chemical and thermal stability of In$_2$S$_3$ under reaction conditions. In recent days, defect-engineered In$_2$S$_3$ has also been reported as an emerging technology for enhancing hydrogen production [486]. A recent article combined band-gap structure optimization and vacancy modulation through a one-step hydrothermal method, In$_2$O$_3$ containing oxygen vacancy (O$_v$/In$_2$O$_3$) and introduced into In$_2$S$_3$ to promote photocatalytic hydrogen evolution. Under light irradiation, In$_2$S$_3$@Ov/In$_2$O$_3$-0.1 nanosheets hold a remarkable average H$_2$ evolution rate up to 4.04 mmol g$^{-1}$ h$^{-1}$, which was 32.14, 11.91, and 2.25-fold better than those of pristine In$_2$S$_3$, In$_2$S$_3$@O$_v$/In$_2$O$_3$-0.02, and In$_2$S$_3$@O$_v$/In$_2$O$_3$-0.25 nanosheets, respectively [487]. In-situ construction of In$_2$O$_3$/In$_2$S$_3$-CdIn$_2$S$_4$ (IOSC) Z-scheme heterojunction nanotubes for enhanced photocatalytic hydrogen production has also been reported. The rationally designed IOSC NTs displayed significantly enhanced photocatalytic H$_2$ production of 723 μmol g$^{-1}$ h$^{-1}$ under visible light irradiation (≥ 420 nm), which is much higher than that of pristine In$_2$S$_3$ and In$_2$O$_3$/In$_2$S$_3$ NTs [488]. An S-scheme heterojunction using P-doped ultrathin g-C$_3$N$_4$ and In$_2$S$_3$ has also been synthesized and tested for enhanced photocatalytic hydrogen production and degradation of ofloxacin. The P doping formed a new Fermi energy level and reduced the forbidden bandwidth. The ultrathin lamellar structure facilitated rapid charge transfer, and the formation of S-scheme heterojunction, the presence of the interfacial electric field, and band bending effectively separated strong redox-capable electron-hole pairs, thus significantly improving the photocatalytic performance. Ultimately, the optimized 50UP-C$_3$N$_4$/In$_2$S$_3$ photocatalyst with 50% UP-C$_3$N$_4$ mass fraction showed a hydrogen production rate of 12,387 μmol h$^{-1}$g$^{-1}$ [489]. Several other heterojunctions with In$_2$O$_3$ [490], [491], [492], ZnO [493], CuInS$_2$/g-C$_3$N$_4$ [494] and WO$_3$ [495] have also been reported in recent days for photocatalytic hydrogen production. Yang *et al.* designed an S-scheme In$_2$S$_3$–ZnIn$_2$S$_4$/Au heterojunction that facilitated directional spatial charge separation towards highly efficient photocatalytic hydrogen evolution. The well-designed S-scheme In$_2$S$_3$–ZnIn$_2$S$_4$/Au heterojunction showed a remarkable H$_2$ production rate of 12,369 μmol g$^{-1}$h$^{-1}$ [496]. Another, In$_2$S$_3$-modified ZnIn$_2$S$_4$ has also been reported for enhanced photogenerated carrier separation and photocatalytic hydrogen evolution, reaching 5690 μmol h$^{-1}$g$^{-1}$ under visible light, which was 8.4 times that of pure ZnIn$_2$S$_4$ [497]. A recent overview can be further assessed for In$_2$S$_3$ and In$_2$S$_3$-based photocatalysts regarding characteristics, synthesis, modifications, design strategies, and catalytic environmental application [498]. We have also added a detailed summary of In$_2$S$_3$-based photocatalysts for H$_2$ production in Table 10.

*Table 10: Summary of In$_2$S$_3$-based photocatalysts for H$_2$ production*

| Photocatalysts | Co-Catalyst/ | Light source | Sacrificial reagent | H$_2$ evolution Activity (μmol h$^{-1}$ g$^{-1}$) | Quantum yield (%) | Ref. |
|---|---|---|---|---|---|---|
| Bi, Mo co-doped In$_2$S$_3$ | Pt | 300 W Xenon lamp (λ > 420 nm) | TEOA | 5450 | 13.17 (400 nm) | [473] |
| In$_2$O$_3$/In$_2$S$_3$ | | 500 W Xenon lamp (λ > 420 nm) | Formaldehyde and NaOH | 6160 | 13.57 (450 nm) | [110] |
| In(OH)$_3$–In$_2$S$_3$–Cu$_2$O | | 5 W Blue light LED (λ$_{max}$ = 420 nm) | Na$_2$S | 1786.5 | | [481] |
| CuS/In$_2$S$_3$ | | 300 W Xenon lamp ($\lambda$ > 420 nm) | Na$_2$S/Na$_2$SO$_3$ | 14950 | 9.3 (420 nm) | [356] |
| MnS/In$_2$S$_3$ | | 300 W Xenon lamp ($\lambda$ > 420 nm) | Na$_2$S/Na$_2$SO$_3$ | 8360 | 34.2 (450 nm) | [483] |
| MnS/In$_2$S$_3$/PdS | | 300 W Xenon lamp ($\lambda$ > 420 nm) | Na$_2$S/Na$_2$SO$_3$ | 22700 | 34 (395 nm) | [484] |
| In$_2$S$_3$@Ov/In$_2$O$_3$ | 1 wt % Pt | 300 W Xenon lamp ($\lambda$ > 420 nm) | 10 % v/v lactic acid | 4040 | | [487] |
| In$_2$O$_3$/In$_2$S$_3$-CdIn$_2$S$_4$ | 3 wt % Pt | 300 W Xenon lamp ($\lambda$ > 420 nm) | 20 vol% TEOA | 723 | | [488] |
| g-C$_3$N$_4$/In$_2$S$_3$ | 1 wt % Pt | 300 W Xenon lamp (780 > $\lambda$ > 320 nm) | 10 vol% TEOA | 12387 | 6.35 (420 nm) | [489] |
| In$_2$O$_3$/In$_2$S$_3$ | | 5 W Blue light LED ($\lambda_{max}$ = 420 nm) | Na$_2$S/Na$_2$SO$_3$ | 1604 | 5.06 | [492] |
| In,S$_3$–ZnIn$_2$S$_4$/Au | | 300 W Xenon lamp ($\lambda$ > 420 nm) | Na$_2$S/Na$_2$SO$_3$ | 12369 | 4.31 (520 nm) | [496] |
| In,S$_3$–ZnIn$_2$S$_4$ | | 300 W Xenon lamp ($\lambda$ > 420 nm) | 20 vol% TEOA | 5690 | 6.79 (420 nm) | [497] |

### 2.1.12. Bismuth sulfide (Bi$_2$S$_3$)

Bismuth sulfide (Bi$_2$S$_3$) is a promising photocatalyst due to its suitable band gap (around 1.3 eV) that allows for effective absorption of visible light and good chemical stability under photocatalytic conditions, which is essential for hydrogen production, particularly in water-splitting applications [499], [500]. Over the years, several reports already exist that summarize its application in hydrogen production [501]. Recent advancements have focused on incorporating elements like Yb [502] in $Bi_2S_3$ to enhance charge separation and light absorption. Doping helps to create localized states in the band structure that facilitate better electron mobility. Other strategies include combining $Bi_2S_3$ with other semiconductors like $SrTiO_3$ [109], $TiO_2$ [502], [503], [504], [505], $ZrO_2$ [506], $WO_3$ [507], $Co_3O_4$ [508], ZnS [509], CdS [510], [511], CdZnS [512], ZnSe [513], $MoS_2$/P25 [514] or carbon-based materials like g-$C_3N_4$ [515], C [516], CNTs [507] which have improved photocatalytic performance by enhancing charge transfer and broadening the light absorption spectrum. Advances in synthesizing nanostructured forms of $Bi_2S_3$ like nanoflakes [517], nanorods [517], [518], [519], and quantum dots [502], [520] have increased surface area and active sites, leading to improved photocatalytic activity [521]. Adding materials like $MoS_2$ [514] and $Ni(OH)_2$ [522] as cocatalysts has been effective in enhancing hydrogen evolution by facilitating the reduction reaction at the surface of $Bi_2S_3$. Additionally, surface modifications using various coatings or layers have been explored to improve stability and photocatalytic efficiency [523]. Recently, a theoretical study revealed that two-dimensional $PtI_2/Bi_2S_3$ and $PtI_2/Bi_2Se_3$ heterostructures can show high solar-to-hydrogen efficiency for photocatalytic hydrogen production [524]. The coupling of the piezoelectric effect and vacancy engineering into the photocatalytic reaction process synergistically promoted carrier separation, thereby promoting the improvement of hydrogen production performance [525]. A dual piezoelectric $Bi_2S_3/Bi_{0.5}Na_{0.5}TiO_3$ (BS-12/BNT) piezo photocatalyst rich in S vacancies was synthesized by an impregnation method. The hydrogen generation rate of 5% BS-12/BNT under the combined impact of light and ultrasound was up to 1019.39 $\mu mol\ g^{-1}h^{-1}$, which was 9.5 times higher than that of pure BNT [108]. In Table 11, a summary of $Bi_2S_3$-based photocatalysts for $H_2$ production can be found further.

*Table 11: Summary of $Bi_2S_3$-based photocatalysts for $H_2$ production*

| Photocatalysts | Co-Catalyst/ | Light source | Sacrificial reagent | $H_2$ evolution Activity ($\mu mol\ h^{-1}\ g^{-1}$) | Quantum yield (%) | Ref. |
|---|---|---|---|---|---|---|
| $Bi_2S_3/Bi_{0.5}Na_{0.5}TiO_3$ | | 45 W metal halide lamp | Lactic acid | 1019.39 | | [108] |
| $SrTiO_3/Bi_2S_3$ | | UV light | $Na_2S/Na_2SO_3$ | 2566 | | [109] |
| $TiO_2/Bi_2S_3$ | | High-pressure $Hg$ pen-lamp $\lambda = 254$ nm | Aq. Methanol | 2460 | | [503] |
| $ZrO_2/Bi_2S_3$ | | High-pressure $Hg$ pen-lamp $\lambda = 254$ nm | Aq. Methanol | 4440 | | [506] |
| $CdS/Bi_2S_3$ | | 300 W Xenon lamp ($\lambda > 400$ nm) | $Na_2S/Na_2SO_3$ | 5500 | | [510] |
| $CdZnS/Bi_2S_3$ | | 300 W Xenon lamp ($\lambda > 400$ nm) | $Na_2S/Na_2SO_3$ | 16300 | 19.6 (425 nm) | [512] |
| $Bi_2S_3/MoS_2/P25$ | | 300 W Xenon lamp | Methyl alcohol | 13000 | | [514] |
| $Bi_2S_3$ FRs | | Natural sunlight | $Na_2S/Na_2SO_3$ | 8880 | | [517] |
| $Bi_2S_3$ NRs | | Natural sunlight | $Na_2S/Na_2SO_3$ | 7080 | | [517] |
| $Bi_2S_3$/glass | | 300 W Xenon lamp | Aq. KoH + $H_2S$ gas | 6418.8 | | [520] |

Although binary metal sulfides are promising photocatalysts for hydrogen production, they still suffer from some drawbacks, like photo corrosion, fast charge carrier recombination, and particle aggregation. Several strategies like defect engineering [96], heterojunction formation [526], [527], doping [42], and surface modification [528], [529] have been employed to overcome these limitations. However, in recent years, ternary metal sulfide compounds have emerged as alternative and promising ways to enhance the overall efficiency of sulfide-based catalysts. We will discuss ternary metal sulfide compounds in the subsequent sections.

## 2.2. Ternary Metal Sulfide Compounds

Bimetallic sulfides are promising materials for photocatalytic hydrogen production due to their unique properties like improved light absorption, enhanced charge separation, and tailored bandgap that can enhance performance compared to their monometallic counterparts. Additionally, two different metals can create synergistic effects that enhance catalytic activity, stability, and resistance to photo-corrosion [530], [531]. In brief, bimetallic sulfides are commonly referred to as ternary metal sulfide compounds in the form of $AB_xS_y$, where A and B can be metals or metalloids. And S is sulfur. Among other Metal sulfide materials, ternary sulfides such as $Zn_xCd_{1-x}S$, $M_xCd_{1-x}S$, $ZnIn_2S_4$, and $CdIn_2S_4$ have been shown to acquire higher selectivity and significant photocatalytic production rate of hydrogen [4], [87]. The exploration of such ternary semiconductor materials has gained particular interest because of their tunable optical properties, which are influenced by constituent mole fraction, particle size, and morphology. We will try to discuss commonly explored ternary metal sulfide in the upcoming sections.

### 2.2.1. $Zn_xCd_{1-x}S$

The solid solution of the narrow band gap CdS (2.4 eV) and wide band gap ZnS (3.6 eV) can be referred to as a ternary metal sulfide $Zn_xCd_{1-x}S$ (ZnCdS), which has offered several advantages over its parent constituent BMS [532]. ZnCdS has been widely studied for its flexible band gap and band edge position. Additionally, $Zn_xCd_{1-x}S$ can provide a strong visible light response and high resistance to photo corrosion for photocatalytic hydrogen production. In this regard, Chen *et al.* have synthesized hollow ZnCdS dodecahedral cages as solid solutions for highly efficient visible-light-driven hydrogen generation. The suitable band matching and strong electron coupling in the solid solutions could be simultaneously achieved via ion exchange, featuring the balance between the light absorption ability and the potential of the conduction band of the photocatalyst and achieving a hydrogen production rate of 5.68 mmol $h^{-1}g^{-1}$ under cocatalyst-free and visible-light irradiation ($\lambda > 420$ nm) conditions [533]. In another report, a series of ZnCdS solid solutions were synthesized through a one-step chemical bath co-precipitation route, and a hydrogen evolution rate of 27.004 mmol $h^{-1}g^{-1}$ was achieved even without any cocatalysts [534]. On the other hand, Hu *et al.* have synthesized ZnCdS using a solid-phase thermal decomposition strategy and obtained a hydrogen evolution rate of 41.211 mmol $h^{-1}g^{-1}$ [535]. Nonetheless, ZnCdS still faces several challenges, like rapid charge carrier recombination and susceptibility to photocorrosion, for practical application. To address this challenge, a ZnCdS heterostructure with other material has emerged as a useful strategy. More recently, a S-scheme heterojunction has been constructed by ZnCdS and $CoWO_4$ and got a reasonable amount of hydrogen of 15.7 mmol $h^{-1}g^{-1}$ with a 5W LED light [536]. Another S-scheme, $Zn_{1-x}Cd_xS$/NiO heterojunction has been formed and tested for efficient photocatalytic hydrogen evolution. Interestingly, the hydrogen evolution activity of the optimal ZCS/NiO was 259.2 mmol $g^{-1}h^{-1}$, with an AQE of 15.8% at 420 nm, which was 4.3 and 1728.0 times that of ZnCdS nanoparticle and NiO nanosheet, respectively [537]. Liu *et al.* fabricated a thiocyanate-capped $CdSe@Zn_{1-X}Cd_XS$ gradient alloyed quantum dots heterojunction and reported a record-breaking photocatalytic hydrogen evolution of 951 mmol/gh at 100 mW/$cm^2$ AM 1.5 illumination without co-catalysts [538]. More recently, a $Cu_{31}S_{16}/Zn_xCd_{1-x}S$ heterostructure has been synthesized and exhibited a high output photocatalytic HER rate of 61.7 mmol g$^{-1}$h$^{-1}$ [539]. Several other heterojunctions, including NiB/ZnCdS [540], PdS QDs/Zn$_{1-x}$Cd$_x$S [541], ZnCdS/ZnS [542], MoSe$_2$/ZnCdS [543], ZnCdS (QDs)/PZH [544], ZnCdS/MoS$_2$ [545], ZnO/ZnCdS [546], ZnCdS/NiMoO$_4$ [547], Mo$_2$TiC$_2$ MXene/ZnCdS [548], ZnCdS/NiCo-LDH [549], [550], Ni/ZnCdS [551], NiS/ZnCdS [552], [553] and so on have also been reported. Recently, the disordered phase has emerged as a new strategy to enhance the photocatalytic performance of ZnCdS-based materials. The defect engineering of ZnCdS with surface disorder layer by simple room temperature Li-ethylenediamine (Li-EDA) treatment with unusual Zn and S dual vacancies served as hole trapping sites and electron trapping sites, respectively. Ultimately, the highest photocatalytic H$_2$ production rate was 33.6 mmol g$^{-1}$h$^{-1}$ under visible light [554]. Wang *et al.* studied the effect of induced dipole moments in amorphous ZnCdS catalysts for photocatalytic H$_2$ evolution. They found out that 1 wt.% of low-cost Co-MoS$_x$ cocatalysts loaded to the ZnCdS material can increase the H$_2$ evolution rate to 70.13 mmol g$^{-1}$h$^{-1}$, which was over 5 times higher than its crystalline counterpart and is stable over the long term up to 160 h [555]. Additionally, a dipole polarization-driven spatial charge separation in defective Se-doped ZnCdS has been studied to boost photocatalytic hydrogen evolution at 85.3 mmol g$^{-1}$h$^{-1}$ [556]. Again, constructing twin crystals with ZnCdS solid solution can help dissociate water by adsorbing it on the Zn$^{2+}$/Cd$^{2+}$ twin boundary. Then, the fast-moving electrons can quickly combine with the protons already attached to S$^{2-}$ to form hydrogen at the twin boundary [557]. Ultimately, the apparent quantum efficiency of photocatalytic water splitting for hydrogen reached 82.5% ($\lambda$ = 420 nm). Also, JI *et al.* synthesized twin-crystal Zn$_{0.5}$Cd$_{0.5}$S solid solutions with rich sulfur vacancies for enhanced photocatalytic hydrogen evolution rate of 31.6 mmol g$^{-1}$h$^{-1}$ under visible light with an AQY reaching 26.33% at 400 nm [558]. We have Summarized ZnCdS-based photocatalysts for H$_2$ production in Table 12.

*Table 12: Summary of ZnCdS-based photocatalytic for H$_2$ production*

| Photocatalysts | Co-Catalyst/ | Light source | Sacrificial reagent | H$_2$ evolution Activity (μmol h$^{-1}$ g$^{-1}$) | Quantum yield (%) | Ref. |
|---|---|---|---|---|---|---|
| Zn$_{0.6}$Cd$_{0.4}$S | | 300 W Xenon lamp (λ > 420 nm) | Na$_2$S/Na$_2$SO$_3$ | 5680 | | [533] |
| Zn$_{0.30}$Cd$_{0.70}$S | | Simulated Sunlight | Na$_2$S/Na$_2$SO$_3$ | 27004 | | [534] |
| ZCS/CW-10 | | 5 W LED (λ ≥ 420) | lactic acid | 15700 | 3.88 (450 nm) | [536] |
| Zn$_{1-x}$Cd$_x$S/NiO | | 300 W Xenon lamp (λ > 420 nm) | Na$_2$S/Na$_2$SO$_3$ | 259200 | 15.8 | [537] |
| Cu$_{31}$S$_{16}$/Zn$_x$Cd$_{1-x}$S | | Visible Light | | 61700 | 67.9 (400 nm) | [539] |
| CdZnS | NiB | LED | lactic acid | 8137 | | [540] |
| Zn$_{0.6}$Cd$_{0.4}$S | 0.5 wt% PdS QDs | Visible light (420 nm < λ < 800 nm) | lactic acid | 27630 | | [541] |
| ZnCdS/ZnS | | 300 W Xenon lamp (λ > 420 nm) | Na$_2$S/Na$_2$SO$_3$ | 1680 | | [542] |
| Zn$_{0.1}$Cd$_{0.9}$S/NiS | | 300 W Xenon lamp (λ > 420 nm) | glucose | 12770 | 12.74 (350 nm) | [552] |
| NiS/ZnCdS | 1.5 wt% GDY | 300 W Xenon lamp (λ > 420 nm) | 10 v/v% lactic acid | 32100 | | [553] |
| ZnCdS | | 300 W Xenon lamp (λ > 420 nm) | Na₂S/Na₂SO₃ | 33600 | 25.4 (350 nm) | [554] |
| ZnCdS | 1 wt% Co-MoSₓ | Natural Sunlight | lactic acid | 70130 | 38.54 (420 nm) | [555] |
| Zn₀.₅Cd₀.₅S | | 300 W Xenon lamp (λ > 400 nm) | Na₂S/Na₂SO₃ | 31600 | 26.33 (400 nm) | [558] |
| Ni₃S₄/ZnCdS | | 300 W Xenon lamp (λ > 400 nm) | Na₂S/Na₂SO₃ | 5030 | 8.322 | [559] |
| ZnCdS/CoMoOS | | 300 W Xenon lamp (λ > 420 nm) | 15 v/v% lactic acid | 52310 | 5.34 | [560] |
| CdS/ZnₓCdᵧS | | 300 W Xenon lamp (λ > 420 nm) | Na₂S/Na₂SO₃ | 29200 | | [561] |
| UiO-66(Ce)/ZnCdS | | 300 W Xenon lamp (λ > 420 nm) | Na₂S/Na₂SO₃ | 3958.8 | | [529] |
| g-C₃N₄/ZnCdS | | 300 W Xenon lamp (λ > 420 nm) | Na₂S/Na₂SO₃ | 1467.23 | 1.2 | [562] |
| Zn₁₋ₓCdₓS/CdS | | 300 W Xenon lamp (λ > 420 nm) | Na₂S/Na₂SO₃ | 2700 | 3.84 (400 nm) | [563] |
| Zn₀.₆Cd₀.₄S | 0.5% Ag₂S | 300 W Xenon lamp (λ > 420 nm) | Na₂S/Na₂SO₃ | 18770 | 23.8 (400 nm) | [564] |
| Cd₀.₉Zn₀.₁S | Ni(OH)₂ | 300 W Xenon lamp (λ > 420 nm) | Na₂S/Na₂SO₃ | 132930 | 76.5 (460 nm) | [565] |
| CdS@ZnₓCd₁₋ₓS | | 300 W Xenon lamp (λ > 420 nm) | Na₂S/Na₂SO₃ | 5170 | | [566] |

### 2.2.2. Mn$_x$Cd$_{1-x}$S

Manganese (Mn)-doped Cadmium Sulfide (CdS), i.e., Mn$_x$Cd$_{1-x}$S (MnCdS), has garnered significant attention as a potential catalyst for photocatalytic H$_2$ production. This interest is largely due to the favorable bandgap properties of CdS, which allow it to efficiently absorb visible light and facilitate water splitting [567]. Doping CdS with Mn can improve its photocatalytic efficiency by altering the electronic structure and enhancing light absorption. Additionally, Mn can act as a co-catalyst, helping to separate photogenerated charge carriers and reduce recombination rates [568], [569]. Ongoing research is focused on optimizing doping levels [570], understanding the fundamental mechanisms at play, and improving the overall efficiency and stability of the photocatalysts in real-world conditions. Further, defect engineering has also been reported over time, such as sulfur vacancy-induced effects [571], zinc vacancy [572], and co-doping. Additionally, combining MnCdS with materials like $Ni_3S_2$ [573], $Co_9S_8/Co_3O_4$ [574], $ZnCo_2O_4$ with HER of 92.1 mmol $g^{-1}h^{-1}$ [575], $CuCo_2S_4$ [576], $NiCo_2S_4$ [577], $WO_3$ with HER of 21.25 mmol $g^{-1}h^{-1}$ [578], thiol-UiO-66 [579], Ni-MOF-74 [61], $CoWO_4$ [580], NiCoB [581] have shown promise in further enhancing photocatalytic activity and stability. These composites can improve charge carrier mobility and provide additional active reaction sites. Improved synthesis methods, such as hydrothermal, solvothermal, and electrospinning techniques, have created MnCdS with controlled morphologies like nanorods [582] and nanosheets that enhance light absorption and surface area. Recently, sulfur vacancy-mediated MnCdS photocatalysts have been synthesized and combined with NiS as an S-scheme heterojunction for enhanced $H_2$ evolution at a rate of 4099.55 μmol $g^{-1}h^{-1}$, which was 6.75 times higher than pure MnCdS [571]. An interfacial-engineered $Co_3S_4$/MnCdS heterostructure for efficient photocatalytic hydrogen evolution has also been reported. The research claims that it was a rare research on S-scheme heterostructure characteristic of type-I energy band alignment, which showed a superior photocatalytic performance of 7,999.89 μmol $g^{-1}h^{-1}$ with a 4.88% apparent quantum efficiency at 420 nm and a good photostability for many consecutive cycles [583]. Another Mn-doped CdS/$Cu_2O$ S-scheme heterojunction has been synthesized and tested for photocatalytic hydrogen production. Interestingly, the optimized catalyst exhibited a hydrogen release rate of 66.3 mmol

g$^{-1}$h$^{-1}$, which was 3.4 and 54.3 times higher than MnCdS and Cu$_2$O [584]. Recently, Qian *et al.* developed a novel ternary FeWO$_4$/CoP/Mn$_{0.5}$Cd$_{0.5}$S composite photocatalyst for photocatalytic hydrogen evolution in water splitting. The ternary FeWO$_4$/CoP/Mn$_{0.5}$Cd$_{0.5}$S photocatalyst exhibited a maximum hydrogen evolution activity of 30.45 mmol g$^{-1}$h$^{-1}$, which was 6.86 times higher than that of pure Mn$_{0.5}$Cd$_{0.5}$S. The enhancement of the hydrogen production performance of the photocatalyst was attributed to the p-n heterojunction constructed at the interface of FeWO$_4$ and Mn$_{0.5}$Cd$_{0.5}$S, which strengthened the charge separation ability of the system. On the other hand, the decoration of CoP remarkably diminishes the hydrogen evolution potential of the system and supplies richer active sites for the catalytic reaction [585]. Another, CoP nanoparticles decorated with MnCdS nanorods have been reported for photocatalytic hydrogen evolution by water splitting. The testing results indicate that MnCdS/CoP-7% exhibited a significantly enhanced photocatalytic H$_2$ production rate of 40.5 mmol g$^{-1}$h$^{-1}$ [586]. A Zn-doped cadmium selenide-twinned Mn$_{0.65}$Cd$_{0.35}$S has been reported for enhanced hydrogen evolution performance. The photocatalytic hydrogen evolution of Zn-CdSe-1/T-MnCdS-40 could reach approximately 9489.43 μmol g$^{-1}$h$^{-1}$ representing a 15-fold increase compared to Mn$_{0.65}$Cd$_{0.35}$S [587]. A ZIF-67-derived hollow CoS and Mn$_{0.2}$Cd$_{0.8}$S was also formed as a type-II heterojunction for boosting photocatalytic hydrogen evolution. The CoS/Mn$_{0.2}$Cd$_{0.8}$S heterostructure has a strong light-trapping ability, which enhances charge separation and transfer due to its abundant reactive sites and heterostructure. The photocatalytic hydrogen production rate under visible light conditions was 43.7 mmol g$^{-1}$h$^{-1}$, which was 43.7 and 14.1 times higher than that of pure CoS and pure Mn$_{0.2}$Cd$_{0.8}$S, respectively [418]. Solvothermal and hydrothermal methods synthesized NiCo$_2$S$_4$ yolk-shell nanospheres and Mn$_{0.8}$Cd$_{0.2}$S nanoparticles, and then Mn$_{0.8}$Cd$_{0.2}$S nanoparticles were loaded onto the surface of NiCo$_2$S$_4$ to form a dense Z-scheme heterojunction. Without precious metals as cocatalysts, the maximum hydrogen production of the optimized $Mn_{0.8}Cd_{0.2}S/NiCo_2S_4$ reached 21.41 mmol $g^{-1}h^{-1}$ [588]. More recently, a two-step hydrothermal method prepared $MnCdS-V_s/NiCo_2S_4$ (MCSN) Schottky junction nanomaterials with strong electron coupling effects and successfully applied to a square meter hydrogen evolution device. The optimized MCSN material demonstrated high hydrogen evolution activity of 34.28 mmol $g^{-1}h^{-1}$, 9.34, and 685.60 times higher than pure $MnCdS-V_s$ and $NiCo_2S_4$ [589]. We have Summarized MnCdS-based photocatalysts for $H_2$ production in Table 13.

*Table 13: Summary of MnCdS-based photocatalytic for $H_2$ production*

| Photocatalysts | Co-Catalyst/ | Light source | Sacrificial reagent | $H_2$ evolution Activity (μmol $h^{-1}$ $g^{-1}$) | Quantum yield (%) | Ref. |
|---|---|---|---|---|---|---|
| $Mn_{0.6}Cd_{0.4}S$ | 1 wt % pt | 300 W Xenon lamp (λ > 420 nm) | 10 v/v% lactic acid | 2253 | | [568] |
| $Zn_{0.2}Cd_{0.8}S/CoN$ | | 5 W xenon lamp | $Na_2S/Na_2SO_3$ | 14612 | | [569] |
| NiS/MnCdS | | 5 W LED lamp | 10 v/v% lactic acid | 4099.55 | | [571] |
| $Ni_3S_2$/MnCdS | | 300 W Xenon lamp (λ > 420 nm) | $Na_2S/Na_2SO_3$ | 12069.27 | 7.33 (420 nm) | [573] |
| $Mn_{0.5}Cd_{0.5}S/CuCo_2S_4$ | | 300 W Xenon lamp (λ ≥ 420 nm) | $Na_2S/Na_2SO_3$ | 15740 | 41.48 | [576] |
| $Mn_{0.2}Cd_{0.8}S/NiCo_2S_4$ | | 5 W LED lamp | $Na_2S/Na_2SO_3$ | 5677.8 | | [577] |
| Ni-MOF-74/MnCdS | | 5 W LED white light, λ ≥ 420 nm | $Na_2S/Na_2SO_3$ | 7104 | | [61] |
| $Mn_xCd_{1-x}S$@D-$MoSe_yS_{2-y}$ | | 300 W Xenon lamp (λ > 420 nm) | 20 v/v% lactic acid | 12460 | | [582] |
| $Co_3S_4$/MnCdS | | 300 W Xenon lamp (λ > 420 nm) | $Na_2S/Na_2SO_3$ | 7999.89 | 4.88 (420 nm) | [583] |

| Catalyst | | Light source | Sacrificial agent | H₂ production (μmol g⁻¹ h⁻¹) | AQY (%) | Ref. |
|---|---|---|---|---|---|---|
| FeWO$_4$/CoP/ Mn$_{0.5}$Cd$_{0.5}$S | | 300 W Xenon lamp (λ > 420 nm) | Na$_2$S/Na$_2$SO$_3$ | 30450 | 9.32 (420 nm) | [585] |
| CoP/MnCdS | | 300 W Xenon lamp (λ > 420 nm) | Na$_2$S/Na$_2$SO$_3$ | 40500 | 16.1 (420 nm) | [586] |
| Zn–CdSe-1/T-MnCdS-40 | | 300 W Xenon lamp | Na$_2$S/Na$_2$SO$_3$ | 9489.43 | 8.22 (420 nm) | [587] |
| CoS/Mn$_{0.2}$Cd$_{0.8}$S | | 5 W LED white light, λ ≥ 420 nm | Na$_2$S/Na$_2$SO$_3$ | 43700 | 21.55 (420 nm) | [418] |
| Mn$_{0.8}$Cd$_{0.2}$S/NiCo$_2$S$_4$ | | 300 W Xenon lamp | Na$_2$S/Na$_2$SO$_3$ | 21410 | | [588] |
| MnCds/NiCo$_2$S$_4$ | | 5 W lamp (Simulated sunlight) | Na$_2$S/Na$_2$SO$_3$ | 34280 | 24.69 (475 nm) | [589] |

### 2.2.3. Zinc indium sulfide (ZnIn$_2$S$_4$)

Zinc indium sulfide (ZnIn$_2$S$_4$) has attracted extensive research interest in photocatalytic hydrogen evolution applications due to its suitable band gap, high stability, nontoxicity, and higher charge carrier dynamics [590], [591]. Several reviews explaining the basic semiconductor properties, mainly including crystal structures, different synthesis conditions, and band structures for photocatalytic applications of ZnIn$_2$S$_4$ (ZIS), have been reported earlier [590], [592], [593], [594], [595]. Recent advancements in ZIS photocatalysts for hydrogen production have focused on improving efficiency and stability. Strategies like morphology modulation [596], structure design, doping [597], [598], [599], and vacancy engineering [600], [601] have been employed for improved light absorption and enhanced catalytic hydrogen production [602], [603], [604]. Nevertheless, pure ZIS exhibits negligible photocatalytic hydrogen production efficiency due to the photoexcited charges' limited separation efficiency and slow carrier transfer kinetics [605].

Approaches like cocatalyst loading like CoNi [606], Pt [607], [608], ZnS [609], MoS$_2$, NiS, and WS$_2$ [610], La [611] and heterojunction construction to fabricate advanced ZIS-based photocatalysts are considered to be emerging technologies that improve charge separation and photocatalytic hydrogen production [612]. ZnCr$_2$O$_4$ [601], g-C$_3$N$_4$ [613], [614], [615], [616], [617], MXene [605], [618], MoS$_2$ [619], [620], [621], [622], [623], ZnS [624], N−Fe$_3$C [625], r-GO [626], MOF [627], [628], WC [629], NiWO$_4$ [630], carbon nanodots [631] have been reported for enhanced photocatalytic hydrogen production [632]. Also, single-atom catalytic sites have been reported to be an advantage of ZIS in accelerating photocatalytic hydrogen evolution [633]. Recently, with a Mo-doped ZnIn$_2$S$_4$ (Mo-ZIS) photocatalyst, the hydrogen production reaction is significantly facilitated through synergistic integration with the kinetically favorable oxidation of benzyl alcohol (BA). Consequently, Mo-ZIS efficiently produces both high-value-added H$_2$ and benzaldehyde (BAD) products, exhibiting a remarkable hydrogen evolution rate (16,353 µmol g$^{-1}$ h$^{-1}$) and BAD yield (13,942 µmolg$^{-1}$ h$^{-1}$) without the need for additional noble-metal cocatalysts [597]. As a highly active photocatalytic, another single-atomic Mo-modified ZnIn$_2$S$_4$ (Mo-ZIS) nanosheet has been prepared for hydrogen evolution. Mo substituting for a portion of In atoms in ZIS nanosheets induces spatial charge redistribution, which promotes the separation of photogenerated charge carriers and optimizes the Gibbs free energy of adsorbing H* on S atoms at basal planes. As a result, Mo-ZIS exhibited an impressive photocatalytic hydrogen production rate of 6.71 mmol g$^{-1}$h$^{-1}$, over 10 times that of the pristine ZIS, with an apparent quantum efficiency (AQE) up to 38.8% at 420 nm [634]. N-doped ZnIn2S4 and tungsten-based polyoxometalate (TPA) composites with S-scheme heterojunction have recently been synthesized. The optimized sample showed a photocatalytic hydrogen production rate of 17345.53 µmol g$^{-1}$ h$^{-1}$, which is 2.92 times that of N-ZIS (5940.08 µmol g$^{-1}$ h$^{-1}$) and 8.03 times that of ZIS (2159.22 µmol g$^{-1}$ h$^{-1}$) [635]. A noble metal-free ZnIn$_2$S$_4$ photocatalyst with geometric defects has been constructed, in which [In-S]$_4$ tetrahedral vacancies (named as (InS)$_v$) on the surface are created by removing the In and S atoms. Further mechanistic investigation has revealed that the (InS)$_v$ is the active site for reductive reactions and improves photo-generated electron-hole separation. As a result, the solar-driven hydrogen evolution rate reaches 10.70 mmol/gh without a co-catalyst after optimizing the densities of the (InS)$_v$ [636]. Photocatalytic activity of ZIS-based material in H$_2$ generation with respective parameters has been summarized earlier [602], [603], [612]. Again, we have summarized ZIS-based material for H$_2$ generation with an updated report in Table 14.

*Table 14: Summary of ZnIn$_2$S$_4$-based photocatalytic for H$_2$ production*

| Photocatalysts | Co-Catalyst/ | Light source | Sacrificial reagent | H$_2$ evolution Activity (µmol h$^{-1}$ g$^{-1}$) | Quantum yield (%) | Ref. |
|---|---|---|---|---|---|---|
| ZnCr$_2$O$_4$/ZnIn$_2$S$_4$ | | 300 W Xenon lamp (λ ≥ 420 nm) | 10 vol% TEOA | 3421 | | [601] |
| MXene/ZnIn$_2$S$_4$ | 3 wt % Pt | 300 W Xenon lamp (λ ≥ 420 nm) | TEOA | 3475 | 20.41 (400 nm) | [605] |
| ZnIn$_2$S$_4$/ZnO | | 300 W Xenon lamp | Ethanol/water | 13638 | 39.32 (365 nm) | [637] |
| BiVO$_4$/ZnIn$_2$S$_4$ | | 300 W Xenon lamp (λ > 420 nm) | TEOA | 2243 | | [638] |
| SnFe$_2$O$_4$/ZnIn$_2$S$_4$/PVDF | | 300 W Xenon lamp (λ ≥ 420 nm) | Na$_2$S/Na$_2$SO$_3$ | 1652.7 | 6.94 (420 nm) | [639] |
| ZnIn$_2$S$_4$/g-C$_3$N$_4$ | | 300 W Xenon lamp (λ > 420 nm) | Na$_2$S/Na$_2$SO$_3$ | 14799 | 0.516 (400 nm) | [613] |

| Catalyst | Cocatalyst | Light source | Sacrificial agent | Activity (μmol g⁻¹ h⁻¹) | AQE (%) | Ref. |
|---|---|---|---|---|---|---|
| g-C₃N₄@ZnIn₂S₄ | | 5 W blue LED light (λ_max = 420 nm) | 50 vol% TEOA | 2377.6 | | [614] |
| ZnIn₂S₄/g-C₃N₄ | 1wt% Pt | 300 W Xenon lamp (λ > 420 nm) | TEOA | 9189.8 | 8.71 (420 nm) | [617] |
| Vs-MoS₂-ZnIn₂S₄ | | 300 W Xenon lamp (λ > 300 nm) | 10 % v/v lactic acid | 6884 | 63.87 (420 nm) | [620] |
| MoS₂-ZnIn₂S₄ | | 300 W Xenon lamp (λ > 420 nm) | Na₂S/Na₂SO₃ | 3891.6 | | [623] |
| ZnIn₂S₄@N−Fe₃C | | 150 W Xe lamp AM 1.5G | TEOA | 9600 | 3.6 | [625] |
| ZnIn₂S₄/NiWO₄ | | 300 W Xenon lamp (Simulated solar light) | 10 vol% TEOA | 16300 | 9.69 (420 nm) | [630] |
| CDs/ZnIn₂S₄ | | 300 W Xenon lamp (λ > 420 nm) | Na₂S/Na₂SO₃ | 4150 | | [631] |
| Single atom Pt-ZnIn₂S₄ | | 300 W Xenon lamp (λ > 420 nm) | 10 vol% TEOA | 17500 | 50.4 (420 nm) | [633] |
| Single atom Pt-ZnIn₂S₄ | | 300 W Xenon lamp (Simulated solar light) | 10 vol% TEOA | 29200 | 50.4 (420 nm) | [633] |
| Mo-ZnIn₂S₄ | | 300 W Xenon lamp AM 1.5 G | 10 vol% TEOA | 6710 | 38.8 (420 nm) | [634] |
| N-ZnIn₂S₄/TPA | | 300 W Xenon lamp (λ > 420 nm) | ascorbic acid | 17345.53 | 3.98 (430 nm) | [635] |
| ZnIn$_2$S$_4$ | | 300 W Xenon lamp (λ > 420 nm) | ascorbic acid | 10700 | 14.0 (380 nm) | [636] |
| WO$_3$·xH$_2$O/ZnIn$_2$S$_4$ | | 300 W Xenon lamp (λ > 420 nm) | 15 % v/v lactic acid | 7200 | 9.3 (420 nm) | [640] |

### 2.2.4. Cadmium Indium Sulfide (CdIn$_2$S$_4$)

CdIn$_2$S$_4$ (CIS) is a ternary semiconductor composed of cadmium (Cd), indium (In), and sulfur (S). It typically crystallizes in a chalcopyrite or spinel structure, influencing its optical and electronic properties [641], [642]. The band gap of CIS is around 1.8 to 2.2 eV, making it suitable for visible light absorption and easy charge carrier generation for hydrogen production [643]. Recent advancements in CIS photocatalysts for hydrogen production have focused on enhancing their efficiency, stability, and overall performance. Researchers have explored doping CIS with elements like nitrogen, phosphorus, or transition metals (e.g., Co, Ni) to modify its electronic properties and improve the efficiency of light absorption and charge separation. Different morphology-based nanostructures, such as nanosheets, are also used [644], [645]. Combining CIS with other semiconductors like MoS$_2$ [646], [647], [648], ZnIn$_2$S$_3$ [649], [650], Mo$_2$C [651], In$_2$O$_3$/In$_2$S$_3$ [652], ZnSn(OH)$_6$ [653], CNFs/Co$_4$S$_3$ [654], CuCo$_2$S$_4$ [655], ReS$_2$ [656], SnS$_2$ [456], MoB [657], Co$_2$P [658], MoP [659], Ni$_{12}$P$_5$ [660], NiCo$_2$O$_4$ [661], MOF [662], Carbon QDs [663], Carbon QDs/CdS [664], C-doped g-C$_3$N$_4$ [665], WC [666], Cobalt Phosphate [667] has shown to enhance photocatalytic activity. These heterostructures facilitate better charge transfer and reduce electron-hole recombination. Techniques such as loading noble metal co-catalysts Ru [668] on the surface of CIS have enhanced photocatalytic efficiency by providing active sites for hydrogen evolution and improving charge separation. New strategies for enhancing the stability of CIS under photocatalytic conditions have emerged, including protective coatings and the development of more robust composite materials that can withstand reaction conditions. Construction of CIS/ZnS type-I band alignment heterojunctions by decorating CIS on ZnS microspheres for efficient photocatalytic $H_2$ evolution has been reported recently. Interestingly, the 5% CIS/ZnS sample showed the highest photocatalytic hydrogen production rate of 10,799 µmol $g^{-1}$ $h^{-1}$, 2.6 and 5.4 times higher than that of ZnS and CIS, respectively [669]. Another Z-scheme CIS/ZnS heterojunction has been fabricated, but the photocatalytic hydrogen evolution rate was only 3743 µmol $g^{-1}$ $h^{-1}$ [670]. 0D/3D $NiTiO_3$/CIS heterostructure photocatalyst prepared via hydrothermal method and 20 wt% $NiTiO_3$/CIS showed superior photocatalytic activity with the $H_2$ generation rate of 5168.6 µmol $g^{-1}$ $h^{-1}$ with an AQY of 5.14% at 420 nm [671]. More recently, an amorphous $CoMoS_4$ was successfully deposited on the flower-like CIS microspheres. The remarkably promoted photocatalytic activity with the optimum 5%-CCIS catalyst at a rate of 2196 µmol $g^{-1}$ $h^{-1}$ was 8.4 folds larger than that for CIS [672]. Enhanced visible-light-driven hydrogen evolution in non-precious metal $Ni_2P$/CIS S-type heterojunction via rapid interfacial charge transfer has also been reported. The hydrogen production of $Ni_2P$/CIS reached 8.7 mmol $g^{-1}h^{-1}$, while AQY reached 7.7 % at 420 nm [673]. A ternary rh/c-$In_2O_3$/CIS heterostructure photocatalyst has been synthesized, and the hydrogen evolution rate of 4.7 mmol $g^{-1}h^{-1}$, which was 3.6 folds and 7.8 folds higher than that of CIS and rh/c-$In_2O_3$, respectively [674]. An efficient Z scheme-type II charge transfer on PAN/ZnO/CIS interfaces for enhanced photocatalytic hydrogen generation has been reported. The optimized catalyst exhibited an H$_2$ production rate of 4.31 mmol g$^{-1}$h$^{-1}$ [675]. Table 15 summarizes recent CdIn$_2$S$_4$-based photocatalytic for H$_2$ production

*Table 15: Summary of CdIn$_2$S$_4$-based photocatalytic for H$_2$ production*

| Photocatalysts | Co-Catalyst/ | Light source | Sacrificial reagent | H$_2$ evolution Activity (μmol h$^{-1}$ g$^{-1}$) | Quantum yield (%) | Ref. |
|---|---|---|---|---|---|---|
| CIS | 0.5 wt% Pt | 300 W Xenon lamp | Na$_2$S/Na$_2$SO$_3$, lactic acid | 7515 | | [644] |
| CIS | | 300 W Xenon lamp (λ ≥ 420 nm) | NaOH + H$_2$S | 3238 | | [645] |
| MoS$_2$/CIS | | 300 W Xenon lamp | Na$_2$S/Na$_2$SO$_3$ | 1868.19 | | [648] |
| ZnIn$_2$S$_4$/CIS | | 300 W Xenon lamp | Na$_2$S/Na$_2$SO$_3$ | 12670 | 18.73 (420 nm) | [649] |
| Mo$_2$C/CIS | | 300 W Xenon lamp (λ > 300 nm) | 10 % v/v lactic acid | 1178.32 | | [651] |
| CIS/CNFs /Co$_4$S$_3$ | | 300 W Xenon lamp (λ > 420 nm) | 20 % v/v lactic acid | 25870 | 19.56 (365 nm) | [654] |
| CuCo$_2$S$_4$/ CIS | | 300 W Xenon lamp (λ ≥ 420 nm) | TEOA | 1320 | 11.12 (420 nm) | [655] |
| Ni$_{12}$P$_5$/CIS | | 300 W Xenon lamp (λ ≥ 420 nm) | 10 vol% TEOA | 5010 | 23.5 (400 nm) | [660] |
| MOF/CIS | | 300 W Xenon lamp (λ ≥ 420 nm) | Na$_2$S/Na$_2$SO$_3$ | 2550 | | [662] |
| C-g-C$_3$N$_4$/CIS | | 500 W Xenon lamp (λ ≥ 420 nm) | MeOH | 2985 | | [665] |
| CoH$_x$PO$_y$/ CIS | | 500 W Xenon lamp (λ ≥ 420 nm) | 20 vol% MeOH | 7280 | 14.08 (405 nm) | [667] |
| CIS/ZnS | | 300 W Xenon lamp | Na$_2$S/Na$_2$SO$_3$ | 10799 | | [669] |
| NiTiO$_3$/CIS | | 300 W Xenon lamp (λ ≥ 420 nm) | 10 vol% TEOA | 5168.6 | 5.14 (420 nm) | [671] |
| CoMoS$_4$/ CIS | | 300 W Xenon lamp (λ ≥ 420 nm) | 10 vol% TEOA | 2196 | 9.14 (420 nm) | [672] |
| Ni$_2$P/CIS | | 300 W Xenon lamp (λ > 300 nm) | 10 % v/v lactic acid | 8700 | 7.7 (420 nm) | [673] |
| rh/c-In$_2$O$_3$/CIS | | 300 W Xenon lamp (λ ≥ 420 nm) | Na$_2$S/Na$_2$SO$_3$ | 4700 | 11.4 (380 nm) | [674] |
| PAN/ZnO /CIS | | 300 W Xenon lamp | Na$_2$S/Na$_2$SO$_3$ | 4310 | 7.32 (365 nm) | [675] |

3. **Future perspective/outlook**

The subject of a review is challenging in the context of a hotly debated topic that continues to develop rapidly. Therefore, rather than recapitulating current knowledge about MSs for photocatalytic hydrogen production, we decided to focus on some key points that we believe can be considered and helpful in future research. A standard protocol and recipe can be constructed more organized than a trial-and-error approach, as follows:

I. Selection and design of material using AI/ML: After reviewing more than 600 journal papers in this chapter, authors realized that there should be a systematic selection of materials using the potential of artificial intelligence (AI) and machine learning (ML), which is crucial given the time constraints for the researcher to review the existing literature and find the gaps with future development. Selection of a particular photocatalyst material (here sulfides) synthesis and material design can be possible by accurately extracting the underlying mechanism in the model that converts readily available data and pre-catalysts into their promising and useful ones [676], [677], [678], [679], [680], [681], [682].

II. Cost-effective material and simple methods for highly pure-scale photocatalyst production:

   A. The synthesis method becomes costly when expensive elements, such as Ag, In, and Ga, have to be used. Cheaper alternatives may be available, but the choice often balances cost against performance and suitability for overall performance.

   B. The synthesis and application of sulfide catalysts can be complicated when scaling up from laboratory to industrial processes. Sulfides can be sensitive to moisture or oxygen contamination, leading to product degradation or altering its properties. The synthesis process may also generate undesirable byproducts, complicating purification and decreasing yield. Optimizing reaction conditions to minimize these byproducts can be challenging.

   C. Many sulfide compounds can be toxic or harmful to the environment, necessitating careful handling and disposal. Cadmium is classified as a human carcinogen. Thus, cadmium-based sulfides are subject to strict regulations in many countries. Proper handling, usage, and disposal protocols are critical to mitigate risks associated with cadmium-based sulfides.

III. Advanced in situ/ex-situ measurements and theoretical understanding using DFT and MD simulations for high efficiency and stability:

   A. Over the years, density functional theory (DFT) studies have been employed over many decades for heterogeneous catalytic studies that, while powerful, have several limitations when applied to metal sulfide-based photocatalysts for hydrogen production [683]. The synthesis of metal sulfides often lacks mechanistic understanding and detailed control over the chemical reaction processes, limiting the flexibility, cost-effectiveness, and efficiency of attaining high-purity metal sulfides. Researchers could employ molecular dynamics simulations to elucidate the reaction mechanisms [684] better.

   B. Nevertheless, in addition to the conventional theoretical and experimental study methods mentioned in this chapter, recent advances in machine learning [685], [686], in-situ spectroscopy and microscopy techniques for investigating the surface active sites, electronic states, and chemical/physical changes of the photocatalysts, together with intermediates and the mechanism of the photocatalytic reactions, have great potential to further promote the research of photocatalysts for the understanding of photocatalytic hydrogen evolution (PHE) mechanisms, especially the charge transfer behavior in heterojunction photocatalysts.

IV. Targeting for industrial-scale production for real-life end use:

A. Coupling hydrogen production with selective organic synthesis can be a more viable and attractive option than overall water-splitting [687]. The organic compounds can be converted into value-added oxidation products via single-electron processes, which are much easier than the sluggish four-electron oxygen evolution reaction. It's worth mentioning that most organic compounds commonly used in photocatalytic hydrogen productions called sacrificial agents (SA) are expensive for practical industrial-scale $H_2$ production. So, a minimum or zero amount of SA in PHE technology should be considered. Alternative sulfide can be discovered with the higher stability of extracting holes to avoid photo corrosion [688], making a protective coating strategy [689].

B. The photocatalytic reaction is complex, including light absorption, charge transfer and separation, and surface chemical reaction. Each strategy is effective for only one or part of the photocatalytic process. For example, heterojunction can significantly improve the charge separation of metal sulfides but does not consider the surface chemical reaction. In future research, integrating several strategies without affecting each other may offer a better way to improve the whole photocatalytic reaction process and stability [690].

C. The design of reactors is pivotal for scaling up photocatalytic hydrogen production [691]. Effective designs optimize light absorption, charge separation, and gas evolution. Despite advancements, efficiency, catalyst lifespan, and system scalability persist. Overcoming these hurdles is crucial for maximizing reactor efficiency and stability, emphasizing the need for careful reactor design and interdisciplinary collaboration for successful scale-up.

4. **Conclusion**

In this chapter, recent research progresses, including the construction of heterojunctions, defect engineering, co-catalyst loading, elemental doping, and single-atom engineering, which improve the electronic structure and charge separation capabilities of metal sulfides for photocatalytic hydrogen production, are summarized and discussed. The photocatalytic hydrogen evolution performance and stability of metal sulfides are significantly improved after modification using the emerging strategies. Based on these considerable advancements, more efficient and stable metal sulfide photocatalysts are expected to be exploited for solar hydrogen production. However, it must be acknowledged that the efficiency and stability of metal sulfides are still far from industrial application requirements at the current stage. Further optimization of the strategies can lead to the commercialization of the technology and towards industrial-scale production. This chapter will give the reader an overall picture that has been explored for many decades to create a pathway to explore it further.

**References**

[1] Intergovernmental Panel on Climate Change (IPCC), Ed., "Framing and Context," in *Global Warming of 1.5°C: IPCC Special Report on Impacts of Global Warming of 1.5°C above Pre-industrial Levels in Context of Strengthening Response to Climate Change, Sustainable Development, and Efforts to Eradicate Poverty*, Cambridge: Cambridge University Press, 2022, pp. 49–92. doi: 10.1017/9781009157940.003.

[2] Q. Hassan, S. Algburi, A. Z. Sameen, H. M. Salman, and M. Jaszczur, "A review of hybrid renewable energy systems: Solar and wind-powered solutions: Challenges, opportunities, and policy implications," *Results Eng.*, vol. 20, p. 101621, Dec. 2023, doi: 10.1016/j.rineng.2023.101621.

[3] S. Impram, S. Varbak Nese, and B. Oral, "Challenges of renewable energy penetration on power system flexibility: A survey," *Energy Strategy Rev.*, vol. 31, p. 100539, Sep. 2020, doi: 10.1016/j.esr.2020.100539.

[4] Z. Mamiyev and N. O. Balayeva, "Metal Sulfide Photocatalysts for Hydrogen Generation: A Review of Recent Advances," *Catalysts*, vol. 12, no. 11, Art. no. 11, Nov. 2022, doi: 10.3390/catal12111316.

[5] K. Shimura and H. Yoshida, "Heterogeneous photocatalytic hydrogen production from water and biomass derivatives," *Energy Environ. Sci.*, vol. 4, no. 7, pp. 2467–2481, Jul. 2011, doi: 10.1039/C1EE01120K.

[6]     B. C. Tashie-Lewis and S. G. Nnabuife, "Hydrogen Production, Distribution, Storage and Power Conversion in a Hydrogen Economy - A Technology Review," *Chem. Eng. J. Adv.*, vol. 8, p. 100172, Nov. 2021, doi: 10.1016/j.ceja.2021.100172.

[7]     M. Balat, "Potential importance of hydrogen as a future solution to environmental and transportation problems," *Int. J. Hydrog. Energy*, vol. 33, no. 15, pp. 4013–4029, Aug. 2008, doi: 10.1016/j.ijhydene.2008.05.047.

[8]     M. Kumar, N. K. Singh, R. S. Kumar, and R. Singh, "Production of Green Hydrogen through Photocatalysis," in *Towards Sustainable and Green Hydrogen Production by Photocatalysis: Insights into Design and Development of Efficient Materials (Volume 2)*, vol. 1468, in ACS Symposium Series, no. 1468, vol. 1468. , American Chemical Society, 2024, pp. 1–24. doi: 10.1021/bk-2024-1468.ch001.

[9]     H. Nishiyama *et al.*, "Photocatalytic solar hydrogen production from water on a 100-m2 scale," *Nature*, vol. 598, no. 7880, pp. 304–307, Oct. 2021, doi: 10.1038/s41586-021-03907-3.

[10]    A. Fujishima and K. Honda, "Electrochemical Photolysis of Water at a Semiconductor Electrode," *Nature*, vol. 238, no. 5358, pp. 37–38, Jul. 1972, doi: 10.1038/238037a0.

[11]    M. Kanai and M. Beller, "Introduction to hybrid catalysis," *Org. Biomol. Chem.*, vol. 19, no. 4, pp. 702–704, 2021, doi: 10.1039/D0OB90177F.

[12]    D. E. Fogg and E. N. dos Santos, "Tandem catalysis: a taxonomy and illustrative review," *Coord. Chem. Rev.*, vol. 248, no. 21, pp. 2365–2379, Dec. 2004, doi: 10.1016/j.ccr.2004.05.012.

[13]    P. Zhou *et al.*, "Solar-to-hydrogen efficiency of more than 9% in photocatalytic water splitting," *Nature*, vol. 613, no. 7942, pp. 66–70, Jan. 2023, doi: 10.1038/s41586-022-05399-1.

[14]    S. C. Baral, M. P, P. K. Mishra, P. Dhanapal, and S. Sen, "Rapid photocatalytic degradation of organic pollutants by $Al^{3+}$ doped CuO powders under low power visible light and natural sunlight," *Opt. Mater.*, vol. 147, p. 114653, Jan. 2024, doi: 10.1016/j.optmat.2023.114653.

[15]    S. C. Baral *et al.*, "Enhanced photocatalytic degradation of organic pollutants in water using copper oxide (CuO) nanosheets for environmental application," *JCIS Open*, vol. 13, p. 100102, Apr. 2024, doi: 10.1016/j.jciso.2024.100102.

[16]    S. C. Baral, P. Maneesha, S. Sen, S. Sen, and S. Sen, "An Introduction to Metal Oxides," in *Optical Properties of Metal Oxide Nanostructures*, V. Kumar, I. Ayoub, V. Sharma, and H. C. Swart, Eds., Singapore: Springer Nature, 2023, pp. 1–34. doi: 10.1007/978-981-99-5640-1_1.

[17]    S. C. Baral, P. Maneesha, E. G. Rini, and S. Sen, "Recent advances in *La2NiMnO*6 double perovskites for various applications; challenges and opportunities," *Prog. Solid State Chem.*, vol. 72, p. 100429, Dec. 2023, doi: 10.1016/j.progsolidstchem.2023.100429.

[18]    P. Maneesha, S. Chandra Baral, E. G. Rini, and S. Sen, "An overview of the recent developments in the structural correlation of magnetic and electrical properties of *Pr2NiMnO*6 double perovskite," *Prog. Solid State Chem.*, vol. 70, p. 100402, Jun. 2023, doi: 10.1016/j.progsolidstchem.2023.100402.

[19]    S. C. Baral *et al.*, "Rare Earth Manganites and Related Multiferroicity," in *Emerging Applications of Low Dimensional Magnets*, CRC Press, 2022.

[20]    P. Maneesha, S. C. Baral, E. G. Rini, and S. Sen, "Nanomagnetic Materials: Structural and Magnetic Properties," in *Fundamentals of Low Dimensional Magnets*, CRC Press, 2022.

[21]    P. Maneesha *et al.*, "Effect of oxygen vacancies and cationic valence state on multiferroicity and magnetodielectric coupling in (1-x)BaTiO3.(x)LaFeO3 solid solution," *J. Alloys Compd.*, vol. 971, p. 172587, Jan. 2024, doi: 10.1016/j.jallcom.2023.172587.

[22]    P. Maneesha, K. S. Samantaray, S. C. Baral, R. Mittal, M. K. Gupta, and S. Sen, "Defect/disorder correlated modification of transport properties from hopping to tunneling processes in BaTiO3–LaFeO3 solid solution," *J. Appl. Phys.*, vol. 135, no. 19, p. 194102, May 2024, doi: 10.1063/5.0195109.

[23]    R. K. Gupta, S. R. Mishra, and T. A. Nguyen, *Fundamentals of Low Dimensional Magnets*. CRC Press, 2022.

[24] R. K. Gupta, S. R. Mishra, and T. A. Nguyen, *Emerging Applications of Low Dimensional Magnets*. CRC Press, 2022.

[25] M. I. Aziz, F. Mughal, H. M. Naeem, A. Zeb, M. A. Tahir, and M. A. Basit, "Evolution of photovoltaic and photocatalytic activity in anatase-TiO2 under visible light *via* simplistic deposition of CdS and PbS quantum-dots," *Mater. Chem. Phys.*, vol. 229, pp. 508–513, May 2019, doi: 10.1016/j.matchemphys.2019.03.042.

[26] F. Mughal, M. Muhyuddin, M. Rashid, T. Ahmed, M. A. Akram, and M. A. Basit, "Multiple energy applications of quantum-dot sensitized TiO2/PbS/CdS and TiO2/CdS/PbS hierarchical nanocomposites synthesized via *p*-SILAR technique," *Chem. Phys. Lett.*, vol. 717, pp. 69–76, Feb. 2019, doi: 10.1016/j.cplett.2019.01.010.

[27] R. D. Shannon, "Revised effective ionic radii and systematic studies of interatomic distances in halides and chalcogenides," *Acta Crystallogr. A*, vol. 32, no. 5, pp. 751–767, Sep. 1976, doi: 10.1107/S0567739476001551.

[28] T. M. Miller and B. Bederson, "Atomic and Molecular Polarizabilities-A Review of Recent Advances," in *Advances in Atomic and Molecular Physics*, vol. 13, D. R. Bates and B. Bederson, Eds., Academic Press, 1978, pp. 1–55. doi: 10.1016/S0065-2199(08)60054-8.

[29] D. R. Lide, G. Baysinger, S. Chemistry, L. I. Berger, R. N. Goldberg, and H. V. Kehiaian, "CRC Handbook of Chemistry and Physics".

[30] T. Weber, R. Prins, and R. A. Santen, Eds., *Transition Metal Sulphides*. Dordrecht: Springer Netherlands, 1998. doi: 10.1007/978-94-017-3577-3.

[31] Q. Zhu *et al.*, "Recent Progress of Metal Sulfide Photocatalysts for Solar Energy Conversion," *Adv. Mater.*, vol. 34, no. 45, p. 2202929, 2022, doi: 10.1002/adma.202202929.

[32] X. Wu, S. Xie, H. Zhang, Q. Zhang, B. F. Sels, and Y. Wang, "Metal Sulfide Photocatalysts for Lignocellulose Valorization," *Adv. Mater.*, vol. 33, no. 50, p. 2007129, 2021, doi: 10.1002/adma.202007129.

[33] S. Zhang, X. Ou, Q. Xiang, S. A. C. Carabineiro, J. Fan, and K. Lv, "Research progress in metal sulfides for photocatalysis: From activity to stability," *Chemosphere*, vol. 303, p. 135085, Sep. 2022, doi: 10.1016/j.chemosphere.2022.135085.

[34] F. Jamal *et al.*, "Review of Metal Sulfide Nanostructures and their Applications," *ACS Appl. Nano Mater.*, vol. 6, no. 9, pp. 7077–7106, May 2023, doi: 10.1021/acsanm.3c00417.

[35] B. Keimer and J. E. Moore, "The physics of quantum materials," *Nat. Phys.*, vol. 13, no. 11, pp. 1045–1055, Nov. 2017, doi: 10.1038/nphys4302.

[36] F. A. Frame and F. E. Osterloh, "CdSe-MoS2: A Quantum Size-Confined Photocatalyst for Hydrogen Evolution from Water under Visible Light," *J. Phys. Chem. C*, vol. 114, no. 23, pp. 10628–10633, Jun. 2010, doi: 10.1021/jp101308e.

[37] K. Sivula and R. van de Krol, "Semiconducting materials for photoelectrochemical energy conversion," *Nat. Rev. Mater.*, vol. 1, no. 2, pp. 1–16, Jan. 2016, doi: 10.1038/natrevmats.2015.10.

[38] L. Amirav and A. P. Alivisatos, "Photocatalytic Hydrogen Production with Tunable Nanorod Heterostructures," *J. Phys. Chem. Lett.*, vol. 1, no. 7, pp. 1051–1054, Apr. 2010, doi: 10.1021/jz100075c.

[39] Y. Shiga, N. Umezawa, N. Srinivasan, S. Koyasu, E. Sakai, and M. Miyauchi, "A metal sulfide photocatalyst composed of ubiquitous elements for solar hydrogen production," *Chem. Commun.*, vol. 52, no. 47, pp. 7470–7473, 2016, doi: 10.1039/C6CC03199D.

[40] J. Bonde, P. G. Moses, T. F. Jaramillo, J. K. Nørskov, and I. Chorkendorff, "Hydrogen evolution on nano-particulate transition metal sulfides," *Faraday Discuss.*, vol. 140, no. 0, pp. 219–231, 2009, doi: 10.1039/B803857K.

[41] K. Maeda *et al.*, "Photocatalyst releasing hydrogen from water," *Nature*, vol. 440, no. 7082, pp. 295–295, Mar. 2006, doi: 10.1038/440295a.

[42] V. N. Rao, M. V. Shankar, and J.-M. Yang, "Cutting-Edge Sulfide-Based Transition Metals as Photocatalysts for Exceptional Hydrogen Production," in *Towards Sustainable and Green Hydrogen Production by Photocatalysis: Insights into Design and Development of Efficient*

*Materials (Volume 2)*, vol. 1468, in ACS Symposium Series, no. 1468, vol. 1468. , American Chemical Society, 2024, pp. 295–331. doi: 10.1021/bk-2024-1468.ch012.

[43] J. Singh and R. K. Soni, "Enhanced sunlight driven photocatalytic activity of In2S3 nanosheets functionalized MoS2 nanoflowers heterostructures," *Sci. Rep.*, vol. 11, no. 1, p. 15352, Jul. 2021, doi: 10.1038/s41598-021-94966-z.

[44] B. Jiang et al., "Mesoporous Metallic Iridium Nanosheets," *J. Am. Chem. Soc.*, vol. 140, no. 39, pp. 12434–12441, Oct. 2018, doi: 10.1021/jacs.8b05206.

[45] Y. Y. Lee, H. S. Jung, J. M. Kim, and Y. T. Kang, "Photocatalytic CO2 conversion on highly ordered mesoporous materials: Comparisons of metal oxides and compound semiconductors," *Appl. Catal. B Environ.*, vol. 224, pp. 594–601, May 2018, doi: 10.1016/j.apcatb.2017.10.068.

[46] N. O. Balayeva and Z. Q. Mamiyev, "Synthesis and characterization of Ag2S/PVA-fullerene (C60) nanocomposites," *Mater. Lett.*, vol. 175, pp. 231–235, Jul. 2016, doi: 10.1016/j.matlet.2016.04.024.

[47] C. Meerbach et al., "General Colloidal Synthesis of Transition-Metal Disulfide Nanomaterials as Electrocatalysts for Hydrogen Evolution Reaction," *ACS Appl. Mater. Interfaces*, vol. 12, no. 11, pp. 13148–13155, Mar. 2020, doi: 10.1021/acsami.9b21607.

[48] Sh. Anju Devi, K. Jugeshwar Singh, and K. Nomita Devi, "Visible light driven photocatalytic activities of metal sulfides synthesized by simple co-precipitation method," *Mater. Today Proc.*, vol. 65, pp. 2819–2824, Jan. 2022, doi: 10.1016/j.matpr.2022.06.270.

[49] Z. Q. Mamiyev and N. O. Balayeva, "CuS nanoparticles synthesized by a facile chemical route under different pH conditions," *Mendeleev Commun.*, vol. 26, no. 3, pp. 235–237, May 2016, doi: 10.1016/j.mencom.2016.05.004.

[50] K. M. Mayer and J. H. Hafner, "Localized Surface Plasmon Resonance Sensors," *Chem. Rev.*, vol. 111, no. 6, pp. 3828–3857, Jun. 2011, doi: 10.1021/cr100313v.

[51] X. Zhou, G. Liu, J. Yu, and W. Fan, "Surface plasmon resonance-mediated photocatalysis by noble metal-based composites under visible light," *J. Mater. Chem.*, vol. 22, no. 40, p. 21337, 2012, doi: 10.1039/c2jm31902k.

[52] J. Li, Z. Lou, and B. Li, "Nanostructured materials with localized surface plasmon resonance for photocatalysis," *Chin. Chem. Lett.*, vol. 33, no. 3, pp. 1154–1168, Mar. 2022, doi: 10.1016/j.cclet.2021.07.059.

[53] S. A. Shah, I. Khan, and A. Yuan, "MoS2 as a Co-Catalyst for Photocatalytic Hydrogen Production: A Mini Review," *Molecules*, vol. 27, no. 10, Art. no. 10, Jan. 2022, doi: 10.3390/molecules27103289.

[54] B. Han and Y. H. Hu, "MoS2 as a co-catalyst for photocatalytic hydrogen production from water," *Energy Sci. Eng.*, vol. 4, no. 5, pp. 285–304, 2016, doi: 10.1002/ese3.128.

[55] Q. Liu, W. Luan, X. Zhang, R. Zhao, J. Han, and L. Wang, "NiMoP2 co-catalyst modified Cu doped ZnS for enhanced photocatalytic hydrogen evolution," *Sep. Purif. Technol.*, vol. 354, p. 128666, Feb. 2025, doi: 10.1016/j.seppur.2024.128666.

[56] R. S, A. M, and S. V S, "Sulfide-Based Photocatalysts for Efficient H2 Production," in *Towards Sustainable and Green Hydrogen Production by Photocatalysis: Insights into Design and Development of Efficient Materials (Volume 2)*, vol. 1468, in ACS Symposium Series, no. 1468, vol. 1468. , American Chemical Society, 2024, pp. 333–362. doi: 10.1021/bk-2024-1468.ch013.

[57] Y. Wang et al., "NiO co-catalyst modification ZnIn2S4 driving efficient hydrogen generation under visible light," *Sep. Purif. Technol.*, vol. 320, p. 124096, Sep. 2023, doi: 10.1016/j.seppur.2023.124096.

[58] X. Huang, J. Song, G. Wu, Z. Miao, Y. Song, and Z. Mo, "Recent progress on the photocatalytic hydrogen evolution reaction over a metal sulfide cocatalyst-mediated carbon nitride system," *Inorg. Chem. Front.*, vol. 11, no. 9, pp. 2527–2552, 2024, doi: 10.1039/D4QI00255E.

[59] Y. Zhao, Y. Lu, L. Chen, X. Wei, J. Zhu, and Y. Zheng, "Redox Dual-Cocatalyst-Modified CdS Double-Heterojunction Photocatalysts for Efficient Hydrogen Production," *ACS Appl. Mater. Interfaces*, vol. 12, no. 41, pp. 46073–46083, Oct. 2020, doi: 10.1021/acsami.0c12790.

[60] Y. Zhang, J. Wan, C. Zhang, and X. Cao, "MoS2 and Fe2O3 co-modify g-C3N4 to improve the performance of photocatalytic hydrogen production," *Sci. Rep.*, vol. 12, no. 1, p. 3261, Feb. 2022, doi: 10.1038/s41598-022-07126-2.

[61] Z. Jin, H. Gong, and H. Li, "Visible-light-driven two dimensional metal-organic framework modified manganese cadmium sulfide for efficient photocatalytic hydrogen evolution," *J. Colloid Interface Sci.*, vol. 603, pp. 344–355, Dec. 2021, doi: 10.1016/j.jcis.2021.06.111.

[62] L. Chen, Z. Yang, and B. Chen, "Uniformly Dispersed Metal Sulfide Nanodots on g-C3N4 as Bifunctional Catalysts for High-Efficiency Photocatalytic H2 and H2O2 Production under Visible-Light Irradiation," *Energy Fuels*, vol. 35, no. 13, pp. 10746–10755, Jul. 2021, doi: 10.1021/acs.energyfuels.1c00688.

[63] M. A. Nazir *et al.*, "Tuning the photocatalytic hydrogen production via co-catalyst engineering," *J. Alloys Compd.*, vol. 990, p. 174378, Jun. 2024, doi: 10.1016/j.jallcom.2024.174378.

[64] Y. Zhang *et al.*, "Zinc-Blende CdS Nanocubes with Coordinated Facets for Photocatalytic Water Splitting," *ACS Catal.*, vol. 7, no. 2, pp. 1470–1477, Feb. 2017, doi: 10.1021/acscatal.6b03212.

[65] Y. Liu *et al.*, "Facet and morphology dependent photocatalytic hydrogen evolution with CdS nanoflowers using a novel mixed solvothermal strategy," *J. Colloid Interface Sci.*, vol. 513, pp. 222–230, Mar. 2018, doi: 10.1016/j.jcis.2017.11.030.

[66] M. Ding, S. Cui, Z. Lin, and X. Yang, "Crystal facet engineering of hollow cadmium sulfide for efficient photocatalytic hydrogen evolution," *Appl. Catal. B Environ. Energy*, vol. 357, p. 124333, Nov. 2024, doi: 10.1016/j.apcatb.2024.124333.

[67] K. Wu, W. E. Rodríguez-Córdoba, Z. Liu, H. Zhu, and T. Lian, "Beyond Band Alignment: Hole Localization Driven Formation of Three Spatially Separated Long-Lived Exciton States in CdSe/CdS Nanorods," *ACS Nano*, vol. 7, no. 8, pp. 7173–7185, Aug. 2013, doi: 10.1021/nn402597p.

[68] M. Micheel, B. Liu, and M. Wächtler, "Influence of Surface Ligands on Charge-Carrier Trapping and Relaxation in Water-Soluble CdSe@CdS Nanorods," *Catalysts*, vol. 10, no. 10, Art. no. 10, Oct. 2020, doi: 10.3390/catal10101143.

[69] J. Schlenkrich *et al.*, "Investigation of the Photocatalytic Hydrogen Production of Semiconductor Nanocrystal-Based Hydrogels," *Small*, vol. 19, no. 21, p. 2208108, 2023, doi: 10.1002/smll.202208108.

[70] H. Li *et al.*, "Construction and Nanoscale Detection of Interfacial Charge Transfer of Elegant Z-Scheme WO3/Au/In2S3 Nanowire Arrays," *Nano Lett.*, vol. 16, no. 9, pp. 5547–5552, Sep. 2016, doi: 10.1021/acs.nanolett.6b02094.

[71] I. Shown *et al.*, "Carbon-doped SnS2 nanostructure as a high-efficiency solar fuel catalyst under visible light," *Nat. Commun.*, vol. 9, no. 1, p. 169, Jan. 2018, doi: 10.1038/s41467-017-02547-4.

[72] S. Wang, B. Y. Guan, and X. W. D. Lou, "Construction of ZnIn2S4–In2O3 Hierarchical Tubular Heterostructures for Efficient CO2 Photoreduction," *J. Am. Chem. Soc.*, vol. 140, no. 15, pp. 5037–5040, Apr. 2018, doi: 10.1021/jacs.8b02200.

[73] S. Sun *et al.*, "Highly correlation of CO2 reduction selectivity and surface electron Accumulation: A case study of Au-MoS2 and Ag-MoS2 catalyst," *Appl. Catal. B Environ.*, vol. 271, p. 118931, Aug. 2020, doi: 10.1016/j.apcatb.2020.118931.

[74] P. Li, X. Zhang, C. Hou, Y. Chen, and T. He, "Highly efficient visible-light driven solar-fuel production over tetra(4-carboxyphenyl)porphyrin iron(III) chloride using CdS/Bi2S3 heterostructure as photosensitizer," *Appl. Catal. B Environ.*, vol. 238, pp. 656–663, Dec. 2018, doi: 10.1016/j.apcatb.2018.07.066.

[75] L. Huang *et al.*, "Fabrication of hierarchical Co 3 O 4 @CdIn 2 S 4 p–n heterojunction photocatalysts for improved CO 2 reduction with visible light," *J. Mater. Chem. A*, vol. 8, no. 15, pp. 7177–7183, 2020, doi: 10.1039/D0TA01817A.

[76] F. Deng *et al.*, "Novel visible-light-driven direct Z-scheme CdS/CuInS2 nanoplates for excellent photocatalytic degradation performance and highly-efficient Cr(VI) reduction," *Chem. Eng. J.*, vol. 361, pp. 1451–1461, Apr. 2019, doi: 10.1016/j.cej.2018.10.176.

[77]   M. Zhou, S. Wang, P. Yang, C. Huang, and X. Wang, "Boron Carbon Nitride Semiconductors Decorated with CdS Nanoparticles for Photocatalytic Reduction of CO2," *ACS Catal.*, vol. 8, no. 6, pp. 4928–4936, Jun. 2018, doi: 10.1021/acscatal.8b00104.

[78]   A. Raza, H. Shen, and A. A. Haidry, "Novel Cu2ZnSnS4/Pt/g-C3N4 heterojunction photocatalyst with straddling band configuration for enhanced solar to fuel conversion," *Appl. Catal. B Environ.*, vol. 277, p. 119239, Nov. 2020, doi: 10.1016/j.apcatb.2020.119239.

[79]   S. Wu *et al.*, "Stabilizing CuGaS 2 by crystalline CdS through an interfacial Z-scheme charge transfer for enhanced photocatalytic CO 2 reduction under visible light," *Nanoscale*, vol. 12, no. 16, pp. 8693–8700, 2020, doi: 10.1039/D0NR00483A.

[80]   X. Li *et al.*, "Selective visible-light-driven photocatalytic CO2 reduction to CH4 mediated by atomically thin CuIn5S8 layers," *Nat. Energy*, vol. 4, no. 8, pp. 690–699, Aug. 2019, doi: 10.1038/s41560-019-0431-1.

[81]   K. Kočí *et al.*, "ZnS/MMT nanocomposites: The effect of ZnS loading in MMT on the photocatalytic reduction of carbon dioxide," *Appl. Catal. B Environ.*, vol. 158–159, pp. 410–417, Oct. 2014, doi: 10.1016/j.apcatb.2014.04.048.

[82]   N. Sharma, T. Das, S. Kumar, R. Bhosale, M. Kabir, and S. Ogale, "Photocatalytic Activation and Reduction of CO2 to CH4 over Single Phase Nano Cu3SnS4: A Combined Experimental and Theoretical Study," *ACS Appl. Energy Mater.*, vol. 2, no. 8, pp. 5677–5685, Aug. 2019, doi: 10.1021/acsaem.9b00813.

[83]   Y. Guo, Y. Ao, P. Wang, and C. Wang, "Mediator-free direct dual-Z-scheme Bi2S3/BiVO4/MgIn2S4 composite photocatalysts with enhanced visible-light-driven performance towards carbamazepine degradation," *Appl. Catal. B Environ.*, vol. 254, pp. 479–490, Oct. 2019, doi: 10.1016/j.apcatb.2019.04.031.

[84]   R. Zeng *et al.*, "Versatile Synthesis of Hollow Metal Sulfides via Reverse Cation Exchange Reactions for Photocatalytic CO2 Reduction," *Angew. Chem. Int. Ed.*, vol. 60, no. 47, pp. 25055–25062, 2021, doi: 10.1002/anie.202110670.

[85]   J. Tang, J. R. Durrant, and D. R. Klug, "Mechanism of Photocatalytic Water Splitting in TiO2. Reaction of Water with Photoholes, Importance of Charge Carrier Dynamics, and Evidence for Four-Hole Chemistry," *J. Am. Chem. Soc.*, vol. 130, no. 42, pp. 13885–13891, Oct. 2008, doi: 10.1021/ja8034637.

[86]   Y. Song *et al.*, "Heterojunction Engineering of Multinary Metal Sulfide-Based Photocatalysts for Efficient Photocatalytic Hydrogen Evolution," *Adv. Mater.*, vol. 36, no. 11, p. 2305835, 2024, doi: 10.1002/adma.202305835.

[87]   Y. Liu *et al.*, "Photocatalytic Hydrogen Evolution Using Ternary-Metal-Sulfide/TiO2 Heterojunction Photocatalysts," *ChemCatChem*, vol. 14, no. 5, p. e202101439, 2022, doi: 10.1002/cctc.202101439.

[88]   K. Qi, C. Imparato, O. Almjasheva, A. Khataee, and W. Zheng, "TiO2-based photocatalysts from type-II to S-scheme heterojunction and their applications," *J. Colloid Interface Sci.*, vol. 675, pp. 150–191, Dec. 2024, doi: 10.1016/j.jcis.2024.06.204.

[89]   H. Liu *et al.*, "Ultrafast interfacial charge evolution of the Type-II cadmium Sulfide/Molybdenum disulfide heterostructure for photocatalytic hydrogen production," *J. Colloid Interface Sci.*, vol. 619, pp. 246–256, Aug. 2022, doi: 10.1016/j.jcis.2022.03.080.

[90]   X. Zhang, Z. Cheng, P. Deng, L. Zhang, and Y. Hou, "NiSe2/Cd0.5Zn0.5S as a type-II heterojunction photocatalyst for enhanced photocatalytic hydrogen evolution," *Int. J. Hydrog. Energy*, vol. 46, no. 29, pp. 15389–15397, Apr. 2021, doi: 10.1016/j.ijhydene.2021.02.018.

[91]   X. Ma, D. Li, P. Su, Z. Jiang, and Z. Jin, "S-scheme W18O49/Mn0.2Cd0.8S Heterojunction for Improved Photocatalytic Hydrogen Evolution," *ChemCatChem*, vol. 13, no. 9, pp. 2179–2190, 2021, doi: 10.1002/cctc.202002069.

[92]   L. Yuan, P. Du, L. Yin, J. Yao, J. Wang, and C. Liu, "Metal–organic framework-based S-scheme heterojunction photocatalysts," *Nanoscale*, vol. 16, no. 11, pp. 5487–5503, Mar. 2024, doi: 10.1039/D3NR06677K.

[93] A. Shabbir, S. Sardar, and A. Mumtaz, "Mechanistic investigations of emerging type-II, Z-scheme and S-scheme heterojunctions for photocatalytic applications – A review," *J. Alloys Compd.*, vol. 1003, p. 175683, Oct. 2024, doi: 10.1016/j.jallcom.2024.175683.

[94] H. Tada, T. Mitsui, T. Kiyonaga, T. Akita, and K. Tanaka, "All-solid-state Z-scheme in CdS–Au–TiO2 three-component nanojunction system," *Nat. Mater.*, vol. 5, no. 10, pp. 782–786, Oct. 2006, doi: 10.1038/nmat1734.

[95] H. Liang *et al.*, "Two-dimensional (2D) oxysulfide nanosheets with sulfur-rich vacancy as an visible-light-driven difunctional photocatalyst for hydrogen and oxygen evolution," *J. Alloys Compd.*, vol. 1004, p. 175898, Nov. 2024, doi: 10.1016/j.jallcom.2024.175898.

[96] N. Zhang, Z. Xing, Z. Li, and W. Zhou, "Sulfur vacancy engineering of metal sulfide photocatalysts for solar energy conversion," *Chem Catal.*, vol. 3, no. 1, Jan. 2023, doi: 10.1016/j.checat.2022.08.021.

[97] M. Rezaei, A. Nezamzadeh-Ejhieh, and A. R. Massah, "A Comprehensive Review on the Boosted Effects of Anion Vacancy in the Photocatalytic Solar Water Splitting: Focus on Sulfur Vacancy," *Energy Fuels*, vol. 38, no. 9, pp. 7637–7664, May 2024, doi: 10.1021/acs.energyfuels.4c00325.

[98] D. G. Moon *et al.*, "A review on binary metal sulfide heterojunction solar cells," *Sol. Energy Mater. Sol. Cells*, vol. 200, p. 109963, Sep. 2019, doi: 10.1016/j.solmat.2019.109963.

[99] M. Ibrahim *et al.*, "Synthesis and characterization of Mo-doped PbS thin films for enhancing the photocatalytic hydrogen production," *Mater. Chem. Phys.*, vol. 315, p. 128962, Mar. 2024, doi: 10.1016/j.matchemphys.2024.128962.

[100] M. Shaban, M. Rabia, A. M. A. El-Sayed, A. Ahmed, and S. Sayed, "Photocatalytic properties of PbS/graphene oxide/polyaniline electrode for hydrogen generation," *Sci. Rep.*, vol. 7, no. 1, p. 14100, Oct. 2017, doi: 10.1038/s41598-017-14582-8.

[101] M. Liu, F. Liu, F. Xue, J. Shi, H. Huang, and N. Li, "Photocatalytic Hydrogen Production Over CdS -based Photocatalysts," in *Water Photo- and Electro-Catalysis*, John Wiley & Sons, Ltd, 2024, pp. 35–106. doi: 10.1002/9783527831005.ch2.

[102] C.-J. Chang, K.-L. Huang, J.-K. Chen, K.-W. Chu, and M.-H. Hsu, "Improved photocatalytic hydrogen production of ZnO/ZnS based photocatalysts by Ce doping," *J. Taiwan Inst. Chem. Eng.*, vol. 55, pp. 82–89, Oct. 2015, doi: 10.1016/j.jtice.2015.04.024.

[103] Z. Liang, R. Shen, Y. H. Ng, P. Zhang, Q. Xiang, and X. Li, "A review on 2D MoS2 cocatalysts in photocatalytic H2 production," *J. Mater. Sci. Technol.*, vol. 56, pp. 89–121, Nov. 2020, doi: 10.1016/j.jmst.2020.04.032.

[104] S. R. Kadam, S. Ghosh, R. Bar-Ziv, and M. Bar-Sadan, "Structural Transformation of SnS2 to SnS by Mo Doping Produces Electro/Photocatalyst for Hydrogen Production," *Chem. – Eur. J.*, vol. 26, no. 29, pp. 6679–6685, 2020, doi: 10.1002/chem.202000366.

[105] P. A. K. Reddy, H. Han, K. C. Kim, and S. Bae, "Rational Synthesis of S-Scheme CdS/SnS2 Photocatalysts with Isolated Redox Cocatalysts for Enhanced H2 Production," *ACS Sustain. Chem. Eng.*, vol. 12, no. 12, pp. 4979–4992, Mar. 2024, doi: 10.1021/acssuschemeng.3c08378.

[106] J. Yu, C.-Y. Xu, F.-X. Ma, S.-P. Hu, Y.-W. Zhang, and L. Zhen, "Monodisperse SnS2 Nanosheets for High-Performance Photocatalytic Hydrogen Generation," *ACS Appl. Mater. Interfaces*, vol. 6, no. 24, pp. 22370–22377, Dec. 2014, doi: 10.1021/am506396z.

[107] L. Jing *et al.*, "Different Morphologies of SnS2 Supported on 2D g-C3N4 for Excellent and Stable Visible Light Photocatalytic Hydrogen Generation," *ACS Sustain. Chem. Eng.*, vol. 6, no. 4, pp. 5132–5141, Apr. 2018, doi: 10.1021/acssuschemeng.7b04792.

[108] P. Shen *et al.*, "Construction of a Bi2S3/Bi0.5Na0.5TiO3 Composite Catalyst with S Vacancies for Efficient Piezo-Photocatalytic Hydrogen Production," *Langmuir*, vol. 40, no. 38, pp. 20228–20239, Sep. 2024, doi: 10.1021/acs.langmuir.4c02578.

[109] M. Ganapathy, Y. Hsu, J. Thomas, L.-Y. Chen, C.-T. Chang, and V. Alagan, "Preparation of SrTiO3/Bi2S3 Heterojunction for Efficient Photocatalytic Hydrogen Production," *Energy Fuels*, vol. 35, no. 18, pp. 14995–15004, Sep. 2021, doi: 10.1021/acs.energyfuels.1c00979.

[110] Z. Peng, H. Kobayashi, N. Lu, J. Zhang, J. Sui, and X. Yan, "Direct Z-Scheme In2O3/In2S3 Heterojunction for Oxygen-Mediated Photocatalytic Hydrogen Production," *Energy Fuels*, vol. 36, no. 24, pp. 15100–15111, Dec. 2022, doi: 10.1021/acs.energyfuels.2c03182.

[111] R. Guan *et al.*, "Reduced mesoporous TiO2 with Cu2S heterojunction and enhanced hydrogen production without noble metal cocatalyst," *Appl. Surf. Sci.*, vol. 507, p. 144772, Mar. 2020, doi: 10.1016/j.apsusc.2019.144772.

[112] Y. Zhou, Y. Lei, D. Wang, C. Chen, Q. Peng, and Y. Li, "Ultra-thin Cu 2 S nanosheets: effective cocatalysts for photocatalytic hydrogen production," *Chem. Commun.*, vol. 51, no. 68, pp. 13305–13308, 2015, doi: 10.1039/C5CC05156H.

[113] T. Song, J. Wang, L. Su, H. Zhao, Y. Liu, and W. Tu, "An efficient NiS-ReS2/CdS nanoparticles with dual cocatalysts for photocatalytic hydrogen production," *Int. J. Hydrog. Energy*, vol. 79, pp. 876–882, Aug. 2024, doi: 10.1016/j.ijhydene.2024.07.074.

[114] L. Wei, Y. Chen, J. Zhao, and Z. Li, "Preparation of NiS/ZnIn2S4 as a superior photocatalyst for hydrogen evolution under visible light irradiation," *Beilstein J. Nanotechnol.*, vol. 4, no. 1, pp. 949–955, Dec. 2013, doi: 10.3762/bjnano.4.107.

[115] Y. Qin, L. Xu, Z. Zhu, and W.-Y. Wong, "Strongly coupled interface facilitating charge separation to the improved visible light-driven hydrogen production on CdS@polydopamine/NiS photocatalyst," *J. Mater. Chem. A*, vol. 11, no. 22, pp. 11840–11848, 2023, doi: 10.1039/D3TA01938A.

[116] B. Wang *et al.*, "Heat Diffusion-Induced Gradient Energy Level in Multishell Bisulfides for Highly Efficient Photocatalytic Hydrogen Production," *Adv. Energy Mater.*, vol. 10, no. 32, p. 2001575, 2020, doi: 10.1002/aenm.202001575.

[117] X. Zheng *et al.*, "Recent Advances in Cadmium Sulfide-Based Photocatalysts for Photocatalytic Hydrogen Evolution," *Renewables*, vol. 1, no. 1, pp. 39–56, Jan. 2023, doi: 10.31635/renewables.022.202200001.

[118] L. Cheng, Q. Xiang, Y. Liao, and H. Zhang, "CdS-Based photocatalysts," *Energy Environ. Sci.*, vol. 11, no. 6, pp. 1362–1391, 2018, doi: 10.1039/C7EE03640J.

[119] U. Abdullah, M. Ali, and E. Pervaiz, "An Inclusive Review on Recent Advancements of Cadmium Sulfide Nanostructures and its Hybrids for Photocatalytic and Electrocatalytic Applications," *Mol. Catal.*, vol. 508, p. 111575, May 2021, doi: 10.1016/j.mcat.2021.111575.

[120] Y.-J. Yuan, D. Chen, Z.-T. Yu, and Z.-G. Zou, "Cadmium sulfide-based nanomaterials for photocatalytic hydrogen production," *J. Mater. Chem. A*, vol. 6, no. 25, pp. 11606–11630, 2018, doi: 10.1039/C8TA00671G.

[121] H. Li *et al.*, "S-Scheme Heterojunction/Single-Atom Dual-Driven Charge Transport for Photocatalytic Hydrogen Production," *ACS Catal.*, vol. 14, no. 10, pp. 7308–7320, May 2024, doi: 10.1021/acscatal.4c00758.

[122] T. Li, N. Tsubaki, and Z. Jin, "S-scheme heterojunction in photocatalytic hydrogen production," *J. Mater. Sci. Technol.*, vol. 169, pp. 82–104, Jan. 2024, doi: 10.1016/j.jmst.2023.04.049.

[123] X. Wang and Z. Jin, "Adjusting inter-semiconductor barrier height via crystal plane engineering: Crystalline face exposed single crystal cadmium sulfide augmentative S-scheme heterojunctions for efficiently photocatalytic hydrogen production," *Appl. Catal. B Environ.*, vol. 342, p. 123373, Mar. 2024, doi: 10.1016/j.apcatb.2023.123373.

[124] Y. Liu *et al.*, "Flower-shaped S-scheme CdS-ZnO nanorods heterojunction assisted with SPR of low-content Au for accelerating photocatalytic hydrogen production," *Int. J. Hydrog. Energy*, vol. 58, pp. 302–309, Mar. 2024, doi: 10.1016/j.ijhydene.2024.01.180.

[125] Q. Sun, N. Wang, J. Yu, and J. C. Yu, "A Hollow Porous CdS Photocatalyst," *Adv. Mater.*, vol. 30, no. 45, p. 1804368, 2018, doi: 10.1002/adma.201804368.

[126] H. Cheng, X.-J. Lv, S. Cao, Z.-Y. Zhao, Y. Chen, and W.-F. Fu, "Robustly photogenerating H2 in water using FeP/CdS catalyst under solar irradiation," *Sci. Rep.*, vol. 6, no. 1, p. 19846, Jan. 2016, doi: 10.1038/srep19846.

[127] W. Zhang et al., "Double Z-Scheme Heterojunctions of CdS and ZnS Nanoparticles on Bimetallic Metal–Organic Frameworks for Photocatalytic H2 Production," *ACS Appl. Nano Mater.*, vol. 7, no. 18, pp. 21993–22001, Sep. 2024, doi: 10.1021/acsanm.4c04005.
[128] K. Wang, N. Zhao, H. Xie, J. Wang, W. Xu, and Z. Jin, "Cocatalyst 1T-WS2 assisted Prussian blue derivatives Ni-CdS to enhance photocatalytic hydrogen production driven by visible light," *Int. J. Hydrog. Energy*, vol. 61, pp. 296–306, Apr. 2024, doi: 10.1016/j.ijhydene.2024.02.265.
[129] P. Varma and D. A. Reddy, "Effect of Phosphorus-Doped Phase-Modulated WS2 Nanosheets on CdS Nanorods for Highly Efficient Photocatalytic Hydrogen Production," *ACS Appl. Energy Mater.*, vol. 7, no. 10, pp. 4581–4593, May 2024, doi: 10.1021/acsaem.4c00650.
[130] Y. Wang et al., "Layered deposited MoS2 nanosheets on acorn leaf like CdS as an efficient anti-photocorrosion photocatalyst for hydrogen production," *Fuel*, vol. 368, p. 131621, Jul. 2024, doi: 10.1016/j.fuel.2024.131621.
[131] D. P. Kumar, S. Hong, D. A. Reddy, and T. K. Kim, "Ultrathin MoS2 layers anchored exfoliated reduced graphene oxide nanosheet hybrid as a highly efficient cocatalyst for CdS nanorods towards enhanced photocatalytic hydrogen production," *Appl. Catal. B Environ.*, vol. 212, pp. 7–14, Sep. 2017, doi: 10.1016/j.apcatb.2017.04.065.
[132] R. Mangiri et al., "Decorating MoS2 and CoSe2 nanostructures on 1D-CdS nanorods for boosting photocatalytic hydrogen evolution rate," *J. Mol. Liq.*, vol. 289, p. 111164, Sep. 2019, doi: 10.1016/j.molliq.2019.111164.
[133] F. Saleem et al., "Synergistic effect of Cu/Ni cocatalysts on CdS for sun-light driven hydrogen generation from water splitting," *Int. J. Hydrog. Energy*, vol. 52, pp. 305–319, Jan. 2024, doi: 10.1016/j.ijhydene.2023.05.048.
[134] W. Tang et al., "CdS nanorods decorated with non-precious metal Bi spheres for photocatalytic hydrogen production," *Mater. Today Chem.*, vol. 37, p. 102000, Apr. 2024, doi: 10.1016/j.mtchem.2024.102000.
[135] Z.-G. Liu et al., "Decorating CdS with cobaltous hydroxide and graphene dual cocatalyst for photocatalytic hydrogen production coupled selective benzyl alcohol oxidation," *Mol. Catal.*, vol. 553, p. 113738, Jan. 2024, doi: 10.1016/j.mcat.2023.113738.
[136] A. Qian et al., "Photocatalytic hydrogen production using novel noble-metal-free NiMo/CdS photocatalysts with hollow nanospheres structure," *Mol. Catal.*, vol. 555, p. 113878, Feb. 2024, doi: 10.1016/j.mcat.2024.113878.
[137] H. Li, J. Pan, W. Zhao, and C. Li, "The 2D nickel-molybdenum bimetals sulfide synergistic modified hollow cubic CdS towards enhanced photocatalytic water splitting hydrogen production," *Appl. Surf. Sci.*, vol. 497, p. 143769, Dec. 2019, doi: 10.1016/j.apsusc.2019.143769.
[138] Y.-A. Chen et al., "Double-Hollow Au@CdS Yolk@Shell Nanostructures as Superior Plasmonic Photocatalysts for Solar Hydrogen Production," *Adv. Funct. Mater.*, vol. n/a, no. n/a, p. 2402392, doi: 10.1002/adfm.202402392.
[139] Z. Sun et al., "The enhancement of CdS ultrathin nanosheets photocatalytic activity for water splitting via activating the (001) polar facet by hydrogenation and its charge separation mechanism," *Catal. Sci. Technol.*, vol. 14, no. 17, pp. 5032–5044, Aug. 2024, doi: 10.1039/D4CY00256C.
[140] E. Hussain et al., "Sunlight-Driven Hydrogen Generation: Acceleration of Synergism between Cu–Ag Cocatalysts on a CdS System," *ACS Appl. Energy Mater.*, vol. 7, no. 5, pp. 1914–1926, Mar. 2024, doi: 10.1021/acsaem.3c03010.
[141] Y. Bi et al., "Efficient metal–organic framework-based dual co-catalysts system assist CdS for hydrogen production from photolysis of water," *J. Colloid Interface Sci.*, vol. 661, pp. 501–511, May 2024, doi: 10.1016/j.jcis.2024.01.158.
[142] L. Chen et al., "Noble-metal-free bimetallic nitride decorated CdS nanorods for photocatalytic hydrogen generation," *CrystEngComm*, vol. 26, no. 20, pp. 2587–2593, May 2024, doi: 10.1039/D4CE00216D.

[143] S. Oros-Ruiz, A. Hernández-Gordillo, C. García-Mendoza, A. A. Rodríguez-Rodríguez, and R. Gómez, "Comparative activity of CdS nanofibers superficially modified by Au, Cu, and Ni nanoparticles as co-catalysts for photocatalytic hydrogen production under visible light," *J. Chem. Technol. Biotechnol.*, vol. 91, no. 8, pp. 2205–2210, 2016, doi: 10.1002/jctb.4992.

[144] H. Chen *et al.*, "Nickel-Based Bifunctional Cocatalyst to Enhance CdS Photocatalytic Hydrogen Production," *Langmuir*, vol. 39, no. 51, pp. 18935–18945, Dec. 2023, doi: 10.1021/acs.langmuir.3c02855.

[145] S. Qiu *et al.*, "CdS nanoflakes decorated by Ni(OH)2 nanoparticles for enhanced photocatalytic hydrogen production," *Int. J. Energy Res.*, vol. 45, no. 10, pp. 14985–14994, 2021, doi: 10.1002/er.6777.

[146] B. Zhang *et al.*, "Simultaneous Ni nanoparticles decoration and Ni doping of CdS nanorods for synergistically promoting photocatalytic H2 evolution," *Appl. Surf. Sci.*, vol. 508, p. 144869, Apr. 2020, doi: 10.1016/j.apsusc.2019.144869.

[147] J. Pan *et al.*, "The enhanced photocatalytic hydrogen production of nickel-cobalt bimetals sulfide synergistic modified CdS nanorods with active facets," *Renew. Energy*, vol. 156, pp. 469–477, Aug. 2020, doi: 10.1016/j.renene.2020.04.053.

[148] T. Di, L. Zhang, B. Cheng, J. Yu, and J. Fan, "CdS nanosheets decorated with Ni@graphene core-shell cocatalyst for superior photocatalytic H2 production," *J. Mater. Sci. Technol.*, vol. 56, pp. 170–178, Nov. 2020, doi: 10.1016/j.jmst.2020.03.032.

[149] 李晨晨那永 and LI Chenchen N. Y., "双功能复合材料g-C3N4/CdS/Ni催化光解水产氢和5-羟甲基糠醛氧化性能," 高等学校化学学报, vol. 42, no. 9, p. 2896, Sep. 2021, doi: 10.7503/cjcu20210255.

[150] H. Shi *et al.*, "NiCS3: A cocatalyst surpassing Pt for photocatalytic hydrogen production," *J. Colloid Interface Sci.*, vol. 659, pp. 878–885, Apr. 2024, doi: 10.1016/j.jcis.2023.12.183.

[151] X. Liu *et al.*, "CdS-based Schottky junctions for efficient visible light photocatalytic hydrogen evolution," *J. Colloid Interface Sci.*, vol. 673, pp. 1–8, Nov. 2024, doi: 10.1016/j.jcis.2024.06.040.

[152] S. Du *et al.*, "NiSe2 as Co-Catalyst with CdS: Nanocomposites for High-Performance Photodriven Hydrogen Evolution under Visible-Light Irradiation," *ChemPlusChem*, vol. 84, no. 7, pp. 999–1010, 2019, doi: 10.1002/cplu.201900380.

[153] R. Mangiri *et al.*, "Boosting solar driven hydrogen evolution rate of CdS nanorods adorned with MoS2 and SnS2 nanostructures," *Colloid Interface Sci. Commun.*, vol. 43, p. 100437, Jul. 2021, doi: 10.1016/j.colcom.2021.100437.

[154] R. Mangiri, K. Sunil kumar, K. Subramanyam, A. Sudharani, D. A. Reddy, and R. P. Vijayalakshmi, "Enhanced solar driven hydrogen evolution rate by integrating dual co-catalysts (MoS2, SeS2) on CdS nanorods," *Colloids Surf. Physicochem. Eng. Asp.*, vol. 625, p. 126852, Sep. 2021, doi: 10.1016/j.colsurfa.2021.126852.

[155] Z. Guo, W. Li, Y. He, G. Li, K. Zheng, and C. Xu, "Effect of Cd source on photocatalytic H2 evolution over CdS/MoS2 composites synthesised via a one-pot hydrothermal strategy," *Appl. Surf. Sci.*, vol. 512, p. 145750, May 2020, doi: 10.1016/j.apsusc.2020.145750.

[156] D. Praveen Kumar, S. Hong, D. Amaranatha Reddy, and T. Kyu Kim, "Noble metal-free ultrathin MoS 2 nanosheet-decorated CdS nanorods as an efficient photocatalyst for spectacular hydrogen evolution under solar light irradiation," *J. Mater. Chem. A*, vol. 4, no. 47, pp. 18551–18558, 2016, doi: 10.1039/C6TA08628D.

[157] L. Yang *et al.*, "Optical Properties of Metal–Molybdenum Disulfide Hybrid Nanosheets and Their Application for Enhanced Photocatalytic Hydrogen Evolution," *ACS Nano*, vol. 8, no. 7, pp. 6979–6985, Jul. 2014, doi: 10.1021/nn501807y.

[158] D. Amaranatha Reddy *et al.*, "Hierarchical dandelion-flower-like cobalt-phosphide modified CdS/reduced graphene oxide-MoS 2 nanocomposites as a noble-metal-free catalyst for efficient hydrogen evolution from water," *Catal. Sci. Technol.*, vol. 6, no. 16, pp. 6197–6206, 2016, doi: 10.1039/C6CY00768F.

[159] H. Lee *et al.*, "Drastic Improvement of 1D-CdS Solar-Driven Photocatalytic Hydrogen Evolution Rate by Integrating with NiFe Layered Double Hydroxide Nanosheets Synthesized by Liquid-Phase Pulsed-Laser Ablation," *ACS Sustain. Chem. Eng.*, vol. 6, no. 12, pp. 16734–16743, Dec. 2018, doi: 10.1021/acssuschemeng.8b04000.
[160] R. Lv *et al.*, "Homologous heterostructure CdSe/CdS nanoflowers to enhance photocatalytic hydrogen production," *Colloids Surf. Physicochem. Eng. Asp.*, vol. 684, p. 133143, Mar. 2024, doi: 10.1016/j.colsurfa.2024.133143.
[161] K. He, "ZnO/ZnS/CdS three-phase composite photocatalyst with a flower cluster structure: Research on its preparation and photocatalytic activity hydrogen production," *Int. J. Hydrog. Energy*, vol. 51, pp. 30–40, Jan. 2024, doi: 10.1016/j.ijhydene.2023.08.050.
[162] X. Gui *et al.*, "Construction of ternary Ni2P/ZIF-8/CdS composite for efficient photocatalytic hydrogen production and pollutant degradation: Accelerating separation of photogenerated carriers," *J. Phys. Chem. Solids*, vol. 190, p. 111983, Jul. 2024, doi: 10.1016/j.jpcs.2024.111983.
[163] P. Varma and D. A. Reddy, "Interfacial Heterojunction Formation Modulating the Charge Carrier's Pathway of CdS@CoSe2/WS2 Nanocomposites for Enhanced Photocatalytic Hydrogen Evolution," *Energy Fuels*, vol. 38, no. 14, pp. 13315–13326, Jul. 2024, doi: 10.1021/acs.energyfuels.4c01648.
[164] N. Li *et al.*, "Efficient, Full Spectrum-Driven H2 Evolution Z-Scheme Co2P/CdS Photocatalysts with Co–S Bonds," *ACS Appl. Mater. Interfaces*, vol. 11, no. 25, pp. 22297–22306, Jun. 2019, doi: 10.1021/acsami.9b03965.
[165] E. H. Kim *et al.*, "Hollow CoSe2 nanocages derived from metal–organic frameworks as efficient non-precious metal co-catalysts for photocatalytic hydrogen production," *Catal. Sci. Technol.*, vol. 9, no. 17, pp. 4702–4710, Aug. 2019, doi: 10.1039/C9CY00843H.
[166] Q. Gai *et al.*, "2D-2D heterostructured CdS–CoP photocatalysts for efficient H2 evolution under visible light irradiation," *Int. J. Hydrog. Energy*, vol. 44, no. 50, pp. 27412–27420, Oct. 2019, doi: 10.1016/j.ijhydene.2019.08.196.
[167] X. Cheng *et al.*, "Interfacial effect between Ni2P/CdS for simultaneously heightening photocatalytic hydrogen production and lignocellulosic biomass photorefining," *J. Colloid Interface Sci.*, vol. 655, pp. 943–952, Feb. 2024, doi: 10.1016/j.jcis.2023.11.031.
[168] J.-S. Hu *et al.*, "Mass Production and High Photocatalytic Activity of ZnS Nanoporous Nanoparticles," *Angew. Chem. Int. Ed.*, vol. 44, no. 8, pp. 1269–1273, 2005, doi: 10.1002/anie.200462057.
[169] G.-J. Lee and J. J. Wu, "Recent developments in ZnS photocatalysts from synthesis to photocatalytic applications — A review," *Powder Technol.*, vol. 318, pp. 8–22, Aug. 2017, doi: 10.1016/j.powtec.2017.05.022.
[170] T. Arai *et al.*, "Cu-Doped ZnS Hollow Particle with High Activity for Hydrogen Generation from Alkaline Sulfide Solution under Visible Light," *Chem. Mater.*, vol. 20, no. 5, pp. 1997–2000, Mar. 2008, doi: 10.1021/cm071803p.
[171] J. Zhang, Y. Wang, J. Zhang, Z. Lin, F. Huang, and J. Yu, "Enhanced Photocatalytic Hydrogen Production Activities of Au-Loaded ZnS Flowers," *ACS Appl. Mater. Interfaces*, vol. 5, no. 3, pp. 1031–1037, Feb. 2013, doi: 10.1021/am302726y.
[172] G. Wang *et al.*, "Synthesis and characterization of ZnS with controlled amount of S vacancies for photocatalytic H2 production under visible light," *Sci. Rep.*, vol. 5, no. 1, p. 8544, Feb. 2015, doi: 10.1038/srep08544.
[173] T. Lange *et al.*, "Zinc sulfide for photocatalysis: White angel or black sheep?," *Prog. Mater. Sci.*, vol. 124, p. 100865, Feb. 2022, doi: 10.1016/j.pmatsci.2021.100865.
[174] Y. Zheng *et al.*, "A dual photocarrier separation channel in CdS/ZnS for outstanding photocatalytic hydrogen evolution," *New J. Chem.*, vol. 48, no. 27, pp. 12227–12234, 2024, doi: 10.1039/D4NJ01184H.

[175] D. Jiang, Z. Sun, H. Jia, D. Lu, and P. Du, "A cocatalyst-free CdS nanorod/ZnS nanoparticle composite for high-performance visible-light-driven hydrogen production from water," *J. Mater. Chem. A*, vol. 4, no. 2, pp. 675–683, 2016, doi: 10.1039/C5TA07420G.

[176] J. Kundu, B. K. Satpathy, and D. Pradhan, "Composition-Controlled CdS/ZnS Heterostructure Nanocomposites for Efficient Visible Light Photocatalytic Hydrogen Generation," *Ind. Eng. Chem. Res.*, vol. 58, no. 51, pp. 22709–22717, Dec. 2019, doi: 10.1021/acs.iecr.9b03764.

[177] Y. Lin *et al.*, "The Evolution from a Typical Type-I CdS/ZnS to Type-II and Z-Scheme Hybrid Structure for Efficient and Stable Hydrogen Production under Visible Light," *ACS Sustain. Chem. Eng.*, vol. 8, no. 11, pp. 4537–4546, Mar. 2020, doi: 10.1021/acssuschemeng.0c00101.

[178] X. Zhang, M. Puttaswamy, H. Bai, B. Hou, and S. Kumar Verma, "CdS/ZnS core–shell nanorod heterostructures co-deposited with ultrathin MoS2 cocatalyst for competent hydrogen evolution under visible-light irradiation," *J. Colloid Interface Sci.*, vol. 665, pp. 430–442, Jul. 2024, doi: 10.1016/j.jcis.2024.03.066.

[179] J. Zhang, J. Yu, Y. Zhang, Q. Li, and J. R. Gong, "Visible Light Photocatalytic H2-Production Activity of CuS/ZnS Porous Nanosheets Based on Photoinduced Interfacial Charge Transfer," *Nano Lett.*, vol. 11, no. 11, pp. 4774–4779, Nov. 2011, doi: 10.1021/nl202587b.

[180] Y. Hong *et al.*, "Enhanced visible light photocatalytic hydrogen production activity of CuS/ZnS nanoflower spheres," *J. Mater. Chem. A*, vol. 3, no. 26, pp. 13913–13919, 2015, doi: 10.1039/C5TA02500A.

[181] G. A. Naikoo, "Boosting green hydrogen production with ZnS@MoS2 2D materials: A structural, electrochemical and photocatalytic analysis," *Mater. Today Commun.*, vol. 41, p. 110175, Dec. 2024, doi: 10.1016/j.mtcomm.2024.110175.

[182] Y.-T. Xiong *et al.*, "Broad Light Absorption and Multichannel Charge Transfer Mediated by Topological Surface State in CdS/ZnS/Bi2Se3 Nanotubes for Improved Photocatalytic Hydrogen Production," *Adv. Funct. Mater.*, vol. n/a, no. n/a, p. 2407819, doi: 10.1002/adfm.202407819.

[183] N. R. Vempuluru *et al.*, "ZnS/ZnSe heterojunction photocatalyst for augmented hydrogen production: Experimental and theoretical insights," *Int. J. Hydrog. Energy*, vol. 51, pp. 524–539, Jan. 2024, doi: 10.1016/j.ijhydene.2023.08.249.

[184] A. Bolatov *et al.*, "Ternary ZnS/ZnO/Graphitic Carbon Nitride Heterojunction for Photocatalytic Hydrogen Production," *Materials*, vol. 17, no. 19, Art. no. 19, Jan. 2024, doi: 10.3390/ma17194877.

[185] J. Li *et al.*, "Fabrication of Hexagonal Prism-Shaped Double S-Scheme Cu2O@CdS/ZnS Heterojunctions Utilizing Zeolite Templates: Enhanced Photocatalytic Hydrogen Production and Mechanism Insight," *Cryst. Growth Des.*, Oct. 2024, doi: 10.1021/acs.cgd.4c00651.

[186] L. Tie *et al.*, "In-situ construction of graphene oxide in microsphere ZnS photocatalyst for high-performance photochemical hydrogen generation," *Int. J. Hydrog. Energy*, vol. 45, no. 33, pp. 16606–16613, Jun. 2020, doi: 10.1016/j.ijhydene.2020.04.158.

[187] K. Yu *et al.*, "Engineering cation defect-mediated Z-scheme photocatalysts for a highly efficient and stable photocatalytic hydrogen production," *J. Mater. Chem. A*, vol. 9, no. 12, pp. 7759–7766, 2021, doi: 10.1039/D0TA12269F.

[188] H. Ren *et al.*, "ZnO@ZnS core–shell nanorods with homologous heterogeneous interface to enhance photocatalytic hydrogen production," *Colloids Surf. Physicochem. Eng. Asp.*, vol. 652, p. 129844, Nov. 2022, doi: 10.1016/j.colsurfa.2022.129844.

[189] C.-J. Chang, Y.-H. Wei, and K.-P. Huang, "Photocatalytic hydrogen production by flower-like graphene supported ZnS composite photocatalysts," *Int. J. Hydrog. Energy*, vol. 42, no. 37, pp. 23578–23586, Sep. 2017, doi: 10.1016/j.ijhydene.2017.04.219.

[190] Z. Liu, J. Xu, C. Xiang, Y. Liu, L. Ma, and L. Hu, "S-scheme heterojunction based on ZnS/CoMoO4 ball-and-rod composite photocatalyst to promote photocatalytic hydrogen production," *Appl. Surf. Sci.*, vol. 569, p. 150973, Dec. 2021, doi: 10.1016/j.apsusc.2021.150973.

[191] C.-J. Chang and K.-W. Chu, "ZnS/polyaniline composites with improved dispersing stability and high photocatalytic hydrogen production activity," *Int. J. Hydrog. Energy*, vol. 41, no. 46, pp. 21764–21773, Dec. 2016, doi: 10.1016/j.ijhydene.2016.07.155.
[192] E. Puentes-Prado *et al.*, "Enhancing the solar photocatalytic hydrogen generation of ZnS films by UV radiation treatment," *Int. J. Hydrog. Energy*, vol. 45, no. 22, pp. 12308–12317, Apr. 2020, doi: 10.1016/j.ijhydene.2020.02.180.
[193] X. Yuan *et al.*, "Facile synthesis of ZnIn2S4@ZnS composites for efficient photocatalytic hydrogen precipitation," *Catal. Sci. Technol.*, vol. 14, no. 3, pp. 630–637, Feb. 2024, doi: 10.1039/D3CY01475D.
[194] B. Radisavljevic, A. Radenovic, J. Brivio, V. Giacometti, and A. Kis, "Single-layer MoS2 transistors," *Nat. Nanotechnol.*, vol. 6, no. 3, pp. 147–150, Mar. 2011, doi: 10.1038/nnano.2010.279.
[195] Y. Li, H. Wang, L. Xie, Y. Liang, G. Hong, and H. Dai, "MoS2 Nanoparticles Grown on Graphene: An Advanced Catalyst for the Hydrogen Evolution Reaction," *J. Am. Chem. Soc.*, vol. 133, no. 19, pp. 7296–7299, May 2011, doi: 10.1021/ja201269b.
[196] Q. Xiang, J. Yu, and M. Jaroniec, "Synergetic Effect of MoS2 and Graphene as Cocatalysts for Enhanced Photocatalytic H2 Production Activity of TiO2 Nanoparticles," *J. Am. Chem. Soc.*, vol. 134, no. 15, pp. 6575–6578, Apr. 2012, doi: 10.1021/ja302846n.
[197] E. S. Kadantsev and P. Hawrylak, "Electronic structure of a single MoS2 monolayer," *Solid State Commun.*, vol. 152, no. 10, pp. 909–913, May 2012, doi: 10.1016/j.ssc.2012.02.005.
[198] H. S. Lee *et al.*, "MoS2 Nanosheet Phototransistors with Thickness-Modulated Optical Energy Gap," *Nano Lett.*, vol. 12, no. 7, pp. 3695–3700, Jul. 2012, doi: 10.1021/nl301485q.
[199] K. F. Mak, C. Lee, J. Hone, J. Shan, and T. F. Heinz, "Atomically Thin ${\mathrm{MoS}}_{2}$: A New Direct-Gap Semiconductor," *Phys. Rev. Lett.*, vol. 105, no. 13, p. 136805, Sep. 2010, doi: 10.1103/PhysRevLett.105.136805.
[200] C. Lee, H. Yan, L. E. Brus, T. F. Heinz, J. Hone, and S. Ryu, "Anomalous Lattice Vibrations of Single- and Few-Layer MoS2," *ACS Nano*, vol. 4, no. 5, pp. 2695–2700, May 2010, doi: 10.1021/nn1003937.
[201] H. Li *et al.*, "From Bulk to Monolayer MoS2: Evolution of Raman Scattering," *Adv. Funct. Mater.*, vol. 22, no. 7, pp. 1385–1390, 2012, doi: 10.1002/adfm.201102111.
[202] H. J. Conley, B. Wang, J. I. Ziegler, R. F. Jr. Haglund, S. T. Pantelides, and K. I. Bolotin, "Bandgap Engineering of Strained Monolayer and Bilayer MoS2," *Nano Lett.*, vol. 13, no. 8, pp. 3626–3630, Aug. 2013, doi: 10.1021/nl4014748.
[203] H. Li, J. Wu, Z. Yin, and H. Zhang, "Preparation and Applications of Mechanically Exfoliated Single-Layer and Multilayer MoS2 and WSe2 Nanosheets," *Acc. Chem. Res.*, vol. 47, no. 4, pp. 1067–1075, Apr. 2014, doi: 10.1021/ar4002312.
[204] D. Merki and X. Hu, "Recent developments of molybdenum and tungsten sulfides as hydrogen evolution catalysts," *Energy Environ. Sci.*, vol. 4, no. 10, pp. 3878–3888, 2011, doi: 10.1039/C1EE01970H.
[205] A. B. Laursen, S. Kegnæs, S. Dahl, and I. Chorkendorff, "Molybdenum sulfides—efficient and viable materials for electro - and photoelectrocatalytic hydrogen evolution," *Energy Environ. Sci.*, vol. 5, no. 2, pp. 5577–5591, 2012, doi: 10.1039/C2EE02618J.
[206] J. D. Benck, T. R. Hellstern, J. Kibsgaard, P. Chakthranont, and T. F. Jaramillo, "Catalyzing the Hydrogen Evolution Reaction (HER) with Molybdenum Sulfide Nanomaterials," *ACS Catal.*, vol. 4, no. 11, pp. 3957–3971, Nov. 2014, doi: 10.1021/cs500923c.
[207] C. G. Morales-Guio and X. Hu, "Amorphous Molybdenum Sulfides as Hydrogen Evolution Catalysts," *Acc. Chem. Res.*, vol. 47, no. 8, pp. 2671–2681, Aug. 2014, doi: 10.1021/ar5002022.
[208] C. G. Morales-Guio, L.-A. Stern, and X. Hu, "Nanostructured hydrotreating catalysts for electrochemical hydrogen evolution," *Chem. Soc. Rev.*, vol. 43, no. 18, pp. 6555–6569, 2014, doi: 10.1039/C3CS60468C.

[209] Y. Yan, B. Xia, Z. Xu, and X. Wang, "Recent Development of Molybdenum Sulfides as Advanced Electrocatalysts for Hydrogen Evolution Reaction," *ACS Catal.*, vol. 4, no. 6, pp. 1693–1705, Jun. 2014, doi: 10.1021/cs500070x.

[210] X. Zou and Y. Zhang, "Noble metal-free hydrogen evolution catalysts for water splitting," *Chem. Soc. Rev.*, vol. 44, no. 15, pp. 5148–5180, 2015, doi: 10.1039/C4CS00448E.

[211] I. Alfa, H. Y. Hafeez, J. Mohammed, S. Abdu, A. B. Suleiman, and C. E. Ndikilar, "A recent progress and advancement on MoS2-based photocatalysts for efficient solar fuel (hydrogen) generation *via* photocatalytic water splitting," *Int. J. Hydrog. Energy*, vol. 71, pp. 1006–1025, Jun. 2024, doi: 10.1016/j.ijhydene.2024.05.203.

[212] Y. Li *et al.*, "MoO2-C@MoS2: A unique cocatalyst with LSPR effect for enhanced quasi-full-spectrum photocatalytic hydrogen evolution of CdS," *Appl. Catal. B Environ.*, vol. 343, p. 123543, Apr. 2024, doi: 10.1016/j.apcatb.2023.123543.

[213] G. Zeng *et al.*, "Ingenious regulation and activation of sites in the 2H-MoS2 basal planes by oxygen incorporation for enhanced photocatalytic hydrogen evolution of CdS," *Chem. Eng. J.*, vol. 499, p. 156367, Nov. 2024, doi: 10.1016/j.cej.2024.156367.

[214] W. Wang *et al.*, "Fast carrier separation induced by the metal-like O-doped MoS2/CoS cocatalyst for achieving photocatalytic and photothermal hydrogen production," *Chem. Eng. J.*, vol. 493, p. 152516, Aug. 2024, doi: 10.1016/j.cej.2024.152516.

[215] M. Yin, C. Wu, F. Jia, L. Wang, P. Zheng, and Y. Fan, "Efficient photocatalytic hydrogen production over eosin Y-sensitized MoS 2," *RSC Adv.*, vol. 6, no. 79, pp. 75618–75625, 2016, doi: 10.1039/C6RA14710K.

[216] Y.-S. Huang *et al.*, "Vastly improved solar-light induced water splitting catalyzed by few-layer MoS2 on Au nanoparticles utilizing localized surface plasmon resonance," *Nano Energy*, vol. 77, p. 105267, Nov. 2020, doi: 10.1016/j.nanoen.2020.105267.

[217] L. Yin, X. Hai, K. Chang, F. Ichihara, and J. Ye, "Synergetic Exfoliation and Lateral Size Engineering of MoS2 for Enhanced Photocatalytic Hydrogen Generation," *Small*, vol. 14, no. 14, p. 1704153, 2018, doi: 10.1002/smll.201704153.

[218] Y. Li, L. Ding, Z. Liang, Y. Xue, H. Cui, and J. Tian, "Synergetic effect of defects rich MoS2 and Ti3C2 MXene as cocatalysts for enhanced photocatalytic H2 production activity of TiO2," *Chem. Eng. J.*, vol. 383, p. 123178, Mar. 2020, doi: 10.1016/j.cej.2019.123178.

[219] S. Kumar, A. Kumar, V. Navakoteswara Rao, A. Kumar, M. V. Shankar, and V. Krishnan, "Defect-Rich MoS2 Ultrathin Nanosheets-Coated Nitrogen-Doped ZnO Nanorod Heterostructures: An Insight into in-Situ-Generated ZnS for Enhanced Photocatalytic Hydrogen Evolution," *ACS Appl. Energy Mater.*, vol. 2, no. 8, pp. 5622–5634, Aug. 2019, doi: 10.1021/acsaem.9b00790.

[220] G. Hu, T. Guo, C. Wang, J. Liu, Y. Liu, and Q. Guo, "High-performance photocatalytic hydrogen evolution in a Zn0.5Cd0.5S/MoS2 p–n heterojunction," *Vacuum*, vol. 227, p. 113451, Sep. 2024, doi: 10.1016/j.vacuum.2024.113451.

[221] Y. Wang, C. Liu, C. Kong, and F. Zhang, "Defect MoS2 and Ti3C2 nanosheets co-assisted CdS to enhance visible-light driven photocatalytic hydrogen production," *Colloids Surf. Physicochem. Eng. Asp.*, vol. 652, p. 129746, Nov. 2022, doi: 10.1016/j.colsurfa.2022.129746.

[222] C. An *et al.*, "NiS nanoparticle decorated MoS2 nanosheets as efficient promoters for enhanced solar H2 evolution over Zn:XCd1-xS nanorods," *Inorg. Chem. Front.*, vol. 4, no. 6, pp. 1042–1047, 2017, doi: 10.1039/c7qi00170c.

[223] H. Dong *et al.*, "High-throughput Production of ZnO-MoS2-Graphene Heterostructures for Highly Efficient Photocatalytic Hydrogen Evolution," *Materials*, vol. 12, no. 14, Art. no. 14, Jan. 2019, doi: 10.3390/ma12142233.

[224] Z. Mou *et al.*, "Significantly enhanced photocatalytic hydrogen production performance of MoS2/CNTs/CdS with carbon nanotubes as the charge mediators," *Int. J. Hydrog. Energy*, vol. 51, pp. 748–757, 2024, doi: 10.1016/j.ijhydene.2023.07.060.

[225] M. Yin *et al.*, "Insight into the factors influencing the photocatalytic H2 evolution performance of molybdenum sulfide," *New J. Chem.*, vol. 43, no. 3, pp. 1230–1237, Jan. 2019, doi: 10.1039/C8NJ04639E.

[226] C. Liu, J. Ma, F.-J. Zhang, Y.-R. Wang, and C. Kong, "Facile formation of Mo-vacancy defective MoS2/CdS nanoparticles enhanced efficient hydrogen production," *Colloids Surf. Physicochem. Eng. Asp.*, vol. 643, p. 128743, Jun. 2022, doi: 10.1016/j.colsurfa.2022.128743.

[227] G.-C. Lee *et al.*, "Induction of a piezo-potential improves photocatalytic hydrogen production over ZnO/ZnS/MoS2 Heterostructures," *Nano Energy*, vol. 93, p. 106867, Mar. 2022, doi: 10.1016/j.nanoen.2021.106867.

[228] Y. Wang *et al.*, "Plate-on-plate structured MoS2/Cd0.6Zn0.4S Z-scheme heterostructure with enhanced photocatalytic hydrogen production activity *via* hole sacrificial agent synchronously strengthen half-reactions," *J. Colloid Interface Sci.*, vol. 630, pp. 341–351, Jan. 2023, doi: 10.1016/j.jcis.2022.10.053.

[229] C. Zhang, C. Zheng, and X. Cao, "Preparation of MoS2/Ni@NiO/g-C3N4 composite catalyst and its photocatalytic performance for hydrogen production," *Int. J. Hydrog. Energy*, vol. 51, pp. 1078–1086, Jan. 2024, doi: 10.1016/j.ijhydene.2023.07.151.

[230] W. Zhao *et al.*, "Evolution of Electronic Structure in Atomically Thin Sheets of WS2 and WSe2," *ACS Nano*, vol. 7, no. 1, pp. 791–797, Jan. 2013, doi: 10.1021/nn305275h.

[231] A. L. Elías *et al.*, "Controlled Synthesis and Transfer of Large-Area WS2 Sheets: From Single Layer to Few Layers," *ACS Nano*, vol. 7, no. 6, pp. 5235–5242, Jun. 2013, doi: 10.1021/nn400971k.

[232] F. Reale *et al.*, "High-Mobility and High-Optical Quality Atomically Thin WS 2," *Sci. Rep.*, vol. 7, no. 1, p. 14911, Nov. 2017, doi: 10.1038/s41598-017-14928-2.

[233] M. R. Molas, K. Nogajewski, A. O. Slobodeniuk, J. Binder, M. Bartos, and M. Potemski, "Optical response of monolayer, few-layer and bulk tungsten disulfide," Jun. 28, 2017, *arXiv*: arXiv:1706.09285. Accessed: Oct. 25, 2024. [Online]. Available: http://arxiv.org/abs/1706.09285

[234] H. R. Gutiérrez *et al.*, "Extraordinary Room-Temperature Photoluminescence in Triangular WS2 Monolayers," *Nano Lett.*, vol. 13, no. 8, pp. 3447–3454, Aug. 2013, doi: 10.1021/nl3026357.

[235] C. Cong, J. Shang, Y. Wang, and T. Yu, "Optical Properties of 2D Semiconductor WS2," *Adv. Opt. Mater.*, vol. 6, no. 1, p. 1700767, 2018, doi: 10.1002/adom.201700767.

[236] D. Ovchinnikov, A. Allain, Y.-S. Huang, D. Dumcenco, and A. Kis, "Electrical Transport Properties of Single-Layer WS2," *ACS Nano*, vol. 8, no. 8, pp. 8174–8181, Aug. 2014, doi: 10.1021/nn502362b.

[237] Z. Wu, B. Fang, A. Bonakdarpour, A. Sun, D. P. Wilkinson, and D. Wang, "WS2 nanosheets as a highly efficient electrocatalyst for hydrogen evolution reaction," *Appl. Catal. B Environ.*, vol. 125, pp. 59–66, Aug. 2012, doi: 10.1016/j.apcatb.2012.05.013.

[238] Y. Hou, Y. Zhu, Y. Xu, and X. Wang, "Photocatalytic hydrogen production over carbon nitride loaded with WS2 as cocatalyst under visible light," *Appl. Catal. B Environ.*, vol. 156–157, pp. 122–127, Sep. 2014, doi: 10.1016/j.apcatb.2014.03.002.

[239] Y. Zhou, X. Ye, and D. Lin, "One-pot synthesis of non-noble metal WS2/g-C3N4 photocatalysts with enhanced photocatalytic hydrogen production," *Int. J. Hydrog. Energy*, vol. 44, no. 29, pp. 14927–14937, Jun. 2019, doi: 10.1016/j.ijhydene.2019.04.091.

[240] M. Zhu, C. Zhai, M. Fujitsuka, and T. Majima, "Noble metal-free near-infrared-driven photocatalyst for hydrogen production based on 2D hybrid of black Phosphorus/WS2," *Appl. Catal. B Environ.*, vol. 221, pp. 645–651, Feb. 2018, doi: 10.1016/j.apcatb.2017.09.063.

[241] D. Jing and L. Guo, "WS2 sensitized mesoporous TiO2 for efficient photocatalytic hydrogen production from water under visible light irradiation," *Catal. Commun.*, vol. 8, no. 5, pp. 795–799, May 2007, doi: 10.1016/j.catcom.2006.09.009.

[242] M. S. Akple, J. Low, S. Wageh, Ahmed. A. Al-Ghamdi, J. Yu, and J. Zhang, "Enhanced visible light photocatalytic H2-production of g-C3N4/WS2 composite heterostructures," *Appl. Surf. Sci.*, vol. 358, pp. 196–203, Dec. 2015, doi: 10.1016/j.apsusc.2015.08.250.

[243] Y. Zou *et al.*, "WS2/Graphitic Carbon Nitride Heterojunction Nanosheets Decorated with CdS Quantum Dots for Photocatalytic Hydrogen Production," *ChemSusChem*, vol. 11, no. 7, pp. 1187–1197, 2018, doi: 10.1002/cssc.201800053.
[244] M. Jiang, J. Xu, P. Munroe, and Z.-H. Xie, "1D/2D CdS/WS2 heterojunction photocatalyst: First-principles insights for hydrogen production," *Mater. Today Commun.*, vol. 35, p. 105991, Jun. 2023, doi: 10.1016/j.mtcomm.2023.105991.
[245] D. Lin, Y. Zhou, X. Ye, and M. Zhu, "Construction of sandwich structured photocatalyst using monolayer WS2 embedded g-C3N4 for highly efficient H2 production," *Ceram. Int.*, vol. 46, no. 9, pp. 12933–12941, Jun. 2020, doi: 10.1016/j.ceramint.2020.02.061.
[246] Q. Xiang, F. Cheng, and D. Lang, "Hierarchical Layered WS2/Graphene-Modified CdS Nanorods for Efficient Photocatalytic Hydrogen Evolution," *ChemSusChem*, vol. 9, no. 9, pp. 996–1002, 2016, doi: 10.1002/cssc.201501702.
[247] B. Mahler, V. Hoepfner, K. Liao, and G. A. Ozin, "Colloidal Synthesis of 1T-WS2 and 2H-WS2 Nanosheets: Applications for Photocatalytic Hydrogen Evolution," *J. Am. Chem. Soc.*, vol. 136, no. 40, pp. 14121–14127, Oct. 2014, doi: 10.1021/ja506261t.
[248] J. Yi *et al.*, "Solvothermal synthesis of metallic 1T-WS2: A supporting co-catalyst on carbon nitride nanosheets toward photocatalytic hydrogen evolution," *Chem. Eng. J.*, vol. 335, pp. 282–289, Mar. 2018, doi: 10.1016/j.cej.2017.10.125.
[249] A. Subramani *et al.*, "Harnessing Ti3C2-WS2 nanostructures as efficient energy scaffoldings for photocatalytic hydrogen generation," *Mater. Today Sustain.*, vol. 28, p. 100964, Dec. 2024, doi: 10.1016/j.mtsust.2024.100964.
[250] C. Li, X. Zhang, T. Song, Y. Tian, S. Wang, and P. Yang, "Interface-engineering derived WS2/S-g-C3N4 nanosheet Z-scheme heterostructures towards enhanced photocatalytic redox and H2O2 generation," *J. Environ. Chem. Eng.*, vol. 12, no. 5, p. 113396, Oct. 2024, doi: 10.1016/j.jece.2024.113396.
[251] D.-B. Seo *et al.*, "Edge-Rich 3D Structuring of Metal Chalcogenide/Graphene with Vertical Nanosheets for Efficient Photocatalytic Hydrogen Production," *ACS Appl. Mater. Interfaces*, vol. 16, no. 22, pp. 28613–28624, Jun. 2024, doi: 10.1021/acsami.4c04329.
[252] R. Cao, H. Yuan, N. Yang, Q. Lu, Y. Xue, and X. Zeng, "Enhanced photocatalytic hydrogen production utilizing few-layered 1T-WS2/g-C3N4 heterostructures prepared with one-step calcination route," *Fuel*, vol. 357, p. 129808, Feb. 2024, doi: 10.1016/j.fuel.2023.129808.
[253] J. Ren *et al.*, "First-principles study of the electronic, optical adsorption, and photocatalytic water-splitting properties of a strain-tuned SiC/WS2 heterojunction," *Int. J. Hydrog. Energy*, vol. 87, pp. 554–565, Oct. 2024, doi: 10.1016/j.ijhydene.2024.09.036.
[254] K. Liang, M. Yin, D. Ma, Y. Fan, and Z. Li, "Facile preparation and photocatalytic hydrogen production of WS2 and its composites," *Int. J. Hydrog. Energy*, vol. 47, no. 91, pp. 38622–38634, Nov. 2022, doi: 10.1016/j.ijhydene.2022.09.058.
[255] T.-M. Tien, Y.-J. Chung, C.-T. Huang, and E. L. Chen, "Fabrication of WS2/WSe2 Z-Scheme Nano-Heterostructure for Efficient Photocatalytic Hydrogen Production and Removal of Congo Red under Visible Light," *Catalysts*, vol. 12, no. 8, Art. no. 8, Aug. 2022, doi: 10.3390/catal12080852.
[256] D. Ma, M. Yin, K. Liang, M. Xue, Y. Fan, and Z. Li, "Simple synthesis and efficient photocatalytic hydrogen production of WO3-WS2 and WO3–WS2–MoS2," *Mater. Sci. Semicond. Process.*, vol. 167, p. 107788, Nov. 2023, doi: 10.1016/j.mssp.2023.107788.
[257] D. A. Reddy, H. Park, R. Ma, D. P. Kumar, M. Lim, and T. K. Kim, "Heterostructured WS2-MoS2 Ultrathin Nanosheets Integrated on CdS Nanorods to Promote Charge Separation and Migration and Improve Solar-Driven Photocatalytic Hydrogen Evolution," *ChemSusChem*, vol. 10, no. 7, pp. 1563–1570, 2017, doi: 10.1002/cssc.201601799.
[258] T. Padma, D. K. Gara, A. N. Reddy, S. V. P. Vattikuti, and C. M. Julien, "MoSe2-WS2 Nanostructure for an Efficient Hydrogen Generation under White Light LED Irradiation," *Nanomaterials*, vol. 12, no. 7, Art. no. 7, Jan. 2022, doi: 10.3390/nano12071160.

[259] D. Xu *et al.*, "High Yield Exfoliation of WS2 Crystals into 1–2 Layer Semiconducting Nanosheets and Efficient Photocatalytic Hydrogen Evolution from WS2/CdS Nanorod Composites," *ACS Appl. Mater. Interfaces*, vol. 10, no. 3, pp. 2810–2818, Jan. 2018, doi: 10.1021/acsami.7b15614.
[260] S. Wei, C. Guo, L. Wang, J. Xu, and H. Dong, "Bacterial synthesis of PbS nanocrystallites in one-step with l-cysteine serving as both sulfur source and capping ligand," *Sci. Rep.*, vol. 11, no. 1, p. 1216, Jan. 2021, doi: 10.1038/s41598-020-80450-7.
[261] W. Cui, S. Ma, L. Liu, and Y. Liang, "PbS-sensitized K2Ti4O9 composite: Preparation and photocatalytic properties for hydrogen evolution under visible light irradiation," *Chem. Eng. J.*, vol. 204–206, pp. 1–7, Sep. 2012, doi: 10.1016/j.cej.2012.07.075.
[262] O. A. Carrasco-Jaim, O. Ceballos-Sanchez, L. M. Torres-Martínez, E. Moctezuma, and C. Gómez-Solís, "Synthesis and characterization of PbS/ZnO thin film for photocatalytic hydrogen production," *J. Photochem. Photobiol. Chem.*, vol. 347, pp. 98–104, Oct. 2017, doi: 10.1016/j.jphotochem.2017.07.016.
[263] W.-Q. Zhao *et al.*, "WO3–x/PbS/Au Ternary Heterojunction Nanostructures for Visible-Light-Driven Photocatalytic Hydrogen Generation," *ACS Appl. Nano Mater.*, vol. 5, no. 11, pp. 16440–16450, Nov. 2022, doi: 10.1021/acsanm.2c03508.
[264] C.-J. Chang, Y.-G. Lin, J. Chen, C.-Y. Huang, S.-C. Hsieh, and S.-Y. Wu, "Ionic liquid/surfactant-hydrothermal synthesis of dendritic PbS@CuS core-shell photocatalysts with improved photocatalytic performance," *Appl. Surf. Sci.*, vol. 546, p. 149106, Apr. 2021, doi: 10.1016/j.apsusc.2021.149106.
[265] M. Ganapathy, C. T. Chang, and V. Alagan, "Facile preparation of amorphous SrTiO3- crystalline PbS heterojunction for efficient photocatalytic hydrogen production," *Int. J. Hydrog. Energy*, vol. 47, no. 64, pp. 27555–27565, Jul. 2022, doi: 10.1016/j.ijhydene.2022.06.086.
[266] Amika, P. E. Lokhande, R. U. Bhaskar, D. Kumar, S. Awasthi, and S. K. Pandey, "Experimental and DFT insights on hydrothermally synthesized PbS doped bismuth titanate perovskites: An Outperforming photocatalytic hydrogen production performance," *Int. J. Hydrog. Energy*, vol. 78, pp. 534–546, Aug. 2024, doi: 10.1016/j.ijhydene.2024.06.217.
[267] C.-J. Chang, S.-C. Hsieh, J. Chen, Y.-C. Wang, C.-L. Chiang, and Y.-G. Lin, "Electron-transfer dynamics and photocatalytic H2-production activity of PbS@Cu2S nanocomposites," *J. Taiwan Inst. Chem. Eng.*, vol. 162, p. 105587, Sep. 2024, doi: 10.1016/j.jtice.2024.105587.
[268] W. Cui, Y. Qi, L. Liu, D. Rana, J. Hu, and Y. Liang, "Synthesis of PbS–K2La2Ti3O10 composite and its photocatalytic activity for hydrogen production," *Prog. Nat. Sci. Mater. Int.*, vol. 22, no. 2, pp. 120–125, Apr. 2012, doi: 10.1016/j.pnsc.2012.03.002.
[269] M.-H. Chiu, C.-C. Kuo, C.-W. Huang, and W.-D. Yang, "Preparation of CuS/PbS/ZnO Heterojunction Photocatalyst for Application in Hydrogen Production," *Catalysts*, vol. 12, no. 12, Art. no. 12, Dec. 2022, doi: 10.3390/catal12121677.
[270] H. Liu *et al.*, "Electronic Structure Reconfiguration toward Pyrite NiS2 via Engineered Heteroatom Defect Boosting Overall Water Splitting," *ACS Nano*, vol. 11, no. 11, pp. 11574–11583, Nov. 2017, doi: 10.1021/acsnano.7b06501.
[271] M. Mahanthappa, M. P. Bhat, A. Alshoaibi, and V. R. S, "Exploring the role of redox-active species on NiS-NiS2 incorporated sulfur-doped graphitic carbon nitride nanohybrid as a bifunctional electrocatalyst for overall water splitting," *Int. J. Hydrog. Energy*, vol. 84, pp. 641–649, Sep. 2024, doi: 10.1016/j.ijhydene.2024.08.220.
[272] T. Rasheed *et al.*, "Multi-walled carbon nanotubes with embedded nickel sulphide as an effective electrocatalyst for Bi-functional water splitting," *Int. J. Hydrog. Energy*, vol. 67, pp. 373–380, May 2024, doi: 10.1016/j.ijhydene.2024.04.131.
[273] D. Mondal *et al.*, "Synthesis, characterization and evaluation of unsupported porous NiS2 sub-micrometer spheres as a potential hydrodesulfurization catalyst," *Appl. Catal. Gen.*, vol. 450, pp. 230–236, Jan. 2013, doi: 10.1016/j.apcata.2012.10.030.

[274] P. Luo *et al.*, "Targeted Synthesis of Unique Nickel Sulfide (NiS, NiS2) Microarchitectures and the Applications for the Enhanced Water Splitting System," *ACS Appl. Mater. Interfaces*, vol. 9, no. 3, pp. 2500–2508, Jan. 2017, doi: 10.1021/acsami.6b13984.
[275] X.-L. Luo, G.-L. He, Y.-P. Fang, and Y.-H. Xu, "Nickel sulfide/graphitic carbon nitride/strontium titanate (NiS/g-C3N4/SrTiO3) composites with significantly enhanced photocatalytic hydrogen production activity," *J. Colloid Interface Sci.*, vol. 518, pp. 184–191, May 2018, doi: 10.1016/j.jcis.2018.02.038.
[276] J. Wen *et al.*, "Constructing Multifunctional Metallic Ni Interface Layers in the g-C3N4 Nanosheets/Amorphous NiS Heterojunctions for Efficient Photocatalytic H2 Generation," *ACS Appl. Mater. Interfaces*, vol. 9, no. 16, pp. 14031–14042, Apr. 2017, doi: 10.1021/acsami.7b02701.
[277] X. Zhou, H. Sun, H. Zhang, and W. Tu, "One-pot hydrothermal synthesis of CdS/NiS photocatalysts for high H2 evolution from water under visible light," *Int. J. Hydrog. Energy*, vol. 42, no. 16, pp. 11199–11205, Apr. 2017, doi: 10.1016/j.ijhydene.2017.03.179.
[278] X. Liu *et al.*, "Highly efficient and noble metal-free NiS modified MnxCd1-xS solid solutions with enhanced photocatalytic activity for hydrogen evolution under visible light irradiation," *Appl. Catal. B Environ.*, vol. 203, pp. 282–288, Apr. 2017, doi: 10.1016/j.apcatb.2016.10.040.
[279] Q. Wang *et al.*, "CuS, NiS as co-catalyst for enhanced photocatalytic hydrogen evolution over TiO2," *Int. J. Hydrog. Energy*, vol. 39, no. 25, pp. 13421–13428, Aug. 2014, doi: 10.1016/j.ijhydene.2014.04.020.
[280] M. Wang, J. Cheng, X. Wang, X. Hong, J. Fan, and H. Yu, "Sulfur-mediated photodeposition synthesis of NiS cocatalyst for boosting H2-evolution performance of g-C3N4 photocatalyst," *Chin. J. Catal.*, vol. 42, no. 1, pp. 37–45, Jan. 2021, doi: 10.1016/S1872-2067(20)63633-6.
[281] Y. Lu, D. Chu, M. Zhu, Y. Du, and P. Yang, "Exfoliated carbon nitride nanosheets decorated with NiS as an efficient noble-metal-free visible-light-driven photocatalyst for hydrogen evolution," *Phys. Chem. Chem. Phys.*, vol. 17, no. 26, pp. 17355–17361, 2015, doi: 10.1039/C5CP01657F.
[282] J. Hong, Y. Wang, Y. Wang, W. Zhang, and R. Xu, "Noble-Metal-Free NiS/C3N4 for Efficient Photocatalytic Hydrogen Evolution from Water," *ChemSusChem*, vol. 6, no. 12, pp. 2263–2268, 2013, doi: 10.1002/cssc.201300647.
[283] J. E. Samaniego-Benitez, K. Jimenez-Rangel, L. Lartundo-Rojas, A. García-García, and A. Mantilla, "Enhanced photocatalytic H2 production over g-C3N4/NiS hybrid photocatalyst," *Mater. Lett.*, vol. 290, p. 129476, May 2021, doi: 10.1016/j.matlet.2021.129476.
[284] H. Zhao *et al.*, "A photochemical synthesis route to typical transition metal sulfides as highly efficient cocatalyst for hydrogen evolution: from the case of NiS/g-C3N4," *Appl. Catal. B Environ.*, vol. 225, pp. 284–290, Jun. 2018, doi: 10.1016/j.apcatb.2017.11.083.
[285] P. Senthil, A. Sankar, K. Paramasivaganesh, and S. P. Saravanan, "Au NPs-incorporated NiS/RGO hybrid composites for efficient visible light photocatalytic hydrogen evolution," *J. Mater. Sci. Mater. Electron.*, vol. 35, no. 1, p. 3, Dec. 2023, doi: 10.1007/s10854-023-11684-0.
[286] W. Zhang, Y. Wang, Z. Wang, Z. Zhong, and R. Xu, "Highly efficient and noble metal-free NiS/CdS photocatalysts for H 2 evolution from lactic acid sacrificial solution under visible light," *Chem. Commun.*, vol. 46, no. 40, pp. 7631–7633, 2010, doi: 10.1039/C0CC01562H.
[287] M. Yin, W. Zhang, F. Qiao, J. Sun, Y. Fan, and Z. Li, "Hydrothermal synthesis of MoS2-NiS/CdS with enhanced photocatalytic hydrogen production activity and stability," *J. Solid State Chem.*, vol. 270, pp. 531–538, Feb. 2019, doi: 10.1016/j.jssc.2018.12.022.
[288] Z. Chen, C. Cheng, F. Xing, and C. Huang, "Strong interfacial coupling for NiS thin layer covered CdS nanorods with highly efficient photocatalytic hydrogen production," *New J. Chem.*, vol. 44, no. 44, pp. 19083–19090, 2020, doi: 10.1039/D0NJ04335D.
[289] Q. Zhang *et al.*, "NiS-Decorated ZnO/ZnS Nanorod Heterostructures for Enhanced Photocatalytic Hydrogen Production: Insight into the Role of NiS," *Sol. RRL*, vol. 4, no. 4, p. 1900568, 2020, doi: 10.1002/solr.201900568.

[290] H. Gu, Z. Lin, Y. Li, D. Wang, and H. Feng, "Research Progress of NiS Cocatalysts in Photocatalysis," *Phys. Status Solidi A*, vol. 221, no. 10, p. 2400187, 2024, doi: 10.1002/pssa.202400187.
[291] K. Khan *et al.*, "Boosting visible-light-driven photocatalytic H2 production by optimizing surface hydrogen desorption of NiS/CdS–P photocatalyst," *Int. J. Hydrog. Energy*, vol. 80, pp. 427–434, Aug. 2024, doi: 10.1016/j.ijhydene.2024.07.081.
[292] G. Zhang *et al.*, "Fabrication of CZS NRs/NiSx NPs hybrids with abundant S-vacancies for signally promoting photocatalytic hydrogen production," *Int. J. Hydrog. Energy*, vol. 68, pp. 255–267, May 2024, doi: 10.1016/j.ijhydene.2024.04.218.
[293] Y. Chen, Y. Cheng, T. Zhang, H. Zhang, and S. Zhong, "Photodeposition of NiS thin film enhanced the visible light hydrogen evolution performance of CdS nanoflowers," *Int. J. Hydrog. Energy*, vol. 77, pp. 184–192, Aug. 2024, doi: 10.1016/j.ijhydene.2024.06.130.
[294] T. Zhou *et al.*, "Construction of NiS/CTF heterojunction photocatalyst with an outstanding photocatalytic hydrogen evolution performance," *Chin. Chem. Lett.*, p. 110415, Sep. 2024, doi: 10.1016/j.cclet.2024.110415.
[295] S. Shanmugaratnam, P. Ravirajan, Y. Shivatharsiny, and D. Velauthapillai, "Green hydrogen production through photocatalytic seawater splitting on MS2/TiO2 (M=Ni/Co/Sn) nanocomposites over simulated solar irradiation," *Int. J. Hydrog. Energy*, vol. 91, pp. 673–682, Nov. 2024, doi: 10.1016/j.ijhydene.2024.10.091.
[296] T.-R. Kuo *et al.*, "Extended visible to near-infrared harvesting of earth-abundant FeS2–TiO2 heterostructures for highly active photocatalytic hydrogen evolution," *Green Chem.*, vol. 20, no. 7, pp. 1640–1647, Apr. 2018, doi: 10.1039/C7GC03173D.
[297] H. Qin, J. Jia, L. Lin, H. Ni, M. Wang, and L. Meng, "Pyrite FeS2 nanostructures: Synthesis, properties and applications," *Mater. Sci. Eng. B*, vol. 236–237, pp. 104–124, Oct. 2018, doi: 10.1016/j.mseb.2018.11.003.
[298] G. Kaur, M. Kaur, A. Thakur, and A. Kumar, "Recent Progress on Pyrite FeS2 Nanomaterials for Energy and Environment Applications: Synthesis, Properties and Future Prospects," *J. Clust. Sci.*, vol. 31, no. 5, pp. 899–937, Sep. 2020, doi: 10.1007/s10876-019-01708-3.
[299] Y. Chen *et al.*, "Modification, application and reaction mechanisms of nano-sized iron sulfide particles for pollutant removal from soil and water: A review," *Chem. Eng. J.*, vol. 362, pp. 144–159, Apr. 2019, doi: 10.1016/j.cej.2018.12.175.
[300] D. Heift, "Iron Sulfide Materials: Catalysts for Electrochemical Hydrogen Evolution," *Inorganics*, vol. 7, no. 6, Art. no. 6, Jun. 2019, doi: 10.3390/inorganics7060075.
[301] A. K. Dutta *et al.*, "Synthesis of FeS and FeSe Nanoparticles from a Single Source Precursor: A Study of Their Photocatalytic Activity, Peroxidase-Like Behavior, and Electrochemical Sensing of H2O2," *ACS Appl. Mater. Interfaces*, vol. 4, no. 4, pp. 1919–1927, Apr. 2012, doi: 10.1021/am300408r.
[302] D. Banjara, Y. Malozovsky, L. Franklin, and D. Bagayoko, "First-principles studies of electronic, transport and bulk properties of pyrite FeS2," *AIP Adv.*, vol. 8, no. 2, p. 025212, Feb. 2018, doi: 10.1063/1.4996551.
[303] S. Middya, A. Layek, A. Dey, and P. P. Ray, "Synthesis of Nanocrystalline FeS2 with Increased Band Gap for Solar Energy Harvesting," *J. Mater. Sci. Technol.*, vol. 30, no. 8, pp. 770–775, Aug. 2014, doi: 10.1016/j.jmst.2014.01.005.
[304] J.-Y. Zhao and J.-M. Zhang, "Modulating the Band Gap of the FeS2 by O and Se Doping," *J. Phys. Chem. C*, vol. 121, no. 35, pp. 19334–19340, Sep. 2017, doi: 10.1021/acs.jpcc.7b04568.
[305] G. Zhou *et al.*, "Photoinduced semiconductor-metal transition in ultrathin troilite FeS nanosheets to trigger efficient hydrogen evolution," *Nat. Commun.*, vol. 10, no. 1, p. 399, Jan. 2019, doi: 10.1038/s41467-019-08358-z.
[306] J. Zander and R. Marschall, "Ni 2 FeS 4 as a highly efficient earth-abundant co-catalyst in photocatalytic hydrogen evolution," *J. Mater. Chem. A*, vol. 11, no. 32, pp. 17066–17078, 2023, doi: 10.1039/D3TA02439C.

[307] B. Wang *et al.*, "The bimetallic iron−nickel sulfide modified g-C3N4 nano-heterojunction and its photocatalytic hydrogen production enhancement," *J. Alloys Compd.*, vol. 766, pp. 421–428, Oct. 2018, doi: 10.1016/j.jallcom.2018.06.377.
[308] M. Zhang *et al.*, "Activate Fe3S4 Nanorods by Ni Doping for Efficient Dye-Sensitized Photocatalytic Hydrogen Production," *ACS Appl. Mater. Interfaces*, vol. 13, no. 12, pp. 14198–14206, Mar. 2021, doi: 10.1021/acsami.0c22869.
[309] A. M. Huerta-Flores *et al.*, "Extended visible light harvesting and boosted charge carrier dynamics in heterostructured zirconate–FeS2 photocatalysts for efficient solar water splitting," *J. Mater. Sci. Mater. Electron.*, vol. 29, no. 22, pp. 18957–18970, Nov. 2018, doi: 10.1007/s10854-018-0019-8.
[310] M. Li, J. Sun, G. Chen, S. Yao, B. Cong, and P. Liu, "Construction photothermal/pyroelectric property of hollow FeS2/Bi2S3 nanostructure with enhanced full spectrum photocatalytic activity," *Appl. Catal. B Environ.*, vol. 298, p. 120573, Dec. 2021, doi: 10.1016/j.apcatb.2021.120573.
[311] G. Lee and M. Kang, "Physicochemical properties of core/shell structured pyrite FeS2/anatase TiO2 composites and their photocatalytic hydrogen production performances," *Curr. Appl. Phys.*, vol. 13, no. 7, pp. 1482–1489, Sep. 2013, doi: 10.1016/j.cap.2013.05.002.
[312] M. S. Goh *et al.*, "Sustainable and stable hydrogen production over petal-shaped CdS/FeS2 S-scheme heterojunction by photocatalytic water splitting," *Int. J. Hydrog. Energy*, vol. 47, no. 65, pp. 27911–27929, Jul. 2022, doi: 10.1016/j.ijhydene.2022.06.118.
[313] Z. He *et al.*, "Electron spin-states reconfiguration induced by alternating temperature gradient for boosting photocatalytic hydrogen evolution on hollow core-shell FeS2/CuCo2O4 Z-scheme heterostructure," *Nano Energy*, vol. 124, p. 109483, Jun. 2024, doi: 10.1016/j.nanoen.2024.109483.
[314] M. Niu *et al.*, "Modulating charge spatial separation in S-scheme FeS2/Cd0.3Mn0.7S heterostructure via strong interaction at interface with boosted photocatalytic hydrogen evolution," *Int. J. Hydrog. Energy*, vol. 65, pp. 759–768, May 2024, doi: 10.1016/j.ijhydene.2024.04.057.
[315] B. Yao, S. Wang, H. Yao, X. Pang, Y. Li, and J. Sun, "Photothermal and pyroelectric effects in hollow FeS2/CdS nanocomposites for enhanced photocatalytic hydrogen evolution," *Surf. Interfaces*, vol. 49, p. 104370, Jun. 2024, doi: 10.1016/j.surfin.2024.104370.
[316] Y. Li *et al.*, "A novel noble-metal-free FeS2/Mn0.5Cd0.5S heterojunction for enhancing photocatalytic H2 production activity: Carrier separation, light absorption, active sites," *J. Taiwan Inst. Chem. Eng.*, vol. 162, p. 105572, Sep. 2024, doi: 10.1016/j.jtice.2024.105572.
[317] R. Zhang *et al.*, "Fabricating FeS2/Mn0.3Cd0.7S S-scheme heterojunction for enhanced photothermal-assisted photocatalytic H2 evolution under full-spectrum light," *Sep. Purif. Technol.*, vol. 354, p. 129479, Feb. 2025, doi: 10.1016/j.seppur.2024.129479.
[318] J. Jia, Q. Zhang, Z. Li, X. Hu, E. Liu, and J. Fan, "Lateral heterojunctions within ultrathin FeS–FeSe 2 nanosheet semiconductors for photocatalytic hydrogen evolution," *J. Mater. Chem. A*, vol. 7, no. 8, pp. 3828–3841, 2019, doi: 10.1039/C8TA11456K.
[319] J. Wang *et al.*, "Disruption Symmetric Crystal Structure Favoring Photocatalytic CO2 Reduction: Reduced *COOH Formation Energy Barrier on Al Doped CuS/TiO2," *Adv. Funct. Mater.*, vol. 34, no. 42, p. 2406549, 2024, doi: 10.1002/adfm.202406549.
[320] B. A. S. Souza, D. V. Freitas, V. N. Lima, C. I. S. Filho, G. Machado, and M. Navarro, "A CuBTC/CuS Photocatalyst for Organic Dyes: Degradation Kinetics and Mechanistic Insights," *ChemistrySelect*, vol. 9, no. 36, p. e202401718, 2024, doi: 10.1002/slct.202401718.
[321] S. Xie *et al.*, "Enhanced photocatalytic aerobic oxidative desulfurization of diesel over FeMo6Ox nanoclusters decorated on CuS nanosheets," *Appl. Catal. B Environ. Energy*, vol. 357, p. 124282, Nov. 2024, doi: 10.1016/j.apcatb.2024.124282.
[322] M. Saranya *et al.*, "Hydrothermal growth of CuS nanostructures and its photocatalytic properties," *Powder Technol.*, vol. 252, pp. 25–32, Jan. 2014, doi: 10.1016/j.powtec.2013.10.031.

[323] M. Saranya, R. Ramachandran, E. J. J. Samuel, S. K. Jeong, and A. N. Grace, "Enhanced visible light photocatalytic reduction of organic pollutant and electrochemical properties of CuS catalyst," *Powder Technol.*, vol. 279, pp. 209–220, Jul. 2015, doi: 10.1016/j.powtec.2015.03.041.

[324] C. Lai *et al.*, "Fabrication of CuS/BiVO4 (0 4 0) binary heterojunction photocatalysts with enhanced photocatalytic activity for Ciprofloxacin degradation and mechanism insight," *Chem. Eng. J.*, vol. 358, pp. 891–902, Feb. 2019, doi: 10.1016/j.cej.2018.10.072.

[325] Q. Chen, S. Wu, and Y. Xin, "Synthesis of Au–CuS–TiO2 nanobelts photocatalyst for efficient photocatalytic degradation of antibiotic oxytetracycline," *Chem. Eng. J.*, vol. 302, pp. 377–387, Oct. 2016, doi: 10.1016/j.cej.2016.05.076.

[326] W. Xu *et al.*, "Nanoporous CuS with excellent photocatalytic property," *Sci. Rep.*, vol. 5, no. 1, p. 18125, Dec. 2015, doi: 10.1038/srep18125.

[327] M. Tanveer *et al.*, "Template free synthesis of CuS nanosheet-based hierarchical microspheres: an efficient natural light driven photocatalyst," *CrystEngComm*, vol. 16, no. 24, pp. 5290–5300, 2014, doi: 10.1039/C4CE00090K.

[328] A. Malathi and J. Madhavan, "Synthesis and Characterization of CuS/CdS Photocatalyst with Enhanced Visible Light-Photocatalytic Activity," *J. Nano Res.*, vol. 48, pp. 49–61, 2017, doi: 10.4028/www.scientific.net/JNanoR.48.49.

[329] L. Xiao, H. Chen, and J. Huang, "Visible light-driven photocatalytic H2-generation activity of CuS/ZnS composite particles," *Mater. Res. Bull.*, vol. 64, pp. 370–374, Apr. 2015, doi: 10.1016/j.materresbull.2015.01.008.

[330] M. W. Kadi, R. M. Mohamed, A. A. Ismail, and D. W. Bahnemann, "H2 production using CuS/g-C3N4 nanocomposites under visible light," *Appl. Nanosci.*, vol. 10, no. 1, pp. 223–232, Jan. 2020, doi: 10.1007/s13204-019-01073-7.

[331] R. A. El-Gendy, H. M. El-Bery, M. Farrag, and D. M. Fouad, "Metal chalcogenides (CuS or MoS2)-modified TiO2 as highly efficient bifunctional photocatalyst nanocomposites for green H2 generation and dye degradation," *Sci. Rep.*, vol. 13, no. 1, p. 7994, May 2023, doi: 10.1038/s41598-023-34743-2.

[332] B. Yu *et al.*, "Construction of hollow TiO2/CuS nanoboxes for boosting full-spectrum driven photocatalytic hydrogen evolution and environmental remediation," *Ceram. Int.*, vol. 47, no. 7, Part A, pp. 8849–8858, Apr. 2021, doi: 10.1016/j.ceramint.2020.12.006.

[333] K. Manjunath, V. S. Souza, G. Nagaraju, J. M. L. Santos, J. Dupont, and T. Ramakrishnappa, "Superior activity of the CuS–TiO 2 /Pt hybrid nanostructure towards visible light induced hydrogen production," *New J. Chem.*, vol. 40, no. 12, pp. 10172–10180, 2016, doi: 10.1039/C6NJ02241C.

[334] J. Hou *et al.*, "One-pot hydrothermal synthesis of CdS–CuS decorated TiO2 NTs for improved photocatalytic dye degradation and hydrogen production," *Ceram. Int.*, vol. 47, no. 21, pp. 30860–30868, Nov. 2021, doi: 10.1016/j.ceramint.2021.07.268.

[335] M. Chandra, K. Bhunia, and D. Pradhan, "Controlled Synthesis of CuS/TiO2 Heterostructured Nanocomposites for Enhanced Photocatalytic Hydrogen Generation through Water Splitting," *Inorg. Chem.*, vol. 57, no. 8, pp. 4524–4533, Apr. 2018, doi: 10.1021/acs.inorgchem.8b00283.

[336] Q. Wang *et al.*, "High photocatalytic hydrogen production from methanol aqueous solution using the photocatalysts CuS/TiO2," *Int. J. Hydrog. Energy*, vol. 38, no. 25, pp. 10739–10745, Aug. 2013, doi: 10.1016/j.ijhydene.2013.02.131.

[337] X. Wang *et al.*, "CuS-modified ZnO rod/reduced graphene oxide/CdS heterostructure for efficient visible-light photocatalytic hydrogen generation," *Int. J. Hydrog. Energy*, vol. 45, no. 53, pp. 28394–28403, Oct. 2020, doi: 10.1016/j.ijhydene.2020.07.143.

[338] P. Gomathisankar, K. Hachisuka, H. Katsumata, T. Suzuki, K. Funasaka, and S. Kaneco, "Photocatalytic hydrogen production with CuS/ZnO from aqueous Na2S + Na2SO3 solution," *Int. J. Hydrog. Energy*, vol. 38, no. 21, pp. 8625–8630, Jul. 2013, doi: 10.1016/j.ijhydene.2013.04.131.

[339] T. P. Yendrapati, A. Gautam, S. Bojja, and U. Pal, "Formation of ZnO@CuS nanorods for efficient photocatalytic hydrogen generation," *Sol. Energy*, vol. 196, pp. 540–548, Jan. 2020, doi: 10.1016/j.solener.2019.12.054.

[340] N. R. Vempuluru et al., "Solar hydrogen generation from organic substance using earth abundant CuS–NiO heterojunction semiconductor photocatalyst," *Ceram. Int.*, vol. 47, no. 7, Part B, pp. 10206–10215, Apr. 2021, doi: 10.1016/j.ceramint.2020.12.062.

[341] R. Shen, J. Xie, P. Guo, L. Chen, X. Chen, and X. Li, "Bridging the g-C3N4 Nanosheets and Robust CuS Cocatalysts by Metallic Acetylene Black Interface Mediators for Active and Durable Photocatalytic H2 Production," *ACS Appl. Energy Mater.*, vol. 1, no. 5, pp. 2232–2241, May 2018, doi: 10.1021/acsaem.8b00311.

[342] T. Chen et al., "In-situ fabrication of CuS/g-C3N4 nanocomposites with enhanced photocatalytic H2-production activity via photoinduced interfacial charge transfer," *Int. J. Hydrog. Energy*, vol. 42, no. 17, pp. 12210–12219, Apr. 2017, doi: 10.1016/j.ijhydene.2017.03.188.

[343] F. Xue et al., "CuS nanosheet-induced local hot spots on g-C3N4 boost photocatalytic hydrogen evolution," *Int. J. Hydrog. Energy*, vol. 48, no. 16, pp. 6346–6357, Feb. 2023, doi: 10.1016/j.ijhydene.2022.05.087.

[344] R. Rameshbabu, P. Ravi, and M. Sathish, "Cauliflower-like CuS/ZnS nanocomposites decorated g-C3N4 nanosheets as noble metal-free photocatalyst for superior photocatalytic water splitting," *Chem. Eng. J.*, vol. 360, pp. 1277–1286, Mar. 2019, doi: 10.1016/j.cej.2018.10.180.

[345] J. Lu, J. Zhang, Q. Chen, and H. Liu, "Porous CuS/ZnS microspheres derived from a bimetallic metal-organic framework as efficient photocatalysts for H2 production," *J. Photochem. Photobiol. Chem.*, vol. 380, p. 111853, Jul. 2019, doi: 10.1016/j.jphotochem.2019.111853.

[346] Y.-C. Chang, Y.-C. Chiao, and Y.-X. Fun, "Cu2O/CuS/ZnS Nanocomposite Boosts Blue LED-Light-Driven Photocatalytic Hydrogen Evolution," *Catalysts*, vol. 12, no. 9, Art. no. 9, Sep. 2022, doi: 10.3390/catal12091035.

[347] X. Wang, Y. Xiao, H. Yu, Y. Yang, X. Dong, and L. Xia, "Noble-metal-free MOF Derived ZnS/CeO2 Decorated with CuS Cocatalyst Photocatalyst with Efficient Photocatalytic Hydrogen Production Character," *ChemCatChem*, vol. 12, no. 22, pp. 5669–5678, 2020, doi: 10.1002/cctc.202001118.

[348] Y.-B. Shao, L.-H. Wang, and J.-H. Huang, "ZnS/CuS nanotubes for visible light-driven photocatalytic hydrogen generation," *RSC Adv.*, vol. 6, no. 87, pp. 84493–84499, 2016, doi: 10.1039/C6RA17046C.

[349] D. Lu et al., "NIR-photocatalytic LDHB-mimetic CuS@ZnS nanoenzyme regulates arthritis microenvironment through lactate oxidation and hydrogen evolution," *Chem. Eng. J.*, vol. 490, p. 151772, Jun. 2024, doi: 10.1016/j.cej.2024.151772.

[350] D. V. Markovskaya, S. V. Cherepanova, A. A. Saraev, E. Yu. Gerasimov, and E. A. Kozlova, "Photocatalytic hydrogen evolution from aqueous solutions of Na2S/Na2SO3 under visible light irradiation on CuS/Cd0.3Zn0.7S and Ni$_z$Cd0.3Zn0.7S1+$_z$," *Chem. Eng. J.*, vol. 262, pp. 146–155, Feb. 2015, doi: 10.1016/j.cej.2014.09.090.

[351] J. Guo et al., "Core-shell structure of sulphur vacancies-CdS@CuS: Enhanced photocatalytic hydrogen generation activity based on photoinduced interfacial charge transfer," *J. Colloid Interface Sci.*, vol. 600, pp. 138–149, Oct. 2021, doi: 10.1016/j.jcis.2021.05.013.

[352] E. Hong, D. Kim, and J. H. Kim, "Heterostructured metal sulfide (ZnS–CuS–CdS) photocatalyst for high electron utilization in hydrogen production from solar water splitting," *J. Ind. Eng. Chem.*, vol. 20, no. 5, pp. 3869–3874, Sep. 2014, doi: 10.1016/j.jiec.2013.12.092.

[353] L. Luo et al., "Cu-MOF assisted synthesis of CuS/CdS(H)/CdS(C): Enhanced photocatalytic hydrogen production under visible light," *Int. J. Hydrog. Energy*, vol. 44, no. 59, pp. 30965–30973, Nov. 2019, doi: 10.1016/j.ijhydene.2019.09.136.

[354] X. Xin et al., "In-situ growth of high-content 1T phase MoS2 confined in the CuS nanoframe for efficient photocatalytic hydrogen evolution," *Appl. Catal. B Environ.*, vol. 269, p. 118773, Jul. 2020, doi: 10.1016/j.apcatb.2020.118773.

[355] Y. Xie *et al.*, "Facile synthesis of CuS/MXene nanocomposites for efficient photocatalytic hydrogen generation," *CrystEngComm*, vol. 22, no. 11, pp. 2060–2066, 2020, doi: 10.1039/D0CE00104J.
[356] A. Prakash *et al.*, "In2S3/CuS nanosheet composite: An excellent visible light photocatalyst for H2 production from H2S," *Sol. Energy Mater. Sol. Cells*, vol. 180, pp. 205–212, Jun. 2018, doi: 10.1016/j.solmat.2018.03.011.
[357] J. Liu *et al.*, "P-N Heterojunction Embedded CuS/TiO2 Bifunctional Photocatalyst for Synchronous Hydrogen Production and Benzylamine Conversion," *Small*, vol. 20, no. 10, p. 2306344, 2024, doi: 10.1002/smll.202306344.
[358] X. Du *et al.*, "Rapid microwave preparation of CuS/ZnCdS Z-scheme heterojunction for efficient photocatalytic hydrogen evolution," *Int. J. Hydrog. Energy*, vol. 51, pp. 936–945, Jan. 2024, doi: 10.1016/j.ijhydene.2023.10.264.
[359] M.-X. Zhao *et al.*, "2D CuS/CdS porous hierarchical heterostructures for efficient solar-to-hydrogen conversion," *Mater. Sci. Semicond. Process.*, vol. 185, p. 108919, Jan. 2025, doi: 10.1016/j.mssp.2024.108919.
[360] S. Shen *et al.*, "Cu doping induced synergistic effect of S-vacancies and S-scheme Cu:Mn0.5Cd0.5S@CuS heterojunction for enhanced H2 evolution from photocatalytic seawater splitting," *Int. J. Hydrog. Energy*, vol. 61, pp. 734–742, Apr. 2024, doi: 10.1016/j.ijhydene.2024.02.287.
[361] S. Liu *et al.*, "Green hydrogen and acetaldehyde production via photo-induced thermal dehydrogenation of bioethanol over a highly dispersed CuS/ TiO2 catalyst," *Int. J. Hydrog. Energy*, vol. 86, pp. 1414–1423, Oct. 2024, doi: 10.1016/j.ijhydene.2024.08.413.
[362] C.-J. Chang, Y.-H. Wei, and W.-S. Kuo, "Free-standing CuS–ZnS decorated carbon nanotube films as immobilized photocatalysts for hydrogen production," *Int. J. Hydrog. Energy*, vol. 44, no. 58, pp. 30553–30562, Nov. 2019, doi: 10.1016/j.ijhydene.2018.04.229.
[363] Y. Guo, Z. Liang, Y. Xue, X. Wang, X. Zhang, and J. Tian, "A cation exchange strategy to construct Rod-shell CdS/Cu2S nanostructures for broad spectrum photocatalytic hydrogen production," *J. Colloid Interface Sci.*, vol. 608, pp. 158–163, Feb. 2022, doi: 10.1016/j.jcis.2021.09.190.
[364] V. N. Rao *et al.*, "Titanate quantum dots-sensitized Cu2S nanocomposites for superficial H2 production via photocatalytic water splitting," *Int. J. Hydrog. Energy*, vol. 47, no. 95, pp. 40379–40390, Dec. 2022, doi: 10.1016/j.ijhydene.2022.05.091.
[365] V. Navakoteswara Rao *et al.*, "Photocatalytic recovery of H2 from H2S containing wastewater: Surface and interface control of photo-excitons in Cu2S@TiO2 core-shell nanostructures," *Appl. Catal. B Environ.*, vol. 254, pp. 174–185, Oct. 2019, doi: 10.1016/j.apcatb.2019.04.090.
[366] Z. Li, H. Li, S. Wang, F. Yang, and W. Zhou, "Mesoporous black TiO2/MoS2/Cu2S hierarchical tandem heterojunctions toward optimized photothermal-photocatalytic fuel production," *Chem. Eng. J.*, vol. 427, p. 131830, Jan. 2022, doi: 10.1016/j.cej.2021.131830.
[367] F. Qiao *et al.*, "Cu 2 O/Cu 2 S microstructure regulation towards high efficiency photocatalytic hydrogen production and its theoretical mechanism analysis," *CrystEngComm*, vol. 25, no. 35, pp. 4939–4945, 2023, doi: 10.1039/D3CE00628J.
[368] X. Li, K. Dai, C. Pan, and J. Zhang, "Diethylenetriamine-Functionalized CdS Nanoparticles Decorated on Cu2S Snowflake Microparticles for Photocatalytic Hydrogen Production," *ACS Appl. Nano Mater.*, vol. 3, no. 11, pp. 11517–11526, Nov. 2020, doi: 10.1021/acsanm.0c02616.
[369] Y. Zhang *et al.*, "Simultaneously Efficient Solar Light Harvesting and Charge Transfer of Hollow Octahedral Cu2S/CdS p–n Heterostructures for Remarkable Photocatalytic Hydrogen Generation," *Trans. Tianjin Univ.*, vol. 27, no. 4, pp. 348–357, Aug. 2021, doi: 10.1007/s12209-021-00291-x.
[370] G. Wang, Y. Quan, K. Yang, and Z. Jin, "EDA-assisted synthesis of multifunctional snowflake-Cu2S/CdZnS S-scheme heterojunction for improved the photocatalytic hydrogen evolution," *J. Mater. Sci. Technol.*, vol. 121, pp. 28–39, Sep. 2022, doi: 10.1016/j.jmst.2021.11.073.

[371] Y. Chen, Z. Qin, X. Wang, X. Guo, and L. Guo, "Noble-metal-free Cu 2 S-modified photocatalysts for enhanced photocatalytic hydrogen production by forming nanoscale p–n junction structure," *RSC Adv.*, vol. 5, no. 23, pp. 18159–18166, 2015, doi: 10.1039/C5RA00091B.
[372] Y. Tang, D. Zhang, X. Pu, B. Ge, Y. Li, and Y. Huang, "Snowflake-like Cu2S/Zn0.5Cd0.5S p–n heterojunction photocatalyst for enhanced visible light photocatalytic H2 evolution activity," *J. Taiwan Inst. Chem. Eng.*, vol. 96, pp. 487–495, Mar. 2019, doi: 10.1016/j.jtice.2018.12.021.
[373] P. Ravi, V. Navakoteswara Rao, M. V. Shankar, and M. Sathish, "Heterojunction engineering at ternary Cu2S/Ta2O5/CdS nanocomposite for enhanced visible light-driven photocatalytic hydrogen evolution," *Mater. Today Energy*, vol. 21, p. 100779, Sep. 2021, doi: 10.1016/j.mtener.2021.100779.
[374] M. Sun *et al.*, "CdS@Ni3S2/Cu2S electrode for electrocatalysis and boosted photo-assisted electrocatalysis hydrogen production," *Sep. Purif. Technol.*, vol. 319, p. 124085, Aug. 2023, doi: 10.1016/j.seppur.2023.124085.
[375] H. Yu, H. Liang, J. Bai, and C. Li, "Sulfur vacancy and CdS phase transition synergistically boosting one-dimensional CdS/Cu2S/SiO2 hollow tube for photocatalytic hydrogen evolution," *Int. J. Hydrog. Energy*, vol. 48, no. 42, pp. 15908–15920, May 2023, doi: 10.1016/j.ijhydene.2023.01.120.
[376] J. Zhang, W. Li, Y. Li, L. Zhong, and C. Xu, "Self-optimizing bifunctional CdS/Cu2S with coexistence of light-reduced Cu0 for highly efficient photocatalytic H2 generation under visible-light irradiation," *Appl. Catal. B Environ.*, vol. 217, pp. 30–36, Nov. 2017, doi: 10.1016/j.apcatb.2017.05.074.
[377] Y. Liang, M. Shao, L. Liu, J. G. McEvoy, J. Hu, and W. Cui, "Synthesis of Cu2S/K4Nb6O17 composite and its photocatalytic activity for hydrogen production," *Catal. Commun.*, vol. 46, pp. 128–132, Feb. 2014, doi: 10.1016/j.catcom.2013.12.004.
[378] X. Gong *et al.*, "Construction of Cu2S/Nv-C3N4 p-n heterojunctions for enhanced photocatalytic hydrogen production under visible light," *Appl. Catal. Gen.*, vol. 668, p. 119470, Nov. 2023, doi: 10.1016/j.apcata.2023.119470.
[379] I. Daskalakis *et al.*, "Surface defect engineering of mesoporous Cu/ZnS nanocrystal-linked networks for improved visible-light photocatalytic hydrogen production," *Inorg. Chem. Front.*, vol. 7, no. 23, pp. 4687–4700, 2020, doi: 10.1039/D0QI01013H.
[380] R. J. V. Michael, J. Theerthagiri, J. Madhavan, M. J. Umapathy, and P. T. Manoharan, "Cu 2 S-incorporated ZnS nanocomposites for photocatalytic hydrogen evolution," *RSC Adv.*, vol. 5, no. 38, pp. 30175–30186, 2015, doi: 10.1039/C5RA03621F.
[381] K. S. Ranjith, D. Ranjith Kumar, Y. S. Huh, Y.-K. Han, T. Uyar, and R. T. Rajendra Kumar, "Promotional Effect of Cu2S–ZnS Nanograins as a Shell Layer on ZnO Nanorod Arrays for Boosting Visible Light Photocatalytic H2 Evolution," *J. Phys. Chem. C*, vol. 124, no. 6, pp. 3610–3620, Feb. 2020, doi: 10.1021/acs.jpcc.9b09666.
[382] X. Zhang *et al.*, "Controllable growth of MoS2 nanosheets on novel Cu2S snowflakes with high photocatalytic activity," *Appl. Catal. B Environ.*, vol. 232, pp. 355–364, Sep. 2018, doi: 10.1016/j.apcatb.2018.03.074.
[383] E. Ha *et al.*, "Dual-Modified Cu2S with MoS2 and Reduced Graphene Oxides as Efficient Photocatalysts for H2 Evolution Reaction," *Catalysts*, vol. 11, no. 11, Art. no. 11, Nov. 2021, doi: 10.3390/catal11111278.
[384] Y. Wu, H. Zhang, Y. Li, and Z. Jin, "Partial phosphating of Ni-MOFs and Cu2S snowflakes form 2D/2D structure for efficiently improved photocatalytic hydrogen evolution," *Int. J. Hydrog. Energy*, vol. 47, no. 86, pp. 36530–36542, Oct. 2022, doi: 10.1016/j.ijhydene.2022.08.205.
[385] Z. Jin, X. Wang, Y. Wang, T. Yan, and X. Hao, "Snowflake-like Cu2S Coated with NiAl-LDH Forms a p–n Heterojunction for Efficient Photocatalytic Hydrogen Evolution," *ACS Appl. Energy Mater.*, vol. 4, no. 12, pp. 14220–14231, Dec. 2021, doi: 10.1021/acsaem.1c02982.

[386] A. Shahzad *et al.*, "Escalating the synergism on CdZnS via Ag2S/Cu2S co-catalysts: Boosts hydrogen evolution from water splitting under sunlight," *J. Catal.*, vol. 429, p. 115210, Jan. 2024, doi: 10.1016/j.jcat.2023.115210.
[387] X. Xu *et al.*, "Boosting solar driven hydrogen production rate of Cu2S@CdS p-n heterostructures and CuxCd1-xS nanorods," *Int. J. Hydrog. Energy*, vol. 51, pp. 869–879, Jan. 2024, doi: 10.1016/j.ijhydene.2023.10.227.
[388] W. Wang, T. Li, S. Komarneni, X. Lu, and B. Liu, "Recent advances in Co-based co-catalysts for efficient photocatalytic hydrogen generation," *J. Colloid Interface Sci.*, vol. 608, pp. 1553–1575, Feb. 2022, doi: 10.1016/j.jcis.2021.10.051.
[389] H. Huang *et al.*, "Bridging localized electron states of pyrite-type CoS2 cocatalyst for activated solar H2 evolution," *Nano Res.*, vol. 15, no. 1, pp. 202–208, Jan. 2022, doi: 10.1007/s12274-021-3457-1.
[390] L. Ma, J. Xu, L. Li, M. Mao, and S. Zhao, "Hydrothermal synthesis of WO 3 /CoS 2 n–n heterojunction for Z-scheme photocatalytic H 2 evolution," *New J. Chem.*, vol. 44, no. 42, pp. 18326–18336, 2020, doi: 10.1039/D0NJ04021E.
[391] R. Zhang, K. Gong, S. Cao, and F. Du, "Amorphous sulfur-rich CoSx nanodots as highly efficient cocatalyst to promote photocatalytic hydrogen evolution over TiO2," *Int. J. Hydrog. Energy*, vol. 47, no. 94, pp. 39875–39885, Dec. 2022, doi: 10.1016/j.ijhydene.2022.09.150.
[392] Z. Liang, Y. Xue, X. Wang, X. Zhang, J. Tian, and H. Cui, "The incorporation of cocatalyst cobalt sulfide into graphitic carbon nitride: Boosted photocatalytic hydrogen evolution performance and mechanism exploration," *Nano Mater. Sci.*, vol. 5, no. 2, pp. 202–209, Jun. 2023, doi: 10.1016/j.nanoms.2022.03.001.
[393] H. Sudrajat, A. Susanti, and S. Hartuti, "Efficient electron extraction by CoS2 loaded onto anatase TiO2 for improved photocatalytic hydrogen evolution," *J. Phys. Condens. Matter*, vol. 34, no. 34, p. 344005, Jun. 2022, doi: 10.1088/1361-648X/ac792d.
[394] J. Tang *et al.*, "CdS nanorods anchored with CoS2 nanoparticles for enhanced photocatalytic hydrogen production," *Appl. Catal. Gen.*, vol. 588, p. 117281, Nov. 2019, doi: 10.1016/j.apcata.2019.117281.
[395] B. Q. Qiu *et al.*, "Revealing the size effect of metallic CoS2 on CdS nanorods for photocatalytic hydrogen evolution based on Schottky junction," *Appl. Catal. Gen.*, vol. 592, p. 117377, Feb. 2020, doi: 10.1016/j.apcata.2019.117377.
[396] L. Shen *et al.*, "Fabrication of CoS/CdS heterojunctions for enhanced photocatalytic hydrogen production," *Inorganica Chim. Acta*, vol. 541, p. 121085, Oct. 2022, doi: 10.1016/j.ica.2022.121085.
[397] Y. Li, Y. Li, C. Yang, and L.-H. Gan, "Preparation of Highly Active Catalyst CoS2/CdS and Its Mechanism of Photocatalytic H2 Evolution," *J. Phys. Chem. C*, vol. 127, no. 36, pp. 17732–17741, Sep. 2023, doi: 10.1021/acs.jpcc.3c04031.
[398] S. Sun *et al.*, "Ultrafast Charge Separation in Ternary V2O5/CdS/CoS2 Z-Scheme Heterojunction Enables Efficient Visible-Light-Driven Hydrogen Generation," *Energy Fuels*, vol. 36, no. 4, pp. 2034–2043, Feb. 2022, doi: 10.1021/acs.energyfuels.1c03902.
[399] L. Ma, J. Xu, S. Zhao, L. Li, and Y. Liu, "Construction of CoS2/Zn0.5Cd0.5S S-Scheme Heterojunction for Enhancing H2 Evolution Activity Under Visible Light," *Chem. – Eur. J.*, vol. 27, no. 63, pp. 15795–15805, 2021, doi: 10.1002/chem.202102811.
[400] Y. Li *et al.*, "Tuning electronic structure via CoS clusters for visual photocatalytic H2 production and mechanism insight," *Chem. Eng. J.*, vol. 446, p. 137399, Oct. 2022, doi: 10.1016/j.cej.2022.137399.
[401] P. Wang *et al.*, "Unraveling the Interfacial Charge Migration Pathway at the Atomic Level in a Highly Efficient Z-Scheme Photocatalyst," *Angew. Chem. Int. Ed.*, vol. 58, no. 33, pp. 11329–11334, 2019, doi: 10.1002/anie.201904571.

[402] H. Ma, Y. Tan, Z. Liu, J. Wei, and R. Xiong, "Construction of CoS x –ZnIn 2 S 4 hollow nanocages derived from metal–organic frameworks for efficient photocatalytic hydrogen production," *New J. Chem.*, vol. 45, no. 31, pp. 13860–13868, 2021, doi: 10.1039/D1NJ00973G.
[403] X. Feng et al., "Heterostructured core–shell CoS1.097@ZnIn2S4 nanosheets for enhanced photocatalytic hydrogen evolution under visible light," *Chem. Eng. J.*, vol. 457, p. 141192, Feb. 2023, doi: 10.1016/j.cej.2022.141192.
[404] Y. Cai, Y. Shi, W. Shi, S. Bai, S. Yang, and F. Guo, "A one-photon excitation pathway in 0D/3D CoS2/ZnIn2S4 composite with nanoparticles on micro-flowers structure for boosted visible-light-driven photocatalytic hydrogen evolution," *Compos. Part B Eng.*, vol. 238, p. 109955, Jun. 2022, doi: 10.1016/j.compositesb.2022.109955.
[405] X. Xi, Q. Dang, G. Wang, W. Chen, and L. Tang, "ZIF-67-derived flower-like ZnIn 2 S 4 @CoS 2 heterostructures for photocatalytic hydrogen production," *New J. Chem.*, vol. 45, no. 43, pp. 20289–20295, 2021, doi: 10.1039/D1NJ03625D.
[406] J. Pan et al., "The CoSx-modified g-C3N4 nanosheets towards photocatalytic water splitting hydrogen production enhancement," *J. Mater. Sci. Mater. Electron.*, vol. 32, no. 2, pp. 2385–2394, Jan. 2021, doi: 10.1007/s10854-020-05004-z.
[407] C. Zhang et al., "Construction of Z-scheme heterojunction CoS/CdS@g-C3N4 hollow sphere with spatical charge separation for enhanced photocatalytic hydrogen production," *Appl. Surf. Sci.*, vol. 626, p. 157214, Jul. 2023, doi: 10.1016/j.apsusc.2023.157214.
[408] Z. Fan, X. Guo, Z. Jin, X. Li, and Y. Li, "Bridging Effect of S–C Bond for Boosting Electron Transfer over Cubic Hollow CoS/g-C3N4 Heterojunction toward Photocatalytic Hydrogen Production," *Langmuir*, vol. 38, no. 10, pp. 3244–3256, Mar. 2022, doi: 10.1021/acs.langmuir.1c03379.
[409] H. Li, P. Deng, and Y. Hou, "Cobalt disulfide/graphitic carbon nitride as an efficient photocatalyst for hydrogen evolution reaction under visible light irradiation," *Mater. Lett.*, vol. 229, pp. 217–220, Oct. 2018, doi: 10.1016/j.matlet.2018.07.004.
[410] Z. Bi et al., "Direct Z-scheme CoS/g-C3N4 heterojunction with NiS co-catalyst for efficient photocatalytic hydrogen generation," *Int. J. Hydrog. Energy*, vol. 47, no. 81, pp. 34430–34443, Sep. 2022, doi: 10.1016/j.ijhydene.2022.08.028.
[411] J. Fu, C. Bie, B. Cheng, C. Jiang, and J. Yu, "Hollow CoSx Polyhedrons Act as High-Efficiency Cocatalyst for Enhancing the Photocatalytic Hydrogen Generation of g-C3N4," *ACS Sustain. Chem. Eng.*, vol. 6, no. 2, pp. 2767–2779, Feb. 2018, doi: 10.1021/acssuschemeng.7b04461.
[412] H. Liang, Q. Zhang, J. Bai, T. Xu, and C. Li, "One-pot fabrication CoS2 modified MoS2-g-C3N4 ternary heterostructure composites with enhanced photocatalytic hydrogen production from water," *Diam. Relat. Mater.*, vol. 134, p. 109764, Apr. 2023, doi: 10.1016/j.diamond.2023.109764.
[413] S. Shanmugaratnam, D. Velauthapillai, P. Ravirajan, A. Christy, and Y. Shivatharsiny, "CoS2/TiO2 Nanocomposites for Hydrogen Production under UV Irradiation," *Materials*, vol. 12, no. 23, p. 3882, Nov. 2019, doi: 10.3390/ma12233882.
[414] L. Ma, J. Xu, J. Zhang, Z. Liu, and X. Liu, "Rare earth material CeO 2 modified CoS 2 nanospheres for efficient photocatalytic hydrogen evolution," *New J. Chem.*, vol. 45, no. 46, pp. 21795–21806, 2021, doi: 10.1039/D1NJ04196G.
[415] J. Xu, Y. Liu, X. Li, and Y. Li, "Hydrothermal Synthesis of Mn3O4/CoS2 as a Promising Photocatalytic Material for Boosting Visible-Light Photocatalytic Hydrogen Production," *Phys. Status Solidi A*, vol. 218, no. 11, p. 2100025, 2021, doi: 10.1002/pssa.202100025.
[416] X. Hong, X. Yu, L. Wang, Q. Liu, J. Sun, and H. Tang, "Lattice-Matched CoP/CoS2 Heterostructure Cocatalyst to Boost Photocatalytic H2 Generation," *Inorg. Chem.*, vol. 60, no. 16, pp. 12506–12516, Aug. 2021, doi: 10.1021/acs.inorgchem.1c01716.
[417] Z. Liu et al., "Synergistic strategy of Z-scheme heterojunction and defect engineering to construct ZnS/CoSx nanospheres for excellent photocatalytic H2 evolution performance," *Mater. Res. Bull.*, vol. 180, p. 113044, Dec. 2024, doi: 10.1016/j.materresbull.2024.113044.

[418] M. Shen *et al.*, "ZIF-67-derived hollow CoS and Mn0·2Cd0·8S to form a type-II heterojunction for boosting photocatalytic hydrogen evolution," *Int. J. Hydrog. Energy*, vol. 55, pp. 365–374, Feb. 2024, doi: 10.1016/j.ijhydene.2023.11.278.
[419] C. Wang *et al.*, "CoNi-layered double hydroxide derived CoS2/NiS2 dodecahedron decorated with ReS2 Z-scheme heterojunction for efficient hydrogen evolution," *J. Colloid Interface Sci.*, vol. 679, pp. 21–30, Feb. 2025, doi: 10.1016/j.jcis.2024.09.200.
[420] W. Fan *et al.*, "Rational design of Z-scheme CoS@NC/CdS heterojunction with N doped carbon (NC) as an electron mediator for enhanced photocatalytic H2 evolution," *Sep. Purif. Technol.*, vol. 354, p. 128358, Feb. 2025, doi: 10.1016/j.seppur.2024.128358.
[421] J. Li *et al.*, "Understanding the unique Ohmic-junction for enhancing the photocatalytic activity of CoS2/MgIn2S4 towards hydrogen production," *Appl. Catal. B Environ. Energy*, vol. 351, p. 123950, Aug. 2024, doi: 10.1016/j.apcatb.2024.123950.
[422] J. Li *et al.*, "Coupling CoS2 and CaIn2S4 for efficient electron trapping and improved surface catalysis to promote solar hydrogen evolution," *Int. J. Hydrog. Energy*, vol. 51, pp. 314–326, Jan. 2024, doi: 10.1016/j.ijhydene.2023.08.156.
[423] K. Kumar Mandari and M. Kang, "CuNi-LDH sheets and CoS nanoflakes decorated on graphitic carbon nitride heterostructure catalyst for efficient photocatalytic H2 production," *Appl. Surf. Sci.*, vol. 655, p. 159550, May 2024, doi: 10.1016/j.apsusc.2024.159550.
[424] X. Fan, Y. Ma, X. Wang, Y. Cao, W. Chen, and Y. Bai, "*In-situ* construction of CoS on porous g-C3N4 for fluent charge transfer in photocatalytic hydrogen production," *Appl. Surf. Sci.*, vol. 660, p. 160018, Jul. 2024, doi: 10.1016/j.apsusc.2024.160018.
[425] X. Liu *et al.*, "CoS nanoparticles anchored on g-C3N4 via a facile and green strategy for boosting visible-light-driven photocatalytic hydrogen production," *J. Alloys Compd.*, vol. 1000, p. 175068, Sep. 2024, doi: 10.1016/j.jallcom.2024.175068.
[426] W. Wang *et al.*, "Metal organic framework derived La/Gd-doped CoS for enhanced photocatalytic H2 evolution," *Colloids Surf. Physicochem. Eng. Asp.*, vol. 685, p. 133219, Mar. 2024, doi: 10.1016/j.colsurfa.2024.133219.
[427] M. Zheng, Y. Ding, L. Yu, X. Du, and Y. Zhao, "In Situ Grown Pristine Cobalt Sulfide as Bifunctional Photocatalyst for Hydrogen and Oxygen Evolution," *Adv. Funct. Mater.*, vol. 27, no. 11, p. 1605846, 2017, doi: 10.1002/adfm.201605846.
[428] X. Guo, F. Zhang, Y. Zhang, and J. Hu, "Review on the advancement of SnS 2 in photocatalysis," *J. Mater. Chem. A*, vol. 11, no. 14, pp. 7331–7343, 2023, doi: 10.1039/D2TA09741A.
[429] G. Li *et al.*, "Band gap narrowing of SnS 2 superstructures with improved hydrogen production," *J. Mater. Chem. A*, vol. 4, no. 1, pp. 209–216, 2016, doi: 10.1039/C5TA07283B.
[430] Y. Liu *et al.*, "Cu doped SnS2 nanostructure induced sulfur vacancy towards boosted photocatalytic hydrogen evolution," *Chem. Eng. J.*, vol. 407, p. 127180, Mar. 2021, doi: 10.1016/j.cej.2020.127180.
[431] W. Tian *et al.*, "Enhanced piezocatalytic activity in ion-doped SnS2 via lattice distortion engineering for BPA degradation and hydrogen production," *Nano Energy*, vol. 107, p. 108165, Mar. 2023, doi: 10.1016/j.nanoen.2023.108165.
[432] K. M. Kim, B. S. Kwak, S. Kang, and M. Kang, "Synthesis of Submicron Hexagonal Plate-Type SnS2 and Band Gap-Tuned Sn1−TiS2 Materials and Their Hydrogen Production Abilities on Methanol/Water Photosplitting," *Int. J. Photoenergy*, vol. 2014, no. 1, p. 479508, 2014, doi: 10.1155/2014/479508.
[433] L. Sun *et al.*, "Role of SnS2 in 2D–2D SnS2/TiO2 Nanosheet Heterojunctions for Photocatalytic Hydrogen Evolution," *ACS Appl. Nano Mater.*, vol. 2, no. 4, pp. 2144–2151, Apr. 2019, doi: 10.1021/acsanm.9b00122.
[434] I. Barba-Nieto, K. C. Christoforidis, M. Fernández-García, and A. Kubacka, "Promoting H2 photoproduction of TiO2-based materials by surface decoration with Pt nanoparticles and SnS2 nanoplatelets," *Appl. Catal. B Environ.*, vol. 277, p. 119246, Nov. 2020, doi: 10.1016/j.apcatb.2020.119246.

[435] E. Liu *et al.*, "Fabrication of 2D SnS2/g-C3N4 heterojunction with enhanced H2 evolution during photocatalytic water splitting," *J. Colloid Interface Sci.*, vol. 524, pp. 313–324, Aug. 2018, doi: 10.1016/j.jcis.2018.04.038.
[436] Y. Liu, C. Lv, J. Sun, X. Zhou, Y. Zhou, and G. Chen, "g-C3N4/SnS2 van der Waals Heterostructures Enabling High-Efficiency Photocatalytic Hydrogen Evolution," *Adv. Mater. Interfaces*, vol. 9, no. 19, p. 2200153, 2022, doi: 10.1002/admi.202200153.
[437] R. Zhang *et al.*, "Investigation on various photo-generated carrier transfer processes of SnS2/g-C3N4 heterojunction photocatalysts for hydrogen evolution," *J. Colloid Interface Sci.*, vol. 578, pp. 431–440, Oct. 2020, doi: 10.1016/j.jcis.2020.04.033.
[438] D. Shang, D. Li, B. Chen, B. Luo, Y. Huang, and W. Shi, "2D–2D SnS2/Covalent Organic Framework Heterojunction Photocatalysts for Highly Enhanced Solar-Driven Hydrogen Evolution without Cocatalysts," *ACS Sustain. Chem. Eng.*, vol. 9, no. 42, pp. 14238–14248, Oct. 2021, doi: 10.1021/acssuschemeng.1c05162.
[439] Y.-Y. Li, J.-G. Wang, H.-H. Sun, W. Hua, and X.-R. Liu, "Heterostructured SnS2/SnO2 nanotubes with enhanced charge separation and excellent photocatalytic hydrogen production," *Int. J. Hydrog. Energy*, vol. 43, no. 31, pp. 14121–14129, Aug. 2018, doi: 10.1016/j.ijhydene.2018.05.130.
[440] S. Zhu *et al.*, "Nanoflower-like CdS and SnS2 loaded TiO2 nanotube arrays for photocatalytic wastewater treatment and hydrogen production," *Ceram. Int.*, vol. 49, no. 4, pp. 5893–5904, Feb. 2023, doi: 10.1016/j.ceramint.2022.10.123.
[441] Y. Tian *et al.*, "A direct dual Z-scheme 3DOM SnS2–ZnS/ZrO2 composite with excellent photocatalytic degradation and hydrogen production performance," *Chemosphere*, vol. 279, p. 130882, Sep. 2021, doi: 10.1016/j.chemosphere.2021.130882.
[442] Q. Li *et al.*, "Nickel-based tungstate supported with different forms of SnS2 to achieve improved photocatalytic performance," *J. Alloys Compd.*, vol. 965, p. 171378, Nov. 2023, doi: 10.1016/j.jallcom.2023.171378.
[443] K. K. Mandari, N. Son, T. Kim, and M. Kang, "Highly efficient SnS2@Ag/AgVO3 heterostructures for improved charge carriers in photocatalytic H2 production," *J. Alloys Compd.*, vol. 927, p. 166886, Dec. 2022, doi: 10.1016/j.jallcom.2022.166886.
[444] Z. Liu, J. Xu, Z. Li, S. Xu, and X. Liu, "Compound SnS2 sensitizer in the S-scheme of Ag2Mo2O7/CoMoO4 heterojunction to improve the hydrogen evolution of semiconductor powder," *Int. J. Hydrog. Energy*, vol. 48, no. 53, pp. 20303–20313, Jun. 2023, doi: 10.1016/j.ijhydene.2022.11.311.
[445] A. P. Rangappa *et al.*, "Highly efficient hydrogen generation in water using 1D CdS nanorods integrated with 2D SnS2 nanosheets under solar light irradiation," *Appl. Surf. Sci.*, vol. 508, p. 144803, Apr. 2020, doi: 10.1016/j.apsusc.2019.144803.
[446] X. Chen *et al.*, "Construction of strongly coupled 2D–2D SnS 2 /CdS S-scheme heterostructures for photocatalytic hydrogen evolution," *Sustain. Energy Fuels*, vol. 7, no. 5, pp. 1311–1321, 2023, doi: 10.1039/D2SE01717B.
[447] X. Zhang, H. Yuan, J. Bao, W. Xiao, and G. He, "Interfacial construction of SnS2/Zn0.2Cd0.8S nanopolyhedron heterojunctions for enhanced photocatalytic hydrogen evolution," *J. Colloid Interface Sci.*, vol. 651, pp. 254–263, Dec. 2023, doi: 10.1016/j.jcis.2023.07.136.
[448] H. Hu *et al.*, "Constructing a noble-metal-free 0D/2D CdS/SnS2 heterojunction for efficient visible-light-driven photocatalytic pollutant degradation and hydrogen generation," *Nanotechnology*, vol. 34, no. 50, p. 505712, Oct. 2023, doi: 10.1088/1361-6528/acfaa6.
[449] L. Wang *et al.*, "Co-catalyst-free ZnS-SnS2 porous nanosheets for clean and recyclable photocatalytic H2 generation," *J. Alloys Compd.*, vol. 753, pp. 60–67, Jul. 2018, doi: 10.1016/j.jallcom.2018.04.086.
[450] C. Zhang *et al.*, "Self-assembled ZnIn2S4/SnS2 QDs S-scheme heterojunction for boosted photocatalytic hydrogen evolution: Energy band engineering and mechanism insight," *J. Alloys Compd.*, vol. 960, p. 170932, Oct. 2023, doi: 10.1016/j.jallcom.2023.170932.

[451] Y. Geng, X. Zou, Y. Lu, and L. Wang, "Fabrication of the SnS2/ZnIn2S4 heterojunction for highly efficient visible light photocatalytic H2 evolution," *Int. J. Hydrog. Energy*, vol. 47, no. 22, pp. 11520–11527, Mar. 2022, doi: 10.1016/j.ijhydene.2022.01.176.
[452] A. Raja, N. Son, S. Pandey, and M. Kang, "Fabrication of solar-driven hierarchical ZnIn2S4/rGO/SnS2 heterojunction photocatalyst for hydrogen generation and environmental pollutant elimination," *Sep. Purif. Technol.*, vol. 293, p. 121119, Jul. 2022, doi: 10.1016/j.seppur.2022.121119.
[453] S. R. Damkale et al., "Two-dimensional hexagonal SnS2 nanostructures for photocatalytic hydrogen generation and dye degradation," *Sustain. Energy Fuels*, vol. 3, no. 12, pp. 3406–3414, Nov. 2019, doi: 10.1039/C9SE00235A.
[454] W. Fu, J. Wang, S. Zhou, R. Li, and T. Peng, "Controllable Fabrication of Regular Hexagon-Shaped SnS2 Nanoplates and Their Enhanced Visible-Light-Driven H2 Production Activity," *ACS Appl. Nano Mater.*, vol. 1, no. 6, pp. 2923–2933, Jun. 2018, doi: 10.1021/acsanm.8b00563.
[455] N. Jawale, S. Arbuj, G. Umarji, M. Shinde, B. Kale, and S. Rane, "Ni loaded SnS 2 hexagonal nanosheets for photocatalytic hydrogen generation via water splitting," *RSC Adv.*, vol. 13, no. 4, pp. 2418–2426, 2023, doi: 10.1039/D2RA07954B.
[456] T. Wu Ren, Y. Liu, and S. Yu, "Synthesis of core–shell SnS2/CdIn2S4 heterojunction photocatalyst for visible light driven hydrogen evolution," *Mater. Lett.*, vol. 304, p. 130611, Dec. 2021, doi: 10.1016/j.matlet.2021.130611.
[457] W. Zhao et al., "Novel Z-scheme Ag-C3N4/SnS2 plasmonic heterojunction photocatalyst for degradation of tetracycline and H2 production," *Chem. Eng. J.*, vol. 405, p. 126555, Feb. 2021, doi: 10.1016/j.cej.2020.126555.
[458] Y. Li, Z. Liu, S. Wu, M. Zhu, and Y. Zhang, "Facile fabrication of Zn3In2S6@SnS2 3D heterostructure for efficient visible-light photocatalytic hydrogen evolution," *Chem. Phys. Lett.*, vol. 812, p. 140248, Feb. 2023, doi: 10.1016/j.cplett.2022.140248.
[459] Y. Cheng, J. He, and P. Yang, "Construction of layered SnS2 and g-C3N4 nanoarchitectonics towards pollution degradation and H2 generation," *Colloids Surf. Physicochem. Eng. Asp.*, vol. 680, p. 132678, Jan. 2024, doi: 10.1016/j.colsurfa.2023.132678.
[460] J. Wang et al., "Rationally constructed CuS1-x/SnS2 Ohmic junction for enhanced photocatalytic H2 evolution under visible-light irradiation," *Int. J. Hydrog. Energy*, vol. 83, pp. 784–791, Sep. 2024, doi: 10.1016/j.ijhydene.2024.08.114.
[461] F. Xing, J. Li, C. Wang, S. Jin, H. Jin, and J. Li, "Efficient photocatalytic hydrogen evolution of g-C 3 N 4 /Vs-SnS 2 /CdS through a sulfur vacancy-rich SnS 2 induced charge storage effect," *Inorg. Chem. Front.*, vol. 11, no. 10, pp. 2884–2893, 2024, doi: 10.1039/D4QI00602J.
[462] J. Wang et al., "Direct Z-scheme GaTe/SnS2 van der Waals heterojunction with tunable electronic properties: A promising highly efficient photocatalyst," *Int. J. Hydrog. Energy*, vol. 54, pp. 979–989, Feb. 2024, doi: 10.1016/j.ijhydene.2023.11.180.
[463] Z. K. Heiba, A. A. Ellatief, M. B. Mohamed, A. M. El-naggar, and H. Elshimy, "Investigation of Sn1-xCoxS nanocomposites as a catalyst for hydrogen production from sodium borohydride methanolysis," *J. Sol-Gel Sci. Technol.*, Oct. 2024, doi: 10.1007/s10971-024-06596-2.
[464] Y. Huang et al., "Hollow cubic SnS2/CdS nanoheterojunction for enhanced photocatalytic hydrogen evolution and degradation via MOFs *in situ* sulfuration," *Int. J. Hydrog. Energy*, vol. 64, pp. 1030–1039, Apr. 2024, doi: 10.1016/j.ijhydene.2024.03.357.
[465] H. Zhao, B. Zhao, H. Liu, and X. Li, "Green synthesis of 3D core–shell SnS 2 /SnS-Cd 0.5 Zn 0.5 S multi-heterojunction for efficient photocatalytic H 2 evolution," *Dalton Trans.*, vol. 53, no. 2, pp. 591–600, 2024, doi: 10.1039/D3DT03533F.
[466] Z. Lei, W. Wang, T. Sun, E. Liu, and T. Gao, "Efficient photocatalytic H2 evolution over SnS2/twinned Mn0.5Cd0.5S hetero-homojunction with double S-scheme charge transfer routes," *J. Mater. Sci. Technol.*, vol. 216, pp. 81–92, May 2025, doi: 10.1016/j.jmst.2024.07.034.

[467] A. Taufik, R. Saleh, and G. Seong, "Enhanced photocatalytic performance of SnS 2 under visible light irradiation: strategies and future perspectives," *Nanoscale*, vol. 16, no. 20, pp. 9680–9709, 2024, doi: 10.1039/D4NR00706A.
[468] S. R. Mishra, V. Gadore, and Md. Ahmaruzzaman, "Shining light on sustainable and clean hydrogen production: Recent developments with In2S3 photocatalysts," *Nano Energy*, vol. 128, p. 109820, Sep. 2024, doi: 10.1016/j.nanoen.2024.109820.
[469] V. Soni *et al.*, "Indium sulfide-based photocatalysts for hydrogen production and water cleaning: a review," *Environ. Chem. Lett.*, vol. 19, no. 2, pp. 1065–1095, Apr. 2021, doi: 10.1007/s10311-020-01148-w.
[470] S. Yang, C.-Y. Xu, B.-Y. Zhang, L. Yang, S.-P. Hu, and L. Zhen, "Ca(II) doped β-In2S3 hierarchical structures for photocatalytic hydrogen generation and organic dye degradation under visible light irradiation," *J. Colloid Interface Sci.*, vol. 491, pp. 230–237, Apr. 2017, doi: 10.1016/j.jcis.2016.12.028.
[471] L. Yang *et al.*, "Spatial charge separated two-dimensional/two-dimensional Cu-In2S3/CdS heterojunction for boosting photocatalytic hydrogen production," *J. Colloid Interface Sci.*, vol. 652, pp. 1503–1511, Dec. 2023, doi: 10.1016/j.jcis.2023.08.149.
[472] R. Zhang *et al.*, "Synergistic manipulation of sulfur vacancies and palladium doping of In2S3 for enhanced photocatalytic H2 production," *J. Colloid Interface Sci.*, vol. 677, pp. 425–434, Jan. 2025, doi: 10.1016/j.jcis.2024.07.242.
[473] M. Han *et al.*, "Bi & Mo co-doped In2S3 nano-foam blocks for boosted photocatalytic hydrogen generation," *Int. J. Hydrog. Energy*, vol. 78, pp. 140–147, Aug. 2024, doi: 10.1016/j.ijhydene.2024.06.296.
[474] X. Ma *et al.*, "A novel noble-metal-free binary and ternary In2S3 photocatalyst with WC and 'W-Mo auxiliary pairs' for highly-efficient visible-light hydrogen evolution," *J. Alloys Compd.*, vol. 875, p. 160058, Sep. 2021, doi: 10.1016/j.jallcom.2021.160058.
[475] B. Chai, T. Peng, P. Zeng, and J. Mao, "Synthesis of floriated In 2 S 3 decorated with TiO 2 nanoparticles for efficient photocatalytic hydrogen production under visible light," *J. Mater. Chem.*, vol. 21, no. 38, pp. 14587–14593, 2011, doi: 10.1039/C1JM11566A.
[476] F. Wang *et al.*, "Probing the charge separation process on In2S3/Pt-TiO2 nanocomposites for boosted visible-light photocatalytic hydrogen production," *Appl. Catal. B Environ.*, vol. 198, pp. 25–31, Dec. 2016, doi: 10.1016/j.apcatb.2016.05.048.
[477] Y.-C. Chang, S.-Y. Syu, and Z.-Y. Wu, "Fabrication of ZnO-In2S3 composite nanofiber as highly efficient hydrogen evolution photocatalyst," *Mater. Lett.*, vol. 302, p. 130435, Nov. 2021, doi: 10.1016/j.matlet.2021.130435.
[478] S. Li, K. Long, X. Sun, H. Yuan, and W. Li, "Activities in photocatalytic hydrogen evolution of In2O3/In2S3 heterostructure and In2O3/In2S3@PAN nanofibers," *Ceram. Int.*, vol. 49, no. 14, Part A, pp. 24093–24099, Jul. 2023, doi: 10.1016/j.ceramint.2023.05.006.
[479] E. Hua, S. Jin, X. Wang, S. Ni, G. Liu, and X. Xu, "Ultrathin 2D type-II p-n heterojunctions La2Ti2O7/In2S3 with efficient charge separations and photocatalytic hydrogen evolution under visible light illumination," *Appl. Catal. B Environ.*, vol. 245, pp. 733–742, May 2019, doi: 10.1016/j.apcatb.2019.01.024.
[480] X. Ma *et al.*, "A novel noble-metal-free Mo2C-In2S3 heterojunction photocatalyst with efficient charge separation for enhanced photocatalytic H2 evolution under visible light," *J. Colloid Interface Sci.*, vol. 582, pp. 488–495, Jan. 2021, doi: 10.1016/j.jcis.2020.08.083.
[481] Y.-C. Chang, S.-Y. Syu, and M.-Y. Lu, "Fabrication of In(OH)3–In2S3–Cu2O nanofiber for highly efficient photocatalytic hydrogen evolution under blue light LED excitation," *Int. J. Hydrog. Energy*, vol. 48, no. 25, pp. 9318–9332, Mar. 2023, doi: 10.1016/j.ijhydene.2022.12.114.
[482] M. Tayyab *et al.*, "One-pot *in-situ* hydrothermal synthesis of ternary In2S3/Nb2O5/Nb2C Schottky/S-scheme integrated heterojunction for efficient photocatalytic hydrogen production," *J. Colloid Interface Sci.*, vol. 628, pp. 500–512, Dec. 2022, doi: 10.1016/j.jcis.2022.08.071.

[483] M. Dan, Q. Zhang, S. Yu, A. Prakash, Y. Lin, and Y. Zhou, "Noble-metal-free MnS/In2S3 composite as highly efficient visible light driven photocatalyst for H2 production from H2S," *Appl. Catal. B Environ.*, vol. 217, pp. 530–539, Nov. 2017, doi: 10.1016/j.apcatb.2017.06.019.

[484] Y. Li *et al.*, "Highly dispersed PdS preferably anchored on In2S3 of MnS/In2S3 composite for effective and stable hydrogen production from H2S," *J. Catal.*, vol. 373, pp. 48–57, May 2019, doi: 10.1016/j.jcat.2019.03.021.

[485] L.-B. Zhan, C.-L. Yang, X.-H. Li, Y.-L. Liu, and W.-K. Zhao, "Newfound 2D In2S3 allotropy monolayers for efficient photocatalytic overall water splitting to produce hydrogen," *Int. J. Hydrog. Energy*, vol. 89, pp. 1185–1195, Nov. 2024, doi: 10.1016/j.ijhydene.2024.09.385.

[486] Y. Liu, Y. He, C. Li, Z. Shi, and S. Feng, "Indium-vacancy-rich In2S3/Ni@C photocatalyst with chemical bonds for producing hydrogen and benzylamine oxidation," *Sep. Purif. Technol.*, vol. 324, p. 124571, Nov. 2023, doi: 10.1016/j.seppur.2023.124571.

[487] C. Dong *et al.*, "Enhanced Photocatalytic Hydrogen Evolution of In2S3 by Decorating In2O3 with Rich Oxygen Vacancies," *Inorg. Chem.*, vol. 63, no. 24, pp. 11125–11134, Jun. 2024, doi: 10.1021/acs.inorgchem.4c00720.

[488] Y. Qi, G. Zhou, Y. Wu, H. Wang, Z. Yan, and Y. Wu, "In-situ construction of In2O3/In2S3-CdIn2S4 Z-scheme heterojunction nanotubes for enhanced photocatalytic hydrogen production," *J. Colloid Interface Sci.*, vol. 664, pp. 107–116, Jun. 2024, doi: 10.1016/j.jcis.2024.03.033.

[489] Y. Li *et al.*, "P-doped ultrathin g-C3N4/In2S3 S-scheme heterojunction enhances photocatalytic hydrogen production and degradation of ofloxacin," *Phys. B Condens. Matter*, vol. 685, p. 416053, Jul. 2024, doi: 10.1016/j.physb.2024.416053.

[490] P. Lu *et al.*, "Heterostructured In2O3/In2S3 hollow fibers enable efficient visible-light driven photocatalytic hydrogen production and 5-hydroxymethylfurfural oxidation," *Chin. J. Struct. Chem.*, vol. 43, no. 8, p. 100361, Aug. 2024, doi: 10.1016/j.cjsc.2024.100361.

[491] Y.-R. Lin, Y.-C. Chang, and F.-H. Ko, "One-pot microwave-assisted synthesis of In2S3/In2O3 nanosheets as highly active visible light photocatalysts for seawater splitting," *Int. J. Hydrog. Energy*, vol. 52, pp. 953–963, Jan. 2024, doi: 10.1016/j.ijhydene.2023.04.036.

[492] Y.-R. Lin, F.-H. Ko, and Y.-C. Chang, "Visible Light-Induced Photocatalytic Hydrogen Generation from Seawater Using Ternary Indium Sulfide/Indium Oxide/Gold Nanocomposites Obtained via Microwave-Assisted Synthesis," *ACS Appl. Nano Mater.*, vol. 7, no. 14, pp. 16831–16841, Jul. 2024, doi: 10.1021/acsanm.4c02913.

[493] Y. Ai *et al.*, "Synergistic interfacial engineering of a S-scheme ZnO/In2S3 photocatalyst with S−O covalent bonds: A dual-functional advancement for tetracycline hydrochloride degradation and H2 evolution," *Appl. Catal. B Environ. Energy*, vol. 353, p. 124098, Sep. 2024, doi: 10.1016/j.apcatb.2024.124098.

[494] S. Golda A, A. R. M. Shaheer, B. Neppolian, and S. K. Lakhera, "Comparative study of interfacial charge transfer at the junction between CuInS2:In2S3/g-C3N4 photocatalysts for sacrificial hydrogen evolution activity," *Int. J. Hydrog. Energy*, vol. 64, pp. 526–534, Apr. 2024, doi: 10.1016/j.ijhydene.2024.03.274.

[495] Y. Lin, L. Chen, J. Zhang, Y. Gui, and L. Liu, "Hierarchical In2S3 microflowers decorated with WO3 quantum dots: Sculpting S-scheme heterostructure for enhanced photocatalytic H2 evolution and nitrobenzene hydrogenation," *J. Mater. Sci. Technol.*, vol. 174, pp. 218–225, Mar. 2024, doi: 10.1016/j.jmst.2023.06.064.

[496] Y. Yang *et al.*, "Directional spatial charge separation in well-designed S-scheme In2S3–ZnIn2S4/Au heterojunction toward highly efficient photocatalytic hydrogen evolution," *Int. J. Hydrog. Energy*, vol. 51, pp. 315–326, Jan. 2024, doi: 10.1016/j.ijhydene.2023.10.026.

[497] J. Ye *et al.*, "In2S3-modified ZnIn2S4 enhanced photogenerated carrier separation efficiency and photocatalytic hydrogen evolution under visible light," *Fuel*, vol. 373, p. 132401, Oct. 2024, doi: 10.1016/j.fuel.2024.132401.

[498] S. R. Mishra, V. Gadore, and Md. Ahmaruzzaman, "An overview of In2S3 and In2S3-based photocatalyst: characteristics, synthesis, modifications, design strategies, and catalytic

environmental application," *J. Environ. Chem. Eng.*, vol. 12, no. 5, p. 113449, Oct. 2024, doi: 10.1016/j.jece.2024.113449.

[499] J. Kim and M. Kang, "High photocatalytic hydrogen production over the band gap-tuned urchin-like Bi2S3-loaded TiO2 composites system," *Int. J. Hydrog. Energy*, vol. 37, no. 10, pp. 8249–8256, May 2012, doi: 10.1016/j.ijhydene.2012.02.057.

[500] Y. Huang et al., "Visible light Bi2S3/Bi2O3/Bi2O2CO3 photocatalyst for effective degradation of organic pollutions," *Appl. Catal. B Environ.*, vol. 185, pp. 68–76, May 2016, doi: 10.1016/j.apcatb.2015.11.043.

[501] Y. Bessekhouad, M. Mohammedi, and M. Trari, "Hydrogen photoproduction from hydrogen sulfide on Bi2S3 catalyst," *Sol. Energy Mater. Sol. Cells*, vol. 73, no. 3, pp. 339–350, Jul. 2002, doi: 10.1016/S0927-0248(01)00218-5.

[502] M. Miodyńska et al., "Urchin-like TiO2 structures decorated with lanthanide-doped Bi2S3 quantum dots to boost hydrogen photogeneration performance," *Appl. Catal. B Environ.*, vol. 272, p. 118962, Sep. 2020, doi: 10.1016/j.apcatb.2020.118962.

[503] C. García-Mendoza et al., "Suitable preparation of BiS nanorods –TiO heterojunction semiconductors with improved photocatalytic hydrogen production from water/methanol decomposition," *J. Chem. Technol. Biotechnol.*, vol. 91, no. 8, pp. 2198–2204, 2016, doi: 10.1002/jctb.4979.

[504] V. NavakoteswaraRao, M. V. Shankar, B. L. Yang, C. W. Ahn, and J. M. Yang, "Effective excitons separation in starfish Bi2S3/TiO2 nanostructures for enhanced hydrogen production," *Mater. Today Chem.*, vol. 26, p. 101096, Dec. 2022, doi: 10.1016/j.mtchem.2022.101096.

[505] Y. Jia, P. Liu, Q. Wang, Y. Wu, D. Cao, and Q.-A. Qiao, "Construction of Bi2S3-BiOBr nanosheets on TiO2 NTA as the effective photocatalysts: Pollutant removal, photoelectric conversion and hydrogen generation," *J. Colloid Interface Sci.*, vol. 585, pp. 459–469, Mar. 2021, doi: 10.1016/j.jcis.2020.10.027.

[506] C. García-Mendoza, S. Oros-Ruiz, S. Ramírez-Rave, G. Morales-Mendoza, R. López, and R. Gómez, "Synthesis of BiS nanorods supported on ZrO semiconductor as an efficient photocatalyst for hydrogen production under UV and visible light," *J. Chem. Technol. Biotechnol.*, vol. 92, no. 7, pp. 1503–1510, 2017, doi: 10.1002/jctb.5262.

[507] U. Bharagav et al., "Z-scheme driven photocatalytic activity of CNTs-integrated Bi2S3/WO3 nanohybrid catalysts for highly efficient hydrogen evolution under solar light irradiation," *Chem. Eng. J.*, vol. 465, p. 142886, Jun. 2023, doi: 10.1016/j.cej.2023.142886.

[508] S. M. Albukhari, "Bismuth sulfide-impregnated cobalt oxide nanocrystals: An enhanced photocatalyst for improved hydrogen production beneath visible light," *Inorg. Chem. Commun.*, vol. 163, p. 112379, May 2024, doi: 10.1016/j.inoche.2024.112379.

[509] M. Nawaz, "Morphology-controlled preparation of Bi2S3-ZnS chloroplast-like structures, formation mechanism and photocatalytic activity for hydrogen production," *J. Photochem. Photobiol. Chem.*, vol. 332, pp. 326–330, Jan. 2017, doi: 10.1016/j.jphotochem.2016.09.005.

[510] L.-X. Hao, G. Chen, Y.-G. Yu, Y.-S. Zhou, Z.-H. Han, and Y. Liu, "Sonochemistry synthesis of Bi2S3/CdS heterostructure with enhanced performance for photocatalytic hydrogen evolution," *Int. J. Hydrog. Energy*, vol. 39, no. 26, pp. 14479–14486, Sep. 2014, doi: 10.1016/j.ijhydene.2014.04.140.

[511] M. Li, H. Yao, S. Yao, G. Chen, and J. Sun, "Vacancy-induced tensile strain of CdS/Bi2S3 as a highly performance and robust photocatalyst for hydrogen evolution," *J. Colloid Interface Sci.*, vol. 630, pp. 224–234, Jan. 2023, doi: 10.1016/j.jcis.2022.10.056.

[512] M. Li, J. Sun, B. Cong, S. Yao, and G. Chen, "Sulphur vacancies modified Cd0.5Zn0.5S/Bi2S3: Engineering localized surface plasma resonance enhanced visible-light-driven hydrogen evolution," *Chem. Eng. J.*, vol. 415, p. 128868, Jul. 2021, doi: 10.1016/j.cej.2021.128868.

[513] I. A. Mkhalid, "Improved photocatalytic performance in Bi2S3-ZnSe nanocomposites for hydrogen production," *Ceram. Int.*, vol. 44, no. 18, pp. 22198–22204, Dec. 2018, doi: 10.1016/j.ceramint.2018.08.338.

[514] Y. Xu, J. Xu, W. Yan, H. Tang, and G. Tang, "Synergistic effect of a noble metal free MoS2 co-catalyst and a ternary Bi2S3/MoS2/P25 heterojunction for enhanced photocatalytic H2 production," *Ceram. Int.*, vol. 47, no. 7, Part A, pp. 8895–8903, Apr. 2021, doi: 10.1016/j.ceramint.2020.12.010.
[515] Y. Li, F. Rao, J. Zhong, and J. Li, "*In-situ* fabrication of Bi2S3/g-C3N4 heterojunctions with boosted H2 production rate under visible light irradiation," *Fuel*, vol. 341, p. 127629, Jun. 2023, doi: 10.1016/j.fuel.2023.127629.
[516] S. V. P. Vattikuti, A. K. R. Police, J. Shim, and C. Byon, "Sacrificial-template-free synthesis of core-shell C@Bi2S3 heterostructures for efficient supercapacitor and H2 production applications," *Sci. Rep.*, vol. 8, no. 1, p. 4194, Mar. 2018, doi: 10.1038/s41598-018-22622-0.
[517] U. V. Kawade *et al.*, "Environmentally benign enhanced hydrogen production via lethal H 2 S under natural sunlight using hierarchical nanostructured bismuth sulfide," *RSC Adv.*, vol. 4, no. 90, pp. 49295–49302, 2014, doi: 10.1039/C4RA07143C.
[518] Z. Li, Q. Zhang, M. Dan, Z. Guo, and Y. Zhou, "A facile preparation route of Bi2S3 nanorod films for photocatalytic H2 production from H2S," *Mater. Lett.*, vol. 201, pp. 118–121, Aug. 2017, doi: 10.1016/j.matlet.2017.05.002.
[519] T. N. S. Trindade, A. M. Mota, R. A. Campos, L. A. Mercante, and L. A. Silva, "Sonochemical synthesis of bismuth sulfide-based nanorods for hydrogen production," *J. Solid State Chem.*, vol. 339, p. 124982, Nov. 2024, doi: 10.1016/j.jssc.2024.124982.
[520] S. R. Kadam, R. P. Panmand, R. S. Sonawane, S. W. Gosavi, and B. B. Kale, "A stable Bi 2 S 3 quantum dot–glass nanosystem: size tuneable photocatalytic hydrogen production under solar light," *RSC Adv.*, vol. 5, no. 72, pp. 58485–58490, 2015, doi: 10.1039/C5RA10244H.
[521] X. Li *et al.*, "Study on the relationship between Bi2S3 with different morphologies and its photocatalytic hydrogen production performance," *J. Anal. Sci. Technol.*, vol. 13, no. 1, p. 19, Jun. 2022, doi: 10.1186/s40543-022-00325-6.
[522] Y. Liu, J. Xu, Z. Ding, M. Mao, and L. Li, "Marigold shaped mesoporous composites Bi2S3/Ni(OH)2 with n-n heterojunction for high efficiency photocatalytic hydrogen production from water decomposition," *Chem. Phys. Lett.*, vol. 766, p. 138337, Mar. 2021, doi: 10.1016/j.cplett.2021.138337.
[523] M. Chahkandi *et al.*, "Graphitic carbon nitride nanosheets decorated with HAp@Bi2S3 core–shell nanorods: Dual S-scheme 1D/2D heterojunction for environmental and hydrogen production solutions," *Chem. Eng. J.*, vol. 499, p. 155886, Nov. 2024, doi: 10.1016/j.cej.2024.155886.
[524] M. Ge, C.-L. Yang, M.-S. Wang, and X.-G. Ma, "Two-dimensional PtI2/Bi2S3 and PtI2/Bi2Se3 heterostructures with high solar-to-hydrogen efficiency," *Colloids Surf. Physicochem. Eng. Asp.*, vol. 666, p. 131286, Jun. 2023, doi: 10.1016/j.colsurfa.2023.131286.
[525] N. Li, B. Zhu, L. Huang, L. Huo, Q. Dong, and J. Ma, "Piezoelectric Polarization and Sulfur Vacancy Enhanced Photocatalytic Hydrogen Evolution Performance of Bi2S3/ZnSn(OH)6 Piezo-photocatalyst," *Inorg. Chem.*, vol. 63, no. 21, pp. 10011–10021, May 2024, doi: 10.1021/acs.inorgchem.4c01213.
[526] S. L. Lee and C.-J. Chang, "Recent Progress on Metal Sulfide Composite Nanomaterials for Photocatalytic Hydrogen Production," *Catalysts*, vol. 9, no. 5, Art. no. 5, May 2019, doi: 10.3390/catal9050457.
[527] F. Wang, F. Huang, F. Yu, X. Kang, Q. Wang, and Y. Liu, "Metal-sulfide photocatalysts for solar-fuel generation across the solar spectrum," *Cell Rep. Phys. Sci.*, vol. 4, no. 6, p. 101450, Jun. 2023, doi: 10.1016/j.xcrp.2023.101450.
[528] Y. Wang, X. Chen, and C. Chen, "Microwave synthesis of Zn, Cd binary metal sulfides with superior photocatalytic H2 evolution performance," *Inorg. Chem. Commun.*, vol. 134, p. 108993, Dec. 2021, doi: 10.1016/j.inoche.2021.108993.
[529] Y. Wang, H. Jin, Y. Li, J. Fang, and C. Chen, "Ce-based organic framework enhanced the hydrogen evolution ability of ZnCdS photocatalyst," *Int. J. Hydrog. Energy*, vol. 47, no. 2, pp. 962–970, Jan. 2022, doi: 10.1016/j.ijhydene.2021.10.090.

[530] W.-K. Chong, B.-J. Ng, L.-L. Tan, and S.-P. Chai, "Recent Advances in Nanoscale Engineering of Ternary Metal Sulfide-Based Heterostructures for Photocatalytic Water Splitting Applications," *Energy Fuels*, vol. 36, no. 8, pp. 4250–4267, Apr. 2022, doi: 10.1021/acs.energyfuels.2c00291.

[531] G. M. Tomboc, B. T. Gadisa, J. Joo, H. Kim, and K. Lee, "Hollow Structured Metal Sulfides for Photocatalytic Hydrogen Generation," *ChemNanoMat*, vol. 6, no. 6, pp. 850–869, 2020, doi: 10.1002/cnma.202000125.

[532] D. Huang et al., "$Zn_xCd_{1-x}S$ based materials for photocatalytic hydrogen evolution, pollutants degradation and carbon dioxide reduction," *Appl. Catal. B Environ.*, vol. 267, p. 118651, Jun. 2020, doi: 10.1016/j.apcatb.2020.118651.

[533] J. Chen, J. Chen, and Y. Li, "Hollow ZnCdS dodecahedral cages for highly efficient visible-light-driven hydrogen generation," *J. Mater. Chem. A*, vol. 5, no. 46, pp. 24116–24125, 2017, doi: 10.1039/C7TA07587A.

[534] L. Lu, Y. Ma, R. Dong, P. Tan, Y. Chen, and J. Pan, "One-step chemical bath co-precipitation method to prepare high hydrogen-producing active $Zn_xCd_{1-x}S$ solid solution with adjustable band structure," *J. Mater. Sci.*, vol. 56, no. 9, pp. 5717–5729, Mar. 2021, doi: 10.1007/s10853-020-05668-2.

[535] "Influence mechanism of ZnCdS solid solution composition regulation on its energy band and photocatalytic hydrogen performance - ScienceDirect." Accessed: Oct. 23, 2024. [Online]. Available: https://www.sciencedirect.com/science/article/abs/pii/S1383586624026728

[536] Q. Xiao, T. Yang, X. Guo, and Z. Jin, "S-scheme heterojunction constructed by ZnCdS and $CoWO_4$ nano-ions promotes photocatalytic hydrogen production," *Surf. Interfaces*, vol. 43, p. 103577, Dec. 2023, doi: 10.1016/j.surfin.2023.103577.

[537] X. Wang, L. Ge, K. Wei, Z. Xing, and S. Feng, "Fabrication of S-scheme $Zn_{1-x}Cd_xS$/NiO heterojunction for efficient photocatalytic hydrogen evolution," *Int. J. Hydrog. Energy*, vol. 55, pp. 718–728, Feb. 2024, doi: 10.1016/j.ijhydene.2023.10.170.

[538] X. Liu, D. Wen, Z. Liu, J. Wei, D. Bu, and S. Huang, "Thiocyanate-capped CdSe@$Zn_{1-X}Cd_XS$ gradient alloyed quantum dots for efficient photocatalytic hydrogen evolution," *Chem. Eng. J.*, vol. 402, p. 126178, Dec. 2020, doi: 10.1016/j.cej.2020.126178.

[539] Z. Lian et al., "Direct Observation of Z-Scheme Route in $Cu_{31}S_{16}$/$Zn_xCd_{1-x}S$ Heteronanoplates for Highly Efficient Photocatalytic Hydrogen Evolution," *Small*, vol. 20, no. 32, p. 2400611, 2024, doi: 10.1002/smll.202400611.

[540] L. Song, S. Zhang, D. Liu, S. Sun, and J. Wei, "High-performance hydrogen evolution of NiB/ZnCdS under visible light irradiation," *Int. J. Hydrog. Energy*, vol. 45, no. 15, pp. 8234–8242, Mar. 2020, doi: 10.1016/j.ijhydene.2020.01.029.

[541] C. Guo et al., "Interfacial electric field construction of hollow PdS QDs/$Zn_{1-x}Cd_xS$ solid solution with enhanced photocatalytic hydrogen evolution," *Nanoscale*, vol. 16, no. 3, pp. 1147–1155, 2024, doi: 10.1039/D3NR05518C.

[542] S. Liu et al., "Hydrogen-bonded organic framework derived ultra-fine ZnCdS/ZnS heterojunction with high-porosity for efficient photocatalytic hydrogen production," *Appl. Surf. Sci.*, vol. 657, p. 159795, Jun. 2024, doi: 10.1016/j.apsusc.2024.159795.

[543] J. Yang, X. Hao, J. Liu, Z. Nan, and Z. Jin, "Strong electron coupling between amorphous WP and $MoSe_2$/ZnCdS S-scheme heterojunction for enhanced wide spectrum photocatalytic hydrogen production," *Sep. Purif. Technol.*, vol. 354, p. 129348, Feb. 2025, doi: 10.1016/j.seppur.2024.129348.

[544] W. Fan et al., "Facile synthesis of ZnCdS quantum dots via a novel photoetching MOF strategy for boosting photocatalytic hydrogen evolution," *Sep. Purif. Technol.*, vol. 330, p. 125258, Feb. 2024, doi: 10.1016/j.seppur.2023.125258.

[545] L. Yu, W. Xie, X. Ji, Y. Zhang, and H. Zhu, "Anchoring $MoS_2$ on ZnCdS to accelerate charge migration to promote photocatalytic water decomposition performance," *Mater. Today Commun.*, vol. 41, p. 110336, Dec. 2024, doi: 10.1016/j.mtcomm.2024.110336.

[546] W. Fan, H. Chang, X. Zheng, J. Lu, G. Yin, and Z. Jiang, "Elaborate construction of a MOF-derived novel morphology Z-scheme ZnO/ZnCdS heterojunction for enhancing photocatalytic H2 evolution and tetracycline degradation," *Sep. Purif. Technol.*, vol. 357, p. 130068, May 2025, doi: 10.1016/j.seppur.2024.130068.
[547] Z. Jin, L. Zhang, and E. Cui, "Molybdate modified ZnCdS to construct fast carrier transfer channels for efficient hydrogen evolution," *J. Photochem. Photobiol. Chem.*, vol. 454, p. 115693, Sep. 2024, doi: 10.1016/j.jphotochem.2024.115693.
[548] Y. Fan, X. Hao, N. Yi, and Z. Jin, "Strong electronic coupling of Mo2TiC2 MXene/ZnCdS ohmic junction for boosting photocatalytic hydrogen evolution," *Appl. Catal. B Environ. Energy*, vol. 357, p. 124313, Nov. 2024, doi: 10.1016/j.apcatb.2024.124313.
[549] P. Bai, Junyu Lang, S. Wang, H. Tong, W. Du, and Z. Chai, "Enhanced interfacial charge transfer by oxygen vacancies in ZnCdS/NiCo-LDH heterojunction for efficient H2 evolution," *Appl. Surf. Sci.*, vol. 643, p. 158715, Jan. 2024, doi: 10.1016/j.apsusc.2023.158715.
[550] B. Wen, X. Guo, Y. Liu, and Z. Jin, "S-Scheme Heterojunction ZnCdS Nanoparticles on Layered Ni–Co Hydroxide for Photocatalytic Hydrogen Evolution," *ACS Appl. Nano Mater.*, vol. 7, no. 6, pp. 6056–6067, Mar. 2024, doi: 10.1021/acsanm.3c05962.
[551] X. Huang, C. Guo, Y. Niu, Y. Ma, and J. Wang, "Photodeposited nickel-loaded ZnCdS nanoparticles for hydrogen production from photocatalytic water splitting," *New J. Chem.*, vol. 48, no. 15, pp. 6765–6770, 2024, doi: 10.1039/D4NJ00293H.
[552] Y.-S. Shen *et al.*, "Zn0.1Cd0.9S/NiS heterojunction photocatalysts for enhanced H2 production and glucose conversion," *Appl. Surf. Sci.*, vol. 626, p. 157237, Jul. 2023, doi: 10.1016/j.apsusc.2023.157237.
[553] X. Hao, J. Yang, Y. Wang, Y. Fan, and Z. Jin, "Gradhdiyne as an electronic bridge to facilitate built-in electric field in NiS/ZnCdS S-Scheme heterostructure for efficient photocatalytic hydrogen evolution," *Int. J. Hydrog. Energy*, vol. 83, pp. 1392–1404, Sep. 2024, doi: 10.1016/j.ijhydene.2024.08.145.
[554] E. Ha *et al.*, "Surface disorder engineering in ZnCdS for cocatalyst free visible light driven hydrogen production," *Nano Res.*, vol. 15, no. 2, pp. 996–1002, Feb. 2022, doi: 10.1007/s12274-021-3587-5.
[555] X. Wang *et al.*, "Induced dipole moments in amorphous ZnCdS catalysts facilitate photocatalytic H2 evolution," *Nat. Commun.*, vol. 15, no. 1, p. 2600, Mar. 2024, doi: 10.1038/s41467-024-47022-z.
[556] Q. Tian, L. Wang, W. Sun, A. Meng, L. Yang, and Z. Li, "Dipole polarization-driven spatial charge separation in defective zinc cadmium sulfoselenide for boosting photocatalytic hydrogen evolution," *Appl. Catal. B Environ. Energy*, vol. 359, p. 124516, Dec. 2024, doi: 10.1016/j.apcatb.2024.124516.
[557] M. Huang *et al.*, "Twin Zn1−CdS Solid Solution: Highly Efficient Photocatalyst for Water Splitting," *Small*, vol. 20, no. 3, p. 2304784, 2024, doi: 10.1002/smll.202304784.
[558] C. Ji *et al.*, "Enhanced Photocatalytic Hydrogen Evolution over Twin-Crystal ZnCdS with Sulfur Vacancies Prepared by High-Gravity Technology," *Ind. Eng. Chem. Res.*, vol. 63, no. 19, pp. 8497–8508, May 2024, doi: 10.1021/acs.iecr.4c00317.
[559] H. Yang *et al.*, "Synthesis of Type-S Ni3S4/ZnCdS Quantum Dots via Constitution Controller l-Cysteine for Photocatalytic H2 Evolution," *ACS Appl. Nano Mater.*, vol. 7, no. 18, pp. 22093–22103, Sep. 2024, doi: 10.1021/acsanm.4c04058.
[560] Q. Liu, C. Zhang, W. Li, S. Ming, L. Li, and Z. Lv, "ZnCdS/CoMoOS Photocatalyst with High Electron Flux at the Ternary Interface Enhances H2 production," *ACS Appl. Nano Mater.*, vol. 7, no. 7, pp. 8278–8288, Apr. 2024, doi: 10.1021/acsanm.4c01243.
[561] X. Cao, L. Zhang, C. Guo, M. Wang, J. Guo, and J. Wang, "Construction of ZnxCdyS with a 3D Hierarchical Structure for Enhanced Photocatalytic Hydrogen Production from Water Splitting," *Inorg. Chem.*, vol. 62, no. 46, pp. 18990–18998, Nov. 2023, doi: 10.1021/acs.inorgchem.3c02638.

[562] T. Bai, X. Shi, M. Liu, H. Huang, J. Zhang, and X.-H. Bu, "g-C 3 N 4 /ZnCdS heterojunction for efficient visible light-driven photocatalytic hydrogen production," *RSC Adv.*, vol. 11, no. 60, pp. 38120–38125, 2021, doi: 10.1039/D1RA05894K.

[563] T. Bai *et al.*, "A metal–organic framework-derived Zn 1−x Cd x S/CdS heterojunction for efficient visible light-driven photocatalytic hydrogen production," *Dalton Trans.*, vol. 50, no. 18, pp. 6064–6070, 2021, doi: 10.1039/D1DT00667C.

[564] L. Wei, L. Wen, Y. Chen, Y. Qin, X. Yang, and J. Yang, "Facile synthesis of Ag2S cocatalyst modified Zn0·6Cd0·4S for efficient photocatalytic hydrogen production," *Int. J. Hydrog. Energy*, vol. 74, pp. 151–160, Jul. 2024, doi: 10.1016/j.ijhydene.2024.06.103.

[565] M. Li, F. Chen, Y. Xu, and M. Tian, "Ni(OH)2 Nanosheet as an Efficient Cocatalyst for Improved Photocatalytic Hydrogen Evolution over Cd0.9Zn0.1S Nanorods under Visible Light," *Langmuir*, vol. 40, no. 7, pp. 3793–3803, Feb. 2024, doi: 10.1021/acs.langmuir.3c03631.

[566] S. Kai, B. Xi, Y. Wang, and S. Xiong, "One-Pot Synthesis of Size-Controllable Core–Shell CdS and Derived CdS@ZnCd1−S Structures for Photocatalytic Hydrogen Production," *Chem. – Eur. J.*, vol. 23, no. 65, pp. 16653–16659, 2017, doi: 10.1002/chem.201703506.

[567] M. Liu, Y. Du, L. Ma, D. Jing, and L. Guo, "Manganese doped cadmium sulfide nanocrystal for hydrogen production from water under visible light," *Int. J. Hydrog. Energy*, vol. 37, no. 1, pp. 730–736, Jan. 2012, doi: 10.1016/j.ijhydene.2011.04.111.

[568] H. Li *et al.*, "Rational synthesis of MnxCd1-xS for enhanced photocatalytic H2 evolution: Effects of S precursors and the feed ratio of Mn/Cd on its structure and performance," *J. Colloid Interface Sci.*, vol. 535, pp. 469–480, Feb. 2019, doi: 10.1016/j.jcis.2018.10.018.

[569] H. Gong, X. Hao, H. Li, and Z. Jin, "A novel materials manganese cadmium sulfide/cobalt nitride for efficiently photocatalytic hydrogen evolution," *J. Colloid Interface Sci.*, vol. 585, pp. 217–228, Mar. 2021, doi: 10.1016/j.jcis.2020.11.088.

[570] Y. Sun *et al.*, "Controlled synthesis of MnxCd1–xS for enhanced visible-light driven photocatalytic hydrogen evolution," *Chin. J. Struct. Chem.*, vol. 42, no. 8, p. 100145, Aug. 2023, doi: 10.1016/j.cjsc.2023.100145.

[571] W. Deng, X. Hao, Y. wang, Y. Fan, and Z. Jin, "Construction of NiS/MnCdS S-scheme heterojunction for efficient photocatalytic overall water splitting: Regulation of surface sulfur vacancy and energy band structure," *Fuel*, vol. 363, p. 130964, May 2024, doi: 10.1016/j.fuel.2024.130964.

[572] Y. Shao, X. Hao, W. Deng, and Z. Jin, "Strong electron coupling effect of Zn-vacancy engineered S-scheme MnCdS/ZnS heterojunction derived from Metal-organic frameworks for highly efficient photocatalytic overall water splitting," *J. Colloid Interface Sci.*, vol. 678, pp. 885–901, Jan. 2025, doi: 10.1016/j.jcis.2024.09.037.

[573] Z. Li *et al.*, "Mn-Cd-S@amorphous-Ni3S2 hybrid catalyst with enhanced photocatalytic property for hydrogen production and electrocatalytic OER," *Appl. Surf. Sci.*, vol. 491, pp. 799–806, Oct. 2019, doi: 10.1016/j.apsusc.2019.05.313.

[574] C. Wang, S. Shi, B. Liu, G. Wang, and Z. Jin, "A novel dual S-scheme Co 9 S 8 /MnCdS/Co 3 O 4 heterojunction for photocatalytic hydrogen evolution under visible light irradiation," *Nanoscale*, vol. 16, no. 36, pp. 17009–17023, 2024, doi: 10.1039/D4NR03195D.

[575] Y. Zhou *et al.*, "Fabrication of novel Mn0.43Cd0.57S/ZnCo2O4 p-n heterojunction photocatalyst with efficient charge separation for highly enhanced visible-light-driven hydrogen evolution," *Ceram. Int.*, vol. 48, no. 21, pp. 31334–31343, Nov. 2022, doi: 10.1016/j.ceramint.2022.06.315.

[576] M. Zhang, N. Fang, X. Song, Y. Chu, S. Shu, and Y. Liu, "p–n Heterojunction Photocatalyst Mn0.5Cd0.5S/CuCo2S4 for Highly Efficient Visible Light-Driven H2 Production," *ACS Omega*, vol. 5, no. 50, pp. 32715–32723, Dec. 2020, doi: 10.1021/acsomega.0c05106.

[577] Y. Shang, J. Xu, Y. Ma, Z. Li, and Q. Li, "3D nanorod-like Mn 0.2 Cd 0.8 S modified by amorphous NiCo 2 S 4 was used for high efficiency photocatalytic hydrogen evolution," *New J. Chem.*, vol. 47, no. 36, pp. 16972–16980, 2023, doi: 10.1039/D3NJ02899B.

[578] Y. Sun, M. Zhang, X. Mou, C. Song, and D. Wang, "Construction of S-scheme Mn0.1Cd0.9S/WO3 1D/0D heterojunction assemblies for visible-light driven high-efficient H2 evolution," *J. Alloys Compd.*, vol. 927, p. 167114, Dec. 2022, doi: 10.1016/j.jallcom.2022.167114.

[579] X. Zhou, S. Liu, C. Yang, J. Qin, and Y. Hu, "Photocatalytic hydrogen energy recovery from sulfide-containing wastewater using thiol-UiO-66 modified Mn0.5Cd0.5S nanocomposites," *Sep. Purif. Technol.*, vol. 316, p. 123772, Jul. 2023, doi: 10.1016/j.seppur.2023.123772.

[580] H. Gong, G. Wang, H. Li, Z. Jin, and Q. Guo, "Mn0.2Cd0.8S nanorods assembled with 0D CoWO4 nanoparticles formed p-n heterojunction for efficient photocatalytic hydrogen evolution," *Int. J. Hydrog. Energy*, vol. 45, no. 51, pp. 26733–26745, Oct. 2020, doi: 10.1016/j.ijhydene.2020.07.059.

[581] Y. Cao, G. Wang, Q. Ma, and Z. Jin, "Amorphous NiCoB nanoalloy modified Mn0.05Cd0.95S for photocatalytic hydrogen evolution," *Mol. Catal.*, vol. 492, p. 111001, Aug. 2020, doi: 10.1016/j.mcat.2020.111001.

[582] Y. Zhao, P. Yang, and J. Li, "Fabrication of one-dimensional MnxCd1-xS@D-MoSeyS2-y heterostructure with enhanced visible-light photocatalytic hydrogen evolution," *Int. J. Hydrog. Energy*, vol. 46, no. 43, pp. 22422–22433, Jun. 2021, doi: 10.1016/j.ijhydene.2021.04.063.

[583] L. Yang, Q. Tian, X. Wang, H. Yang, A. Meng, and Z. Li, "Interfacial-Engineered Co3S4/MnCdS Heterostructure for Efficient Photocatalytic Hydrogen Evolution," *Sol. RRL*, vol. 7, no. 17, p. 2300403, 2023, doi: 10.1002/solr.202300403.

[584] L. Fan *et al.*, "Mn-doped CdS/Cu2O: An S-scheme heterojunction for photocatalytic hydrogen production," *J. Alloys Compd.*, vol. 960, p. 170382, Oct. 2023, doi: 10.1016/j.jallcom.2023.170382.

[585] H. Qian *et al.*, "Development of a novel ternary FeWO4/CoP/Mn0.5Cd0.5S composite photocatalyst for photocatalytic hydrogen evolution in watersplitting," *Colloids Surf. Physicochem. Eng. Asp.*, vol. 682, p. 132935, Feb. 2024, doi: 10.1016/j.colsurfa.2023.132935.

[586] F. Zhao, X. Yang, S. Xiong, J. Li, H. Fu, and X. An, "Efficient Schottky heterojunctions of CoP nanoparticles decorated Mn-doped CdS nanorods for photocatalytic hydrogen evolution by water splitting," *Int. J. Hydrog. Energy*, vol. 89, pp. 434–442, Nov. 2024, doi: 10.1016/j.ijhydene.2024.09.334.

[587] B. Liu, S. Shi, C. Wang, G. Wang, and Z. Jin, "Enhanced hydrogen evolution performance of twinned Mn0.65Cd0.35S with Zn-doped cadmium selenide under visible light irradiation," *Int. J. Hydrog. Energy*, vol. 88, pp. 1072–1084, Oct. 2024, doi: 10.1016/j.ijhydene.2024.09.260.

[588] H. Wang *et al.*, "Preparation of Mn0.8Cd0.2S/NiCo2S4 Z-scheme heterojunction composite for enhanced photocatalytic hydrogen production," *Surf. Interfaces*, vol. 45, p. 103900, Feb. 2024, doi: 10.1016/j.surfin.2024.103900.

[589] W. Deng, X. Hao, J. Yang, and Z. Jin, "Strong electron coupling effect of non-precious metal Schottky junctions enhanced square meter level photocatalytic hydrogen evolution," *Appl. Catal. B Environ. Energy*, vol. 360, p. 124551, Jan. 2025, doi: 10.1016/j.apcatb.2024.124551.

[590] J. Wang *et al.*, "A review: Synthesis, modification and photocatalytic applications of ZnIn2S4," *J. Mater. Sci. Technol.*, vol. 78, pp. 1–19, Jul. 2021, doi: 10.1016/j.jmst.2020.09.045.

[591] N. S. Chaudhari *et al.*, "Ecofriendly hydrogen production from abundant hydrogen sulfide using solar light-driven hierarchical nanostructured ZnIn2S4 photocatalyst," *Green Chem.*, vol. 13, no. 9, pp. 2500–2506, 2011, doi: 10.1039/C1GC15515F.

[592] Y. Song, J. Zhang, X. Dong, and H. Li, "A Review and Recent Developments in Full-Spectrum Photocatalysis using ZnIn2S4-Based Photocatalysts," *Energy Technol.*, vol. 9, no. 5, p. 2100033, 2021, doi: 10.1002/ente.202100033.

[593] R. Yang *et al.*, "ZnIn2S4-Based Photocatalysts for Energy and Environmental Applications," *Small Methods*, vol. 5, no. 10, p. 2100887, 2021, doi: 10.1002/smtd.202100887.

[594] G. Zhang *et al.*, "A mini-review on ZnIn2S4-Based photocatalysts for energy and environmental application," *Green Energy Environ.*, vol. 7, no. 2, pp. 176–204, Apr. 2022, doi: 10.1016/j.gee.2020.12.015.

[595] Y. Kumar *et al.*, "Novel Z-Scheme ZnIn2S4-based photocatalysts for solar-driven environmental and energy applications: Progress and perspectives," *J. Mater. Sci. Technol.*, vol. 87, pp. 234–257, Oct. 2021, doi: 10.1016/j.jmst.2021.01.051.
[596] S. Shen, L. Zhao, and L. Guo, "Cetyltrimethylammoniumbromide (CTAB)-assisted hydrothermal synthesis of ZnIn2S4 as an efficient visible-light-driven photocatalyst for hydrogen production," *Int. J. Hydrog. Energy*, vol. 33, no. 17, pp. 4501–4510, Sep. 2008, doi: 10.1016/j.ijhydene.2008.05.043.
[597] Y. F. Yu *et al.*, "Mo-activated ZnIn2S4 photocatalyst for enhanced hydrogen evolution coupled with benzyl alcohol oxidation," *Sci. China Mater.*, vol. 67, no. 8, pp. 2645–2652, Aug. 2024, doi: 10.1007/s40843-024-2964-3.
[598] X. Feng *et al.*, "Facile Synthesis of P-Doped ZnIn2S4 with Enhanced Visible-Light-Driven Photocatalytic Hydrogen Production," *Molecules*, vol. 28, no. 11, Art. no. 11, Jan. 2023, doi: 10.3390/molecules28114520.
[599] R. Janani *et al.*, "Enhanced solar light driven hydrogen generation and environment remediation through Nd incorporated ZnIn2S4," *Renew. Energy*, vol. 162, pp. 2031–2040, Dec. 2020, doi: 10.1016/j.renene.2020.09.081.
[600] K. You *et al.*, "Efficient photocatalytic hydrogen production over ZnIn2S4 by producing sulfur vacancies and coupling with nickel-based polyoxometalate," *Chem. Commun.*, vol. 59, no. 73, pp. 10972–10975, Sep. 2023, doi: 10.1039/D3CC03329E.
[601] R. Xiong, X. Zhou, K. Chen, Y. Xiao, B. Cheng, and S. Lei, "Oxygen-Defect-Mediated ZnCr2O4/ZnIn2S4 Z-Scheme Heterojunction as Photocatalyst for Hydrogen Production and Wastewater Remediation," *Inorg. Chem.*, vol. 62, no. 8, pp. 3646–3659, Feb. 2023, doi: 10.1021/acs.inorgchem.2c04500.
[602] R. Janani, R. Preethi V, S. Singh, A. Rani, and C.-T. Chang, "Hierarchical Ternary Sulfides as Effective Photocatalyst for Hydrogen Generation Through Water Splitting: A Review on the Performance of ZnIn2S4," *Catalysts*, vol. 11, no. 2, Art. no. 2, Feb. 2021, doi: 10.3390/catal11020277.
[603] X. Zheng *et al.*, "ZnIn2S4-based photocatalysts for photocatalytic hydrogen evolution via water splitting," *Coord. Chem. Rev.*, vol. 475, p. 214898, Jan. 2023, doi: 10.1016/j.ccr.2022.214898.
[604] Y. Ren, J. J. Foo, D. Zeng, and W.-J. Ong, "ZnIn2S4-Based Nanostructures in Artificial Photosynthesis: Insights into Photocatalytic Reduction toward Sustainable Energy Production," *Small Struct.*, vol. 3, no. 11, p. 2200017, 2022, doi: 10.1002/sstr.202200017.
[605] G. Zuo *et al.*, "Ultrathin ZnIn2S4 Nanosheets Anchored on Ti3C2T MXene for Photocatalytic H2 Evolution," *Angew. Chem.*, vol. 132, no. 28, pp. 11383–11388, 2020, doi: 10.1002/ange.202002136.
[606] Z. Li, X. Wang, W. Tian, A. Meng, and L. Yang, "CoNi Bimetal Cocatalyst Modifying a Hierarchical ZnIn2S4 Nanosheet-Based Microsphere Noble-Metal-Free Photocatalyst for Efficient Visible-Light-Driven Photocatalytic Hydrogen Production," *ACS Sustain. Chem. Eng.*, vol. 7, no. 24, pp. 20190–20201, Dec. 2019, doi: 10.1021/acssuschemeng.9b06430.
[607] Q. Li, C. Cui, H. Meng, and J. Yu, "Visible-Light Photocatalytic Hydrogen Production Activity of ZnIn2S4 Microspheres Using Carbon Quantum Dots and Platinum as Dual Co-catalysts," *Chem. – Asian J.*, vol. 9, no. 7, pp. 1766–1770, 2014, doi: 10.1002/asia.201402128.
[608] Y. Su *et al.*, "Defect-driven electroless deposition and activation of platinum sites on ZnIn2S4 nanosheets for accelerated kinetics of photocatalytic hydrogen production," *Appl. Catal. B Environ.*, vol. 334, p. 122827, Oct. 2023, doi: 10.1016/j.apcatb.2023.122827.
[609] Y. Li, J. Wang, S. Peng, G. Lu, and S. Li, "Photocatalytic hydrogen generation in the presence of glucose over ZnS-coated ZnIn2S4 under visible light irradiation," *Int. J. Hydrog. Energy*, vol. 35, no. 13, pp. 7116–7126, Jul. 2010, doi: 10.1016/j.ijhydene.2010.02.017.
[610] C.-L. Tan, M.-Y. Qi, Z.-R. Tang, and Y.-J. Xu, "Cocatalyst decorated ZnIn2S4 composites for cooperative alcohol conversion and H2 evolution," *Appl. Catal. B Environ.*, vol. 298, p. 120541, Dec. 2021, doi: 10.1016/j.apcatb.2021.120541.

[611] F. Tian, R. Zhu, K. Song, F. Ouyang, and G. Cao, "The effects of amount of La on the photocatalytic performance of ZnIn2S4 for hydrogen generation under visible light," *Int. J. Hydrog. Energy*, vol. 40, no. 5, pp. 2141–2148, Feb. 2015, doi: 10.1016/j.ijhydene.2014.12.025.

[612] C. Liu, Q. Zhang, and Z. Zou, "Recent advances in designing ZnIn2S4-based heterostructured photocatalysts for hydrogen evolution," *J. Mater. Sci. Technol.*, vol. 139, pp. 167–188, Mar. 2023, doi: 10.1016/j.jmst.2022.08.030.

[613] M. Tan et al., "Boosting Photocatalytic Hydrogen Production via Interfacial Engineering on 2D Ultrathin Z-Scheme ZnIn2S4/g-C3N4 Heterojunction," *Adv. Funct. Mater.*, vol. 32, no. 14, p. 2111740, 2022, doi: 10.1002/adfm.202111740.

[614] Y.-C. Chang, Y.-C. Chiao, and C.-J. Chang, "Synthesis of g-C3N4@ZnIn2S4 Heterostructures with Extremely High Photocatalytic Hydrogen Production and Reusability," *Catalysts*, vol. 13, no. 8, Art. no. 8, Aug. 2023, doi: 10.3390/catal13081187.

[615] J. Zhang, X. Gu, Y. Zhao, K. Zhang, Y. Yan, and K. Qi, "Photocatalytic Hydrogen Production and Tetracycline Degradation Using ZnIn2S4 Quantum Dots Modified g-C3N4 Composites," *Nanomaterials*, vol. 13, no. 2, Art. no. 2, Jan. 2023, doi: 10.3390/nano13020305.

[616] X. Liu et al., "Construction of Au/g-C3N4/ZnIn2S4 plasma photocatalyst heterojunction composite with 3D hierarchical microarchitecture for visible-light-driven hydrogen production," *Int. J. Hydrog. Energy*, vol. 47, no. 5, pp. 2900–2913, Jan. 2022, doi: 10.1016/j.ijhydene.2021.10.203.

[617] L. Hou et al., "Flower-Like Dual-Defective Z-Scheme Heterojunction g-C3N4/ZnIn2S4 High-Efficiency Photocatalytic Hydrogen Evolution and Degradation of Mixed Pollutants," *Nanomaterials*, vol. 11, no. 10, Art. no. 10, Oct. 2021, doi: 10.3390/nano11102483.

[618] M. Cai et al., "Enhanced Photocatalytic Hydrogen Production of ZnIn2S4 by Using Surface-Engineered Ti3C2Tx MXene as a Cocatalyst," *Materials*, vol. 16, no. 6, Art. no. 6, Jan. 2023, doi: 10.3390/ma16062168.

[619] G. Swain, S. Sultana, and K. Parida, "One-Pot-Architectured Au-Nanodot-Promoted MoS2/ZnIn2S4: A Novel p–n Heterojunction Photocatalyst for Enhanced Hydrogen Production and Phenol Degradation," *Inorg. Chem.*, vol. 58, no. 15, pp. 9941–9955, Aug. 2019, doi: 10.1021/acs.inorgchem.9b01105.

[620] S. Zhang et al., "MoS2 Quantum Dot Growth Induced by S Vacancies in a ZnIn2S4 Monolayer: Atomic-Level Heterostructure for Photocatalytic Hydrogen Production," *ACS Nano*, vol. 12, no. 1, pp. 751–758, Jan. 2018, doi: 10.1021/acsnano.7b07974.

[621] G. Swain and K. Parida, "Designing of a novel p-MoS2@n-ZnIn2S4 heterojunction based semiconducting photocatalyst towards photocatalytic HER," *Mater. Today Proc.*, vol. 35, pp. 268–274, Jan. 2021, doi: 10.1016/j.matpr.2020.05.756.

[622] Y. Xu et al., "MoS2/HCSs/ZnIn2S4 nanocomposites with enhanced charge transport and photocatalytic hydrogen evolution performance," *J. Alloys Compd.*, vol. 895, p. 162504, Feb. 2022, doi: 10.1016/j.jallcom.2021.162504.

[623] Z. Zhang et al., "In situ constructing interfacial contact MoS2/ZnIn2S4 heterostructure for enhancing solar photocatalytic hydrogen evolution," *Appl. Catal. B Environ.*, vol. 233, pp. 112–119, Oct. 2018, doi: 10.1016/j.apcatb.2018.04.006.

[624] H. Song et al., "A facile synthesis of a ZIF-derived ZnS/ZnIn 2 S 4 heterojunction and enhanced photocatalytic hydrogen evolution," *Dalton Trans.*, vol. 49, no. 31, pp. 10816–10823, 2020, doi: 10.1039/D0DT02141E.

[625] P. Varma, M. Gopannagari, K. A. J. Reddy, T. K. Kim, and D. A. Reddy, "Design of ZnIn2S4@N−Fe3C Nanorods-Embedded Nanocages Assemblies for Efficient Photocatalytic Hydrogen Generation," *ChemCatChem*, vol. 15, no. 12, p. e202300316, 2023, doi: 10.1002/cctc.202300316.

[626] R. Yang, K. Song, J. He, Y. Fan, and R. Zhu, "Photocatalytic Hydrogen Production by RGO/ZnIn2S4 under Visible Light with Simultaneous Organic Amine Degradation," *ACS Omega*, vol. 4, no. 6, pp. 11135–11140, Jun. 2019, doi: 10.1021/acsomega.9b01034.

[627] P. Jin *et al.*, "Construction of hierarchical ZnIn2S4@PCN-224 heterojunction for boosting photocatalytic performance in hydrogen production and degradation of tetracycline hydrochloride," *Appl. Catal. B Environ.*, vol. 284, p. 119762, May 2021, doi: 10.1016/j.apcatb.2020.119762.
[628] H. Liu, J. Zhang, and D. Ao, "Construction of heterostructured ZnIn2S4@NH2-MIL-125(Ti) nanocomposites for visible-light-driven H2 production," *Appl. Catal. B Environ.*, vol. 221, pp. 433–442, Feb. 2018, doi: 10.1016/j.apcatb.2017.09.043.
[629] X. Ma *et al.*, "Fabrication of novel noble-metal-free ZnIn2S4/WC Schottky junction heterojunction photocatalyst: Efficient charge separation, increased active sites and low hydrogen production overpotential for boosting visible-light H2 evolution," *J. Alloys Compd.*, vol. 901, p. 163709, Apr. 2022, doi: 10.1016/j.jallcom.2022.163709.
[630] H. Lv *et al.*, "Engineering of direct Z-scheme ZnIn2S4/NiWO4 heterojunction with boosted photocatalytic hydrogen production," *Colloids Surf. Physicochem. Eng. Asp.*, vol. 666, p. 131384, Jun. 2023, doi: 10.1016/j.colsurfa.2023.131384.
[631] K. Lei, M. Kou, Z. Ma, Y. Deng, L. Ye, and Y. Kong, "A comparative study on photocatalytic hydrogen evolution activity of synthesis methods of CDs/ZnIn2S4 photocatalysts," *Colloids Surf. Physicochem. Eng. Asp.*, vol. 574, pp. 105–114, Aug. 2019, doi: 10.1016/j.colsurfa.2019.04.073.
[632] J. Lee, H. Kim, T. Lee, W. Jang, K. H. Lee, and A. Soon, "Revisiting Polytypism in Hexagonal Ternary Sulfide ZnIn2S4 for Photocatalytic Hydrogen Production Within the Z-Scheme," *Chem. Mater.*, vol. 31, no. 21, pp. 9148–9155, Nov. 2019, doi: 10.1021/acs.chemmater.9b03539.
[633] X. Shi *et al.*, "Protruding Pt single-sites on hexagonal ZnIn2S4 to accelerate photocatalytic hydrogen evolution," *Nat. Commun.*, vol. 13, no. 1, p. 1287, Mar. 2022, doi: 10.1038/s41467-022-28995-1.
[634] J. Cui *et al.*, "Single-atomic activation on ZnIn2S4 basal planes boosts photocatalytic hydrogen evolution," *Nano Res.*, vol. 17, no. 7, pp. 5949–5955, Jul. 2024, doi: 10.1007/s12274-024-6617-2.
[635] D. Zeng *et al.*, "Nitrogen-doped ZnIn2S4 and TPA multi-dimensional synergistically enhance photocatalytic hydrogen production," *Fuel*, vol. 380, p. 133151, Jan. 2025, doi: 10.1016/j.fuel.2024.133151.
[636] P. Zhang *et al.*, "ZnIn2S4 Nanosheets with Geometric Defects for Enhanced Solar-Driven Hydrogen Evolution and Wastewater Treatment," *Renew. Energy*, p. 121741, Oct. 2024, doi: 10.1016/j.renene.2024.121741.
[637] Z. Xie *et al.*, "Efficient photocatalytic hydrogen production by space separation of photo-generated charges from S-scheme ZnIn2S4/ZnO heterojunction," *J. Colloid Interface Sci.*, vol. 650, pp. 784–797, Nov. 2023, doi: 10.1016/j.jcis.2023.07.032.
[638] L. Li, Z. Zhang, D. Fang, and D. Yang, "Efficient photocatalytic hydrogen evolution of Z-scheme BiVO4/ZnIn2S4 heterostructure driven by visible light," *Inorg. Chem. Commun.*, vol. 169, p. 112971, Nov. 2024, doi: 10.1016/j.inoche.2024.112971.
[639] P. Su *et al.*, "SnFe$_2$O$_4$/ZnIn$_2$S$_4$/PVDF piezophotocatalyst with improved photocatalytic hydrogen production by synergetic effects of heterojunction and piezoelectricity," *J. Adv. Ceram.*, vol. 12, no. 9, pp. 1685–1700, Sep. 2023, doi: 10.26599/JAC.2023.9220758.
[640] L. Ye, X. Peng, Z. Wen, and H. Huang, "Solid-state Z-scheme assisted hydrated tungsten trioxide/ZnIn2S4 photocatalyst for efficient photocatalytic H2 production," *Mater. Futur.*, vol. 1, no. 3, p. 035103, Aug. 2022, doi: 10.1088/2752-5724/ac7faf.
[641] B. B. Kale, J.-O. Baeg, S. M. Lee, H. Chang, S.-J. Moon, and C. W. Lee, "CdIn2S4 Nanotubes and 'Marigold' Nanostructures: A Visible-Light Photocatalyst," *Adv. Funct. Mater.*, vol. 16, no. 10, pp. 1349–1354, 2006, doi: 10.1002/adfm.200500525.
[642] C. Ling *et al.*, "Solvothermal synthesis of CdIn2S4 photocatalyst for selective photosynthesis of organic aromatic compounds under visible light," *Sci. Rep.*, vol. 7, no. 1, p. 27, Feb. 2017, doi: 10.1038/s41598-017-00055-5.

[643] G. Yadav and Md. Ahmaruzzaman, "CdIn2S4-based advanced composite materials: Structure, properties, and applications in environment and energy – A concise review," *Inorg. Nano-Met. Chem.*, vol. 0, no. 0, pp. 1–15, doi: 10.1080/24701556.2023.2240775.

[644] J. He *et al.*, "Ultra-thin CdIn2S4 nanosheets with nanoholes for efficient photocatalytic hydrogen evolution," *Opt. Mater.*, vol. 108, p. 110231, Oct. 2020, doi: 10.1016/j.optmat.2020.110231.

[645] A. Bhirud *et al.*, "Surfactant tunable hierarchical nanostructures of CdIn2S4 and their photohydrogen production under solar light," *Int. J. Hydrog. Energy*, vol. 36, no. 18, pp. 11628–11639, Sep. 2011, doi: 10.1016/j.ijhydene.2011.06.061.

[646] M. Yu *et al.*, "The construction of three-dimensional CdIn2S4/MoS2 composite materials for efficient hydrogen production," *J. Alloys Compd.*, vol. 892, p. 162168, Feb. 2022, doi: 10.1016/j.jallcom.2021.162168.

[647] Y. Lu, H. Liu, L. Wang, Y. Geng, and M. Zhang, "Preparation of CdIn2S4 nanoparticles@MoS2 microrods heterojunctions for boosted photocatalytic hydrogen production," *J. Alloys Compd.*, vol. 982, p. 173750, Apr. 2024, doi: 10.1016/j.jallcom.2024.173750.

[648] B. Zhang, H. Shi, X. Hu, Y. Wang, E. Liu, and J. Fan, "A novel S-scheme MoS2/CdIn2S4 flower-like heterojunctions with enhanced photocatalytic degradation and H2 evolution activity," *J. Phys. Appl. Phys.*, vol. 53, no. 20, p. 205101, Mar. 2020, doi: 10.1088/1361-6463/ab7563.

[649] X. Dang, M. Xie, F. Dai, J. Guo, J. Liu, and X. Lu, "The in situ construction of ZnIn 2 S 4 /CdIn 2 S 4 2D/3D nano hetero-structure for an enhanced visible-light-driven hydrogen production," *J. Mater. Chem. A*, vol. 9, no. 26, pp. 14888–14896, 2021, doi: 10.1039/D1TA02052H.

[650] Y. Yu, G. Chen, G. Wang, and Z. Lv, "Visible-light-driven ZnIn2S4/CdIn2S4 composite photocatalyst with enhanced performance for photocatalytic H2 evolution," *Int. J. Hydrog. Energy*, vol. 38, no. 3, pp. 1278–1285, Feb. 2013, doi: 10.1016/j.ijhydene.2012.11.020.

[651] H. Qiu *et al.*, "Fabrication of Noble-Metal-Free Mo2C/CdIn2S4 Heterojunction Composites with Elevated Carrier Separation for Photocatalytic Hydrogen Production," *Molecules*, vol. 28, no. 6, Art. no. 6, Jan. 2023, doi: 10.3390/molecules28062508.

[652] Y. Qi, G. Zhou, Y. Wu, H. Wang, and Y. Wu, "In-Situ Construction of In2o3/In2s3-Cdin2s4 Z-Scheme Heterojunction Nanotubes for Enhanced Photocatalytic Hydrogen Production," Jan. 16, 2024, *Social Science Research Network, Rochester, NY*: 4696262. doi: 10.2139/ssrn.4696262.

[653] Y. Geng, Y. Lu, H. Liu, and L. Wang, "Construction of CdIn2S4/ZnSn(OH)6 heterojunctions for efficient photocatalytic hydrogen generation," *Int. J. Hydrog. Energy*, vol. 68, pp. 181–189, May 2024, doi: 10.1016/j.ijhydene.2024.04.203.

[654] S. Guo *et al.*, "Controllable construction of hierarchically CdIn2S4/CNFs/Co4S3 nanofiber networks towards photocatalytic hydrogen evolution," *Chem. Eng. J.*, vol. 419, p. 129213, Sep. 2021, doi: 10.1016/j.cej.2021.129213.

[655] Y. Rao, M. Sun, B. Zhou, Z. Wang, T. Yan, and Y. Shao, "Highly Efficient 2D/3D Structured CuCo2S4/CdIn2S4 Composites with a p–n Heterojunction for Boosting Photocatalytic Hydrogen Evolution," *ACS Appl. Energy Mater.*, vol. 7, no. 13, pp. 5457–5466, Jul. 2024, doi: 10.1021/acsaem.4c00794.

[656] Y. Xu, M. Wu, F. Chen, and M. Tian, "ReS2/CdIn2S4–SV Heterojunctions for Photocatalytic Hydrogen Production," *ACS Appl. Nano Mater.*, vol. 7, no. 14, pp. 16031–16041, Jul. 2024, doi: 10.1021/acsanm.4c01750.

[657] X. Ma, S. Hou, D. Li, Y. Wu, and S. Li, "Novel and noble-metal-free CdIn2S4/MoB Schottky heterojunction photocatalysts with efficient charge separation for boosting photocatalytic H2 production," *Sep. Purif. Technol.*, vol. 354, p. 129057, Feb. 2025, doi: 10.1016/j.seppur.2024.129057.

[658] X. Ma *et al.*, "Novel noble-metal-free Co2P/CdIn2S4 heterojunction photocatalysts for elevated photocatalytic H2 production: Light absorption, charge separation and active site," *J. Colloid Interface Sci.*, vol. 639, pp. 87–95, Jun. 2023, doi: 10.1016/j.jcis.2023.02.062.

[659] X. Ma *et al.*, "Construction of novel noble-metal-free MoP/CdIn2S4 heterojunction photocatalysts: Effective carrier separation, accelerating dynamically H2 release and increased active sites for

enhanced photocatalytic H2 evolution," *J. Colloid Interface Sci.*, vol. 628, pp. 368–377, Dec. 2022, doi: 10.1016/j.jcis.2022.07.184.

[660] W. Yang, S.-S. Xu, Y. Niu, Y. Zhang, and J. Xu, "Ni12P5-Supported Marigold-Shaped CdIn2S4: A 2D/3D Non-Noble-Metal Catalyst for Visible-Light-Driven Hydrogen Production," *J. Phys. Chem. C*, vol. 127, no. 10, pp. 4853–4861, Mar. 2023, doi: 10.1021/acs.jpcc.3c00092.

[661] X. Ma, S. Hou, D. Li, Y. Wu, J. Yin, and S. Li, "Novel noble-metal-free NiCo2O4/CdIn2S4 S-scheme heterojunction photocatalyst with redox center for highly efficient photocatalytic H2 evolution," *Appl. Surf. Sci.*, vol. 672, p. 160895, Nov. 2024, doi: 10.1016/j.apsusc.2024.160895.

[662] Y. Zhou *et al.*, "In situ growth of CdIn2S4 on NH2-MIL-125 as efficient photocatalysts for H2 production under visible-light irradiation," *J. Phys. Chem. Solids*, vol. 173, p. 111096, Feb. 2023, doi: 10.1016/j.jpcs.2022.111096.

[663] Q. Wang *et al.*, "Morphology-engineered carbon quantum dots embedded on octahedral CdIn2S4 for enhanced photocatalytic activity towards pollutant degradation and hydrogen evolution," *Environ. Res.*, vol. 209, p. 112800, Jun. 2022, doi: 10.1016/j.envres.2022.112800.

[664] J. Huang, L. Li, J. Chen, F. Ma, and Y. Yu, "Broad spectrum response flower spherical-like composites CQDs@CdIn2S4/CdS modified by CQDs with up-conversion property for photocatalytic degradation and water splitting," *Int. J. Hydrog. Energy*, vol. 45, no. 3, pp. 1822–1836, Jan. 2020, doi: 10.1016/j.ijhydene.2019.11.078.

[665] J. Chen, C.-J. Mao, H. Niu, and J.-M. Song, "Synthesis of novel C-doped g-C3N4 nanosheets coupled with CdIn2S4 for enhanced photocatalytic hydrogen evolution," *Beilstein J. Nanotechnol.*, vol. 10, no. 1, pp. 912–921, Apr. 2019, doi: 10.3762/bjnano.10.92.

[666] H. Qiu, W. Li, and X. Ma, "Construction of novel and noble-metal-free WC/CdIn2S4 Schottky heterojunction photocatalyst for efficient photocatalytic H2 evolution," *J. Taiwan Inst. Chem. Eng.*, vol. 156, p. 105392, Mar. 2024, doi: 10.1016/j.jtice.2024.105392.

[667] J. Xu *et al.*, "In Situ Photodeposition of Cobalt Phosphate (CoHxPOy) on CdIn2S4 Photocatalyst for Accelerated Hole Extraction and Improved Hydrogen Evolution," *Nanomaterials*, vol. 13, no. 3, Art. no. 3, Jan. 2023, doi: 10.3390/nano13030420.

[668] M. A. Hamza, J. D. Evans, G. G. Andersson, G. F. Metha, and C. J. Shearer, "Ultrathin Ru-CdIn2S4 nanosheets for simultaneous photocatalytic green hydrogen production and selective oxidation of furfuryl alcohol to furfural," *Chem. Eng. J.*, vol. 493, p. 152603, Aug. 2024, doi: 10.1016/j.cej.2024.152603.

[669] M. Li *et al.*, "Constructing CdIn2S4/ZnS type-I band alignment heterojunctions by decorating CdIn2S4 on ZnS microspheres for efficient photocatalytic H2 evolution," *Int. J. Hydrog. Energy*, vol. 48, no. 95, pp. 37224–37233, Dec. 2023, doi: 10.1016/j.ijhydene.2023.06.136.

[670] L. Xie *et al.*, "Construction of a Z-scheme CdIn2S4/ZnS heterojunction for the enhanced photocatalytic hydrogen evolution," *J. Alloys Compd.*, vol. 948, p. 169692, Jul. 2023, doi: 10.1016/j.jallcom.2023.169692.

[671] X. Liu, Z. Jiang, L. Xu, and C. Liu, "Enhanced photocatalytic hydrogen evolution over 0D/3D NiTiO3 nanoparticles/CdIn2S4 microspheres heterostructure photocatalyst," *Int. J. Hydrog. Energy*, vol. 48, no. 58, pp. 22079–22090, Jul. 2023, doi: 10.1016/j.ijhydene.2023.03.119.

[672] Y. Rao *et al.*, "Facile deposition of CoMoS4 on flower-like CdIn2S4 microspheres: Enhanced charge separation and efficient photocatalytic hydrogen evolution," *Int. J. Hydrog. Energy*, vol. 51, pp. 133–144, Jan. 2024, doi: 10.1016/j.ijhydene.2023.08.089.

[673] J. Teng, F. Li, T. Li, M. Huttula, and W. Cao, "Enhanced visible light-driven hydrogen evolution in non-precious metal Ni2P/CdIn2S4 S-type heterojunction via rapid interfacial charge transfer," *Mater. Today Adv.*, vol. 22, p. 100503, Jun. 2024, doi: 10.1016/j.mtadv.2024.100503.

[674] X. Wu, S. Lv, B. Jing, X. Liu, D. Wang, and C. Song, "A ternary rh/c-In2O3/CdIn2S4 heterostructure photocatalyst: In-situ construction and boosting high-efficient visible-light H2 production," *Inorganica Chim. Acta*, vol. 559, p. 121788, Jan. 2024, doi: 10.1016/j.ica.2023.121788.

[675] H. Liu *et al.*, "Efficient Z scheme-type II charge transfer on the interfaces of PAN/ZnO/CdIn2S4 for the enhanced photocatalytic hydrogen generation," *Int. J. Hydrog. Energy*, vol. 63, pp. 36–47, Apr. 2024, doi: 10.1016/j.ijhydene.2024.03.136.

[676] W. Yang, T. T. Fidelis, and W.-H. Sun, "Machine Learning in Catalysis, From Proposal to Practicing," *ACS Omega*, vol. 5, no. 1, pp. 83–88, Jan. 2020, doi: 10.1021/acsomega.9b03673.

[677] A. P. Dmitrieva *et al.*, "AI and ML for selecting viable electrocatalysts: progress and perspectives," *J. Mater. Chem. A*, 2024, doi: 10.1039/D4TA04991H.

[678] H. Xin, T. Mou, H. S. Pillai, S.-H. Wang, and Y. Huang, "Interpretable Machine Learning for Catalytic Materials Design toward Sustainability," *Acc. Mater. Res.*, vol. 5, no. 1, pp. 22–34, Jan. 2024, doi: 10.1021/accountsmr.3c00131.

[679] Z. Yang and W. Gao, "Applications of Machine Learning in Alloy Catalysts: Rational Selection and Future Development of Descriptors," *Adv. Sci.*, vol. 9, no. 12, p. 2106043, 2022, doi: 10.1002/advs.202106043.

[680] D. Chen, C. Shang, and Z.-P. Liu, "Machine-learning atomic simulation for heterogeneous catalysis," *Npj Comput. Mater.*, vol. 9, no. 1, pp. 1–9, Jan. 2023, doi: 10.1038/s41524-022-00959-5.

[681] G. Ramkumar *et al.*, "Enhanced machine learning for nanomaterial identification of photo thermal hydrogen production," *Int. J. Hydrog. Energy*, vol. 52, pp. 696–708, Jan. 2024, doi: 10.1016/j.ijhydene.2023.07.128.

[682] B. Arabacı, R. Bakır, C. Orak, and A. Yüksel, "Integrating experimental and machine learning approaches for predictive analysis of photocatalytic hydrogen evolution using Cu/g-C3N4," *Renew. Energy*, vol. 237, p. 121737, Dec. 2024, doi: 10.1016/j.renene.2024.121737.

[683] M. Y. Akram, B. Hu, J. Jia, C. Li, H. Dong, and H. Lu, "The science behind the scene: Theoretical and experimental foundations of metal sulfide photocatalyst modification," *Int. J. Hydrog. Energy*, vol. 93, pp. 21–42, Dec. 2024, doi: 10.1016/j.ijhydene.2024.10.359.

[684] M. F. C. Andrade, H.-Y. Ko, L. Zhang, R. Car, and A. Selloni, "Free energy of proton transfer at the water–TiO 2 interface from ab initio deep potential molecular dynamics," *Chem. Sci.*, vol. 11, no. 9, pp. 2335–2341, 2020, doi: 10.1039/C9SC05116C.

[685] M. I. Jordan and T. M. Mitchell, "Machine learning: Trends, perspectives, and prospects," *Science*, vol. 349, no. 6245, pp. 255–260, Jul. 2015, doi: 10.1126/science.aaa8415.

[686] H. Mai, T. C. Le, D. Chen, D. A. Winkler, and R. A. Caruso, "Machine Learning for Electrocatalyst and Photocatalyst Design and Discovery," *Chem. Rev.*, vol. 122, no. 16, pp. 13478–13515, Aug. 2022, doi: 10.1021/acs.chemrev.2c00061.

[687] C. Bie, L. Wang, and J. Yu, "Challenges for photocatalytic overall water splitting," *Chem*, vol. 8, no. 6, pp. 1567–1574, Jun. 2022, doi: 10.1016/j.chempr.2022.04.013.

[688] Q. Wang *et al.*, "Oxysulfide photocatalyst for visible-light-driven overall water splitting," *Nat. Mater.*, vol. 18, no. 8, pp. 827–832, Aug. 2019, doi: 10.1038/s41563-019-0399-z.

[689] V. Navakoteswara Rao *et al.*, "Metal chalcogenide-based core/shell photocatalysts for solar hydrogen production: Recent advances, properties and technology challenges," *J. Hazard. Mater.*, vol. 415, p. 125588, Aug. 2021, doi: 10.1016/j.jhazmat.2021.125588.

[690] G. Zhang, Z. Guan, J. Yang, Q. Li, Y. Zhou, and Z. Zou, "Metal Sulfides for Photocatalytic Hydrogen Production: Current Development and Future Challenges," *Sol. RRL*, vol. 6, no. 10, p. 2200587, 2022, doi: 10.1002/solr.202200587.

[691] A. B. *et al.*, "Challenges in photocatalytic hydrogen evolution: Importance of photocatalysts and photocatalytic reactors," *Int. J. Hydrog. Energy*, vol. 81, pp. 1442–1466, Sep. 2024, doi: 10.1016/j.ijhydene.2024.07.262.